\documentclass[twocolumn,showpacs,aps,floatfix,prd,nofootinbib]{revtex4}

\usepackage{graphicx}
\usepackage{dcolumn}
\usepackage{array, longtable}
\usepackage{epsfig}
\usepackage{amsmath}
\usepackage{colordvi}
\usepackage{hhline}

\newcommand{\BaBarYear}{11}
\newcommand{\BaBarNumber}{001}
\newcommand{\SLACPubNumber}{14403}

 \newcommand{\BaBarType}      {PUB}  

\input pubboard/babarsym

\def\Ecm       {\ensuremath {\rm E_{c.m.}}\xspace}
\def\KKKK      {\ensuremath {\Kp\Km\Kp\Km}\xspace}
\def\KKppch    {\ensuremath {\Kp\Km\pip\pim}\xspace}
\def\KKppnt    {\ensuremath {\Kp\Km\pi^0\pi^0}\xspace}

\def\Kppch     {\ensuremath {2K2\pi}\xspace}

\def\mKKpp     {\ensuremath {m(\Kp\Km\pi\pi)}\xspace}
\def\mKKpp0     {\ensuremath {m(\Kp\Km\piz\piz)}\xspace}
\def\chifourpi {\ensuremath {\chi^2_{4\pi}}\xspace}
\def\chiKKppch {\ensuremath {\chi^2_{2K2\pi}}\xspace}
\def\chiKKppnt {\ensuremath {\chi^2_{2K2\piz}}\xspace}
\def\chifourK  {\ensuremath {\chi^2_{4K}}\xspace}

\long\def\inst#1{\par\nobreak\kern 4pt\nobreak
    {\it #1}\par\vskip 10pt plus 3pt minus 3pt}

\begin{document}

\begin{flushleft}
arXiv:1103.3001v2\\
\babar-\BaBarType-\BaBarYear/\BaBarNumber \\
SLAC-PUB-\SLACPubNumber \\
Phys. Rev. {\bf D86}, 012008 (2012)
\end{flushleft}


\title{\large \bf
\boldmath
Cross Sections for the
Reactions $\epem\to K^+ K^- \pipi$, $K^+ K^-  \ppz$, and $K^+ K^- K^+ K^-$    
Measured using Initial-State Radiation Events 
} 

%
\author{J.~P.~Lees}
\author{V.~Poireau}
\author{E.~Prencipe}
\author{V.~Tisserand}
\affiliation{Laboratoire d'Annecy-le-Vieux de Physique des Particules (LAPP), Universit\'e de Savoie, CNRS/IN2P3,  F-74941 Annecy-Le-Vieux, France}
\author{J.~Garra~Tico}
\author{E.~Grauges}
\affiliation{Universitat de Barcelona, Facultat de Fisica, Departament ECM, E-08028 Barcelona, Spain }
\author{M.~Martinelli$^{ab}$}
\author{D.~A.~Milanes$^{a}$}
\author{A.~Palano$^{ab}$ }
\author{M.~Pappagallo$^{ab}$ }
\affiliation{INFN Sezione di Bari$^{a}$; Dipartimento di Fisica, Universit\`a di Bari$^{b}$, I-70126 Bari, Italy }
% ...
\collaboration{The \babar\ Collaboration}
\noaffiliation

\date{\today}

\begin{abstract}

We study the processes $\epem\to K^+ K^- \pipi\gamma$, 
$K^+K^-\ppz\gamma$, and $K^+ K^- K^+ K^-\gamma$,
where the photon is radiated from the initial state.  
About 84000, 8000, and 4200 fully-reconstructed events, respectively, 
are selected from 454~\invfb of \babar\ data. 
The invariant mass of the hadronic final state defines the \epem 
center-of-mass energy, 
so that the $K^+ K^- \pipi\gamma$ data can be compared with direct 
measurements of the $\epem\to K^+K^- \pipi$ reaction.
No direct measurements exist for the 
$\epem\to K^+ K^- \ppz$ or $\epem\to K^+ K^-  K^+ K^-$ reactions, and we present
an update of our previous result 
 based on a data sample that is twice as large.
Studying the structure of these events, we find contributions from a
number of intermediate states, and extract their cross sections.
In particular, we perform a more detailed study of the
$\epem\to\phi(1020)\pi\pi\gamma$ reaction, and confirm the presence of the
$Y(2175)$ resonance in the $\phi(1020) f_{0}(980)$ and $\Kp\Km
f_0(980)$ modes. 
In the charmonium region,
we observe the $J/\psi$ in all three final states and in several
intermediate states, 
as well as the $\psi(2S)$ in some modes,
and measure the corresponding products of branching fraction and electron width.  

\end{abstract}


\vfill
\maketitle

\setcounter{footnote}{0}

\section{Introduction}
\label{sec:Introduction}

Electron-positron annihilation at fixed center-of-mass (c.m.)~energies
has long been a mainstay of research in elementary particle physics.
The idea of utilizing initial-state radiation (ISR) to explore \epem 
reactions below the nominal c.m.~energies was outlined in 
Ref.~\cite{baier},
and discussed in the context of high-luminosity $\rm \phi$ and
$B$ factories in Refs.~\cite{arbus, kuehn, ivanch}.
At high c.m.\ energies, \epem annihilation is dominated by quark-level
processes producing two or more hadronic jets.
Low-multiplicity processes dominate 
below or around 2~\gev, and the region near the charm threshold, 3.0--4.5~\gev,
features a number of resonances~\cite{PDG}.
Thus, studies with ISR events allow us
 to probe a wealth of physics topics,
including cross sections, spectroscopy and form factors.
Charmonium and other states with $J^{PC}=1^{--}$ can be observed,
and intermediate states may contribute to the final state hadronic system.
Measurements of their decay modes and  branching fractions are important
to an understanding of the nature of such states.

Of particular current interest (see Ref.~\cite{y2175theory}) is 
the $Y(2175)$ state 
observed to decay to $\phi(1020)f_0(980)$ 
in our previous study~\cite{isr2k2pi}
and confirmed by the BES~\cite{y2175bes} and Belle~\cite{belle_phif0} Collaborations.
With twice the integrated luminosity 
(compared to Ref.~\cite{isr2k2pi})
in the present analysis,
we perform a more detailed study of this structure.

The study of $\epem\to\,$hadrons reactions in data is also critical to
hadronic-loop corrections to the muon magnetic anomaly, $a_\mu = (g_\mu-2)/2$.
The theoretical predictions of this anomaly rely on these measurements~\cite{dehz}.
Improving this prediction requires not only more precise measurements,
but also measurements from threshold to the highest c.m.~energy possible.
In addition,
all the important sub-processes should be studied
in order to properly incorporate possible
acceptance effects.
Events produced via ISR at $B$ factories provide independent and 
contiguous measurements of
hadronic cross sections from the production threshold to a
c.m.~energy of $\sim$5~\gev. With more data we also are able 
 to reduce systematic uncertainties in the 
cross section measurements.

The cross section for the radiation of a photon of energy $E_{\gamma}$
in the c.m. frame, 
followed by the production of a particular hadronic final state $f$, 
is related to the corresponding direct $\epem\to f$ cross section 
$\sigma_f(s)$ by
\begin{equation}
\frac{d\sigma_{\gamma f}(s_0,x)}{dx} = W(s_0,x)\cdot \sigma_f(s_0(1-x))\ ,
\label{eq1}
\end{equation}
where $\sqrt{s_0}$ is the nominal \epem c.m.\@ energy, 
$x\! =\! 2E_{\gamma}/\sqrt{s_0}$ is the fraction of the beam energy 
carried by the ISR photon, 
and $\Ecm \!\equiv\! \sqrt{s_0(1-{\it x})}\!\equiv\! \sqrt{\it s} $ is
the effective c.m.\@ energy  at   
which the final state $f$ is produced. 
The probability density function $W(s_0,x)$ for ISR photon emission has
been calculated with better than 1\% precision (see, e.g.\, Ref.~\cite{ivanch}).
It falls rapidly as $E_{\gamma}$ increases from zero, but has a long
tail, which in combination with
the increasing $\sigma_f(s_0(1-x))$ produces
a sizable event rate at very low \Ecm.
The angular distribution of the ISR photon peaks along the beam
directions. For a typical $e^+ e^-$
detector, around 10-15\% of the ISR photons fall within the
experimental acceptance~\cite{ivanch} .
 
Experimentally, the measured invariant mass of the hadronic final
state defines \Ecm.
An important feature of ISR data is that a wide range of energies is
scanned continuously in a single experiment, 
so that no structure is missed, and
the relative normalization uncertainties in
data from different experiments are avoided.
Furthermore, for large values of $x$ the hadronic system is collimated, 
reducing acceptance issues and allowing measurements down to 
production threshold.
The mass resolution is not as good as the typical beam energy spread used in 
direct measurements,
but resolution and absolute energy scale can be monitored
by means of the measured values of the width and mass of 
well-known resonances, such as the $J/\psi$
produced in the reaction $\epem \to J/\psi\gamma$. 
Backgrounds from $\epem \!\to\,$hadrons events at the nominal $\sqrt{s_0}$
and from other ISR processes can be 
suppressed by a combination of particle identification and 
kinematic fitting techniques.
Studies of $\epem\to\mumu\gamma$ and several multi-hadron ISR processes using
\babar\ data have been performed~
\cite{isr2k2pi,Druzhinin1,isr3pi,isr4pi,isr6pi,isr5pi,isrkkpi,isr2pi},
demonstrating the viability of such measurements.
These analyses have led to improvements in 
background reduction procedures for more rare ISR processes.

The $K^+K^- \pipi$ final state has been measured directly by the
DM1 Collaboration~\cite{2k2pidm1} for $\sqrt{s} <\! 2.2~\gev$,
and we have previously published ISR measurements of the
$K^+K^- \pipi$ and $K^+K^-K^+K^-$ final states~\cite{isr4pi} for
$\Ecm \!<\! 4.5~\gev$.
Later we reported an updated measurement of the  
$K^+K^- \pipi$ final state with a larger data sample, together with
the first measurement of the $K^+K^- \ppz$ final state, in which we
observed a structure near threshold in the $\phi f_0$ intermediate state~\cite{isr2k2pi}.

In this paper we present a more detailed study of these two final states 
along with an updated measurement of the $K^+K^-K^+K^-$ final state.
In all cases we require the detection of the ISR photon and perform a set
of kinematic fits.
We are able to suppress backgrounds sufficiently to study these
final states from their respective production thresholds up to \Ecm=5~\gev.
In addition to measuring the overall cross sections, 
we study the internal structure of the final states
and measure cross sections for a number of 
intermediate states that contribute to them.
We also study the charmonium region, 
measure several $J/\psi$ and $\psi(2S)$ products of branching fraction 
and electron width,
and set limits on other states.

\section{\boldmath The \babar\ detector and dataset}
\label{sec:babar}

The data used in this analysis were collected with the \babar\ detector at
the \pep2\ asymmetric-energy \epem\ storage rings at the 
SLAC National Accelerator Laboratory.
The total integrated luminosity used is 454.2~\invfb, 
which includes 413.1~\invfb collected at the $\Upsilon(4S)$ peak, 
$\sqrt{s_0}=10.58~\gev$, 
and 41.1~\invfb collected at about $\sqrt{s_0}=10.54~\gev$.

The \babar\ detector is described elsewhere~\cite{babar}. 
In the present work, we use charged-particle tracks reconstructed in the tracking system,
which is comprised of a five double-sided-layer silicon vertex tracker (SVT) 
and a 40-layer drift chamber (DCH) in a 1.5 T axial magnetic field.
Separation of charged pions, kaons, and protons is achieved using a
combination of Cherenkov angles measured in the detector of internally-reflected 
Cherenkov light (DIRC) and specific-ionization measurements in
the SVT and DCH. 
For the present study we use a kaon identification algorithm that provides 90--95\% 
efficiency, depending on momentum, and pion and proton rejection
factors in the 20--100 range.
Photon and electron energies are measured in a CsI(Tl)
electromagnetic calorimeter (EMC).
We use muon identification provided by an instrumented flux return (IFR)
to select the $\mumu\gamma$ final state used for photon 
efficiency studies.

To study the detector acceptance and efficiency, 
we use a simulation package developed for radiative processes.
The simulation of hadronic final states, including
$K^+K^- \pipi \gamma$, $K^+K^- \ppz\gamma$ and $K^+K^- K^+K^-\gamma$,
is based on the approach suggested by Czy\.z and K\"uhn \cite{kuehn2}.  
Multiple soft-photon emission from the initial-state charged particles is
implemented with a structure-function technique~\cite{kuraev, strfun},
and photon radiation from the final-state particles is simulated by
the PHOTOS package~\cite{PHOTOS}.  
The precision of the radiative corrections is about 1\%~\cite{kuraev, strfun}.

We simulate the two $K^+K^-\pi\pi$ ($\pipi, \ppz$) final states
uniformly in phase 
space, and also according to models that include the $\phi(1020)\to K^+K^-$ and/or
$f_{0}(980)\to \pi\pi$ channels.
The $K^+K^- K^+K^-$ final state is simulated according to phase space, and
also including the $\phi \to K^+K^-$ channel.
The generated events are subjected to
a detailed detector simulation~\cite{GEANT4}, 
and we reconstruct them with the same software chain used for the experimental data. 
Variations in detector and background conditions 
over the course of the experiment are taken into account.

We also generate a large number of potential background processes, 
including the ISR reactions
$\epem\to\pipi\pipi\gamma$, $\epem\to\pipi\ppz\gamma$, and $\epem\to K_S K\pi\gamma$,
which can contribute due to particle misidentification.
We also simulate 
$\epem\to\phi\eta\gamma$, $\epem\to\phi\pi^0\gamma$, and $\epem\to\pipi\pi^0\gamma$,
which have larger cross sections and can contribute background via missing or
spurious tracks or photons.
In addition, we study non-ISR backgrounds resulting from
$\epem \!\!\to\! q \qbar$ $(q = u, d, s, c)$ generated using
JETSET~\cite{jetset} and from
$\epem \!\!\to\! \tau^+\tau^-$ generated using KORALB~\cite{koralb}. 
The cross sections for these processes are known to about 10\% accuracy
or better, which is sufficiently precise for the purposes of the
measurements in this paper.
The contribution from 
\Y4S decays is found to be negligible.

\section{\boldmath Event Selection and Kinematic Fit}
\label{sec:Fits}

In the selection of candidate events, we consider
photon candidates in the EMC with energy above 0.03~\gev,
and charged-particle tracks reconstructed in either or both of the DCH
and SVT, that 
extrapolate within 0.25 cm of the collision
 axis in the transverse plane and within 3 cm of the nominal collision point
along this axis.
We require a photon with c.m.~energy 
$E_\gamma > 3~\gev$ in each event,
and either four charged-particle tracks with zero net charge 
and total momentum roughly (within 0.3 radians) opposite to the photon direction,
or two oppositely-charged tracks that combine with
other  photons to roughly balance the high-energy photon momentum.
We assume that the photon with the largest value of $E_\gamma$ 
is the ISR photon.
We fit the set of charged-particle tracks to a common vertex
and use this as the point
of origin in calculating the photon direction(s).
If additional well-reconstructed tracks exist, the nearest 
four (two) to the interaction region are chosen for the four-track (two-track) analysis.
Most events contain additional soft photons due to 
machine background or interactions in the detector material.

We subject each candidate event to a set of constrained 
kinematic fits and use the fit results,
along with charged-particle identification,
both to select the final states of interest and to measure backgrounds
from other processes.
The kinematic fits use the ISR photon direction and energy along with the
four-momenta and covariance matrices of the initial \epem and the
set of selected tracks and photons.
The ISR photon energy and position are additionally aligned 
and calibrated using the $\mumu\gamma$ ISR process, since the two  
well-identified muons predict precisely 
the position and energy of the photon. This
process is also used to identify and measure
data - Monte Carlo (MC) simulation differences in the photon detection efficiency and 
resolution.
The fitted three-momentum for each charged-particle track and the photon are used 
in further kinematical calculations.

For the four-track event candidates the fits have four constraints (4C).
We first fit to the $\pipi\pipi$ hypothesis, obtaining 
the chi-squared value \chifourpi.
If the four tracks include one identified $K^+$ and one identified $K^-$,
we fit to the $K^+K^-\pipi$ hypothesis and retain the event as a
\KKppch candidate.
For events with one identified kaon, we perform fits with each of
the two oppositely charged tracks given the kaon hypothesis, and the
combination with the lower \chiKKppch is retained if its value is less
than $\chifourpi$.
If the event contains three or four identified $K^\pm$, 
we fit to the $K^+K^-K^+K^-$ hypothesis and retain the event as a 
\KKKK candidate with chi-squared value \chifourK.

For the events with two charged-particle tracks and five or more photon candidates, 
we require that both tracks be identified as kaons
to suppress background from ISR $\pipi\ppz$ and $K^\pm\KS\pi^\mp$ events.
We then pair all non-ISR photon candidates and consider combinations with
invariant mass within $\pm$30~\mevcc of the \piz mass~\cite{PDG} as \piz candidates.
We perform a six-constraint (6C) fit to each set of two non-overlapping 
\piz candidates, the ISR photon, the two charged-particle tracks, and
the beam particles.
Both \piz candidates are constrained to the \piz mass, 
and we retain the combination with the lowest chi-squared value, 
\chiKKppnt.

\section{The {\boldmath $K^+ K^- \pipi$} final state}
\subsection{Final Selection and Backgrounds}
\label{sec:selection1}

\begin{figure}[t]
\includegraphics[width=0.9\linewidth]{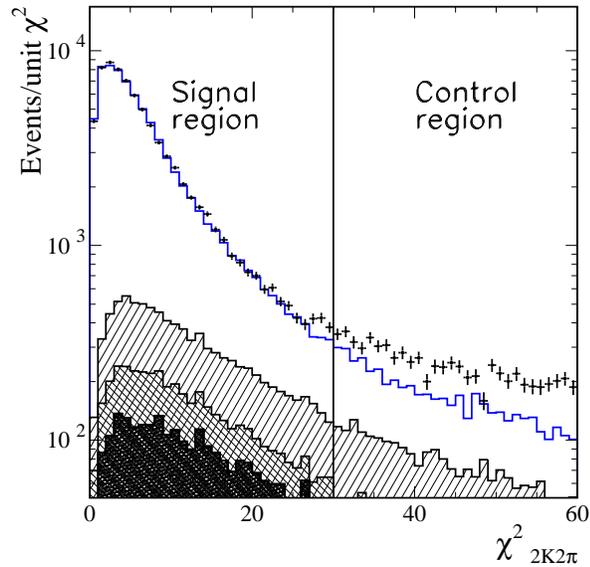}
\vspace{-0.4cm}
\caption{
  Distribution of \chisq from the four-constraint fit for \KKppch candidates
  in the data (points).
  The open histogram is the distribution for simulated signal events, 
  normalized as described in the text.
  The shaded, cross-hatched, and hatched regions
  represent, respectively, the background from  non-ISR events,
from the ISR $K_S K\pi$ process, and backgrounds with
dominant contribution from mis-identified ISR $4\pi$ events.
Signal and control regions are indicated.
}
\label{2k2pi_chi2_all}
\end{figure}

The \chiKKppch distribution in data for the
\KKppch candidates is shown in Fig.~\ref{2k2pi_chi2_all} (points);
the open histogram is the distribution for the simulated \KKppch events.
The distributions are broader than those for a typical 4C \chisq
distribution due to higher order ISR, and
the experimental distribution has contributions from background processes.
The simulated distribution is normalized to the data in the region 
$\chiKKppch\!\! <\! 10$ where the contributions of the
backgrounds and radiative corrections do not exceed 10\%.

The shaded histogram in Fig.~\ref{2k2pi_chi2_all} represents
the background from non-ISR $\epem \!\!\to\! \qqbar$ events
obtained from the JETSET simulation. 
It is dominated by events with a hard $\piz$ that results in a fake ISR
photon. These events otherwise have kinematics similar to the signal,   
resulting in the peaking structure at low values of \chiKKppch. 
We evaluate this background in a number of \Ecm ranges by
combining the ISR photon candidate with another photon candidate in 
both data and simulated events,
and comparing the \piz signals in the resulting $\gamma\gamma$
invariant mass distributions.
The simulation gives an \Ecm-dependence consistent with the data,
so we normalize it using an overall factor.
The cross-hatched region in Fig.~\ref{2k2pi_chi2_all} represents 
$\epem \!\!\to\! K_S K\pi\gamma$ events with $K_S\to\pipi$ decays
close to the interaction region, and one pion mis-identified as a kaon. The process
has similar kinematics to the signal process, and a contribution of about
1\% is estimated using the cross
section measured in our previous study~\cite{isrkkpi}.
The hatched region represents the contribution from
ISR $\epem \!\!\to\! \pipi\pipi$ events with one or two misidentified 
pions; this process contributes mainly at low \chisq values.
We estimate the contribution as a function of \Ecm from a
simulation using the cross section value and shape
from our previous study~\cite{isr4pi}.

All remaining background sources are either negligible or give a
\chiKKppch distribution that is nearly uniform over the range
shown in Fig.~\ref{2k2pi_chi2_all}.
We define the signal region by requiring $\chiKKppch\!\! <\! 30$,
and estimate the sum of the remaining backgrounds from the difference
between the number of data and simulated entries in the control region,
$30\! <\!\chiKKppch\!\! <\!60$, as shown in
Fig.~\ref{2k2pi_chi2_all}.
 The background contribution to any distribution other than \chisq is
estimated as the difference between the distributions in the relevant
quantity for data and MC events from the control region of Fig.~\ref{2k2pi_chi2_all},
normalized to the difference between the number of data and MC events
in the signal region. The non-ISR background is subtracted separately.
The signal region contains 85598 data and 63784 simulated events; 
the control region contains 9684 data and 4315 simulated events.

\begin{figure}[tbh]
\includegraphics[width=0.9\linewidth]{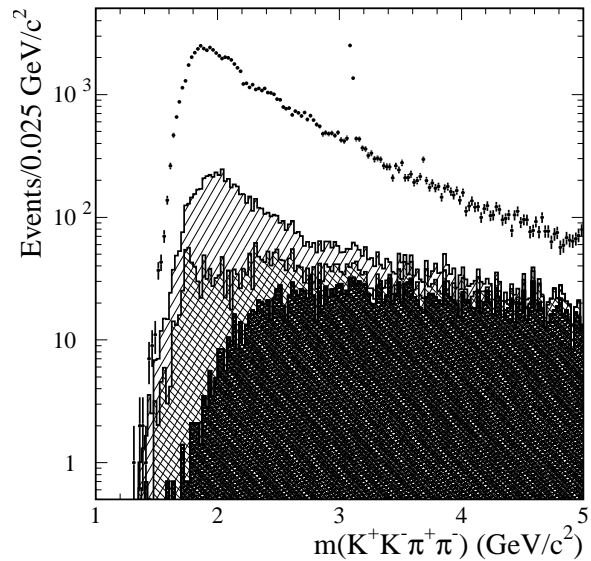}
\vspace{-0.4cm}
\caption{
  The invariant mass distribution for \KKppch candidates in the data (points):
the shaded, cross-hatched and hatched regions show, respectively,  
the non-ISR background from JETSET simulation, the $K_S K\pi$ 
background with a small
contribution from the control 
region of Fig.~\ref{2k2pi_chi2_all}, and the  
dominant
  contribution resulting from ISR mis-identified $\pipi\pipi$ events.
  }
\label{2k2pi_babar}
\end{figure}

Figure~\ref{2k2pi_babar} shows the \KKppch invariant mass distribution 
from threshold up to 5.0~\gevcc for events in the signal region.
Narrow peaks are apparent at the $J/\psi$ and $\psi(2S)$ masses.
The shaded histogram represents the \qqbar background,
which is negligible at low mass but dominates at higher masses.
The cross-hatched region represents the background from
the $K_S K\pi$ channel (which exhibits a $\phi(1680)$ peak~\cite{isrkkpi}) and 
from the \chisq control region.
The hatched region represents the contribution from
mis-identified ISR $\pipi\pipi$, and is dominant for masses below 3.0~\gevcc.
The total background is 6--8\% at low mass, but accounts for 20-25\% of the
observed distribution near 4 \gevcc,
and increases further for higher masses.

We subtract the sum of backgrounds in each mass interval to obtain the number
of signal events.
Considering uncertainties in the cross sections for the background processes, 
the normalization of events in the control region, and 
the simulation statistics,
we estimate a systematic uncertainty on the signal yield that is
2\% or less in the 1.6--3.3~\gevcc mass region, but 
increases linearly to 10\% in the 3.3-5.0~\gevcc region, and is about
20\% for the masses below 1.6~\gevcc.

\subsection{Selection Efficiency}
\label{sec:eff1}

\begin{figure}[t]
\includegraphics[width=0.9\linewidth]{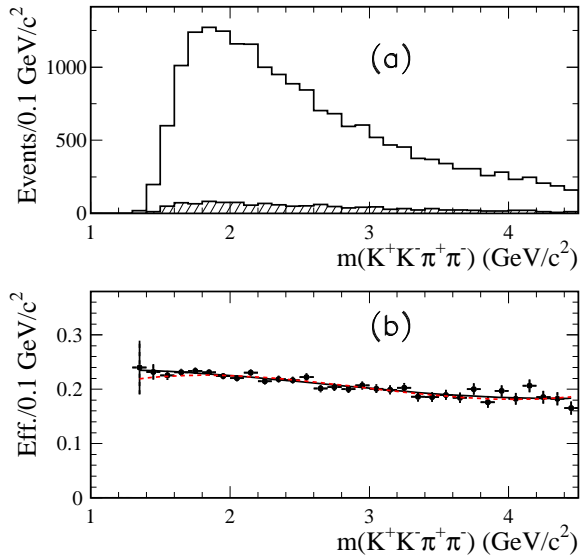}
\vspace{-0.4cm}
\caption{
  (a) 
The invariant mass distributions for \KKppch MC events that are
simulated uniformly in phase space, reconstructed
in the signal (open) and control 
  (hatched) regions of Fig.~\ref{2k2pi_chi2_all};
  (b) net reconstruction and selection efficiency as a function of
  mass obtained from this simulation 
  (the curve represents a third-order polynomial fit). The dashed curve is obtained
 for the $\phi(1020)\pipi$ final state.
}
\label{mc_acc1}
\end{figure}

The selection procedure applied to the data is also applied to the         
simulated signal samples.
The resulting \KKppch invariant-mass distributions in the signal and
control regions are shown in Fig.~\ref{mc_acc1}(a) for the uniform phase space 
simulation. This model reproduces the observed distributions of kaon and pion
momenta and polar angles.
A broad, smooth mass distribution is chosen to facilitate the estimation 
of the efficiency as a function of mass.
We divide the number of reconstructed simulated events in each   
mass interval by the number generated in that interval to obtain the
efficiency shown by the points in Fig.~\ref{mc_acc1}(b).
The result of fitting a third-order polynomial to the points
is used for further calculations.
We simulate events with the ISR photon confined to the
angular range 20--160$^\circ$ with respect to the electron beam in the
\epem c.m.\ frame;  
this angular range is wider than the actual EMC acceptance.
The calculated efficiency is for this fiducial region, 
and includes the acceptance for the final-state hadrons, 
the inefficiencies of the detector subsystems, 
and event loss due to additional soft-photon emission.

The simulations including the $\phi(1020)\pipi$ and/or $\Kp\Km f_0(980)$
channels give very different mass and angular distributions in the  
\KKppch rest frame.
However, the angular acceptance is quite uniform for ISR events (see
Ref.~\cite{isr4pi}), and the efficiencies are within $1\%$ of those 
from the uniform phase space simulation, as shown
by the dashed curve in Fig.~\ref{mc_acc1}(b) for the
$\phi(1020)\pipi$ final state.

To study possible mis-modeling of the acceptance, we repeat the analysis
with tighter requirements. 
All charged tracks are required to lie within
the DIRC acceptance, $0.45 \!<\! \theta_{\rm ch} \!<\! 2.4$ radians,
and the ISR photon must not appear near the edges of the EMC,
$0.35 \!<\! \theta_{\rm ISR} \!<\! 2.4$ radians.
The fraction of selected data events satisfying the tighter
requirements differs from the simulated ratio by 1.5\%.
We take the sum in quadrature of this variation and the
1\% model variation (2\% total) 
as the systematic uncertainty due to acceptance and model dependence.

\begin{table*}
\caption{Summary of the cross section measurements for $\ep\en\to K^+ K^- \pipi$. 
Errors are statistical only.}
\label{2k2pi_tab}
\begin{ruledtabular}
\begin{tabular}{ c c c c c c c c }
\Ecm (GeV) & $\sigma$ (nb)  
& \Ecm (GeV) & $\sigma$ (nb) 
& \Ecm (GeV) & $\sigma$ (nb) 
& \Ecm (GeV) & $\sigma$ (nb)  
\\
\hline

 1.4125 & 0.000 $\pm$ 0.004 & 2.3125 & 1.531 $\pm$ 0.056 & 3.2125 & 0.357 $\pm$ 0.025 & 4.1125 & 0.082 $\pm$ 0.011 \\
 1.4375 & 0.009 $\pm$ 0.008 & 2.3375 & 1.586 $\pm$ 0.056 & 3.2375 & 0.328 $\pm$ 0.023 & 4.1375 & 0.078 $\pm$ 0.011 \\
 1.4625 & 0.018 $\pm$ 0.008 & 2.3625 & 1.496 $\pm$ 0.055 & 3.2625 & 0.339 $\pm$ 0.023 & 4.1625 & 0.065 $\pm$ 0.010 \\
 1.4875 & 0.014 $\pm$ 0.010 & 2.3875 & 1.574 $\pm$ 0.055 & 3.2875 & 0.304 $\pm$ 0.022 & 4.1875 & 0.079 $\pm$ 0.010 \\
 1.5125 & 0.075 $\pm$ 0.017 & 2.4125 & 1.427 $\pm$ 0.053 & 3.3125 & 0.292 $\pm$ 0.022 & 4.2125 & 0.082 $\pm$ 0.011 \\
 1.5375 & 0.078 $\pm$ 0.018 & 2.4375 & 1.407 $\pm$ 0.052 & 3.3375 & 0.295 $\pm$ 0.021 & 4.2375 & 0.065 $\pm$ 0.010 \\
 1.5625 & 0.135 $\pm$ 0.022 & 2.4625 & 1.353 $\pm$ 0.051 & 3.3625 & 0.257 $\pm$ 0.020 & 4.2625 & 0.071 $\pm$ 0.009 \\
 1.5875 & 0.297 $\pm$ 0.030 & 2.4875 & 1.221 $\pm$ 0.048 & 3.3875 & 0.242 $\pm$ 0.020 & 4.2875 & 0.075 $\pm$ 0.010 \\
 1.6125 & 0.550 $\pm$ 0.040 & 2.5125 & 1.203 $\pm$ 0.047 & 3.4125 & 0.245 $\pm$ 0.020 & 4.3125 & 0.076 $\pm$ 0.010 \\
 1.6375 & 0.975 $\pm$ 0.053 & 2.5375 & 1.020 $\pm$ 0.044 & 3.4375 & 0.199 $\pm$ 0.018 & 4.3375 & 0.061 $\pm$ 0.009 \\
 1.6625 & 1.363 $\pm$ 0.061 & 2.5625 & 0.991 $\pm$ 0.043 & 3.4625 & 0.254 $\pm$ 0.019 & 4.3625 & 0.060 $\pm$ 0.009 \\
 1.6875 & 1.808 $\pm$ 0.069 & 2.5875 & 0.986 $\pm$ 0.043 & 3.4875 & 0.212 $\pm$ 0.019 & 4.3875 & 0.068 $\pm$ 0.009 \\
 1.7125 & 2.291 $\pm$ 0.078 & 2.6125 & 0.837 $\pm$ 0.040 & 3.5125 & 0.265 $\pm$ 0.020 & 4.4125 & 0.041 $\pm$ 0.008 \\
 1.7375 & 2.500 $\pm$ 0.083 & 2.6375 & 0.925 $\pm$ 0.041 & 3.5375 & 0.176 $\pm$ 0.018 & 4.4375 & 0.062 $\pm$ 0.009 \\
 1.7625 & 3.376 $\pm$ 0.094 & 2.6625 & 0.886 $\pm$ 0.040 & 3.5625 & 0.186 $\pm$ 0.017 & 4.4625 & 0.065 $\pm$ 0.009 \\
 1.7875 & 3.879 $\pm$ 0.099 & 2.6875 & 0.839 $\pm$ 0.038 & 3.5875 & 0.190 $\pm$ 0.018 & 4.4875 & 0.053 $\pm$ 0.008 \\
 1.8125 & 4.160 $\pm$ 0.101 & 2.7125 & 0.902 $\pm$ 0.039 & 3.6125 & 0.170 $\pm$ 0.016 & 4.5125 & 0.047 $\pm$ 0.008 \\
 1.8375 & 4.401 $\pm$ 0.103 & 2.7375 & 0.768 $\pm$ 0.037 & 3.6375 & 0.173 $\pm$ 0.016 & 4.5375 & 0.055 $\pm$ 0.008 \\
 1.8625 & 4.630 $\pm$ 0.105 & 2.7625 & 0.831 $\pm$ 0.038 & 3.6625 & 0.195 $\pm$ 0.017 & 4.5625 & 0.041 $\pm$ 0.007 \\
 1.8875 & 4.219 $\pm$ 0.101 & 2.7875 & 0.752 $\pm$ 0.036 & 3.6875 & 0.272 $\pm$ 0.019 & 4.5875 & 0.028 $\pm$ 0.008 \\
 1.9125 & 4.016 $\pm$ 0.098 & 2.8125 & 0.689 $\pm$ 0.034 & 3.7125 & 0.161 $\pm$ 0.016 & 4.6125 & 0.050 $\pm$ 0.007 \\
 1.9375 & 4.199 $\pm$ 0.099 & 2.8375 & 0.644 $\pm$ 0.033 & 3.7375 & 0.147 $\pm$ 0.015 & 4.6375 & 0.033 $\pm$ 0.007 \\
 1.9625 & 3.942 $\pm$ 0.095 & 2.8625 & 0.555 $\pm$ 0.031 & 3.7625 & 0.156 $\pm$ 0.015 & 4.6625 & 0.052 $\pm$ 0.008 \\
 1.9875 & 3.611 $\pm$ 0.091 & 2.8875 & 0.559 $\pm$ 0.031 & 3.7875 & 0.133 $\pm$ 0.015 & 4.6875 & 0.043 $\pm$ 0.006 \\
 2.0125 & 3.403 $\pm$ 0.088 & 2.9125 & 0.543 $\pm$ 0.030 & 3.8125 & 0.143 $\pm$ 0.015 & 4.7125 & 0.039 $\pm$ 0.006 \\
 2.0375 & 3.112 $\pm$ 0.085 & 2.9375 & 0.550 $\pm$ 0.030 & 3.8375 & 0.112 $\pm$ 0.013 & 4.7375 & 0.027 $\pm$ 0.006 \\
 2.0625 & 3.249 $\pm$ 0.085 & 2.9625 & 0.508 $\pm$ 0.030 & 3.8625 & 0.121 $\pm$ 0.015 & 4.7625 & 0.032 $\pm$ 0.006 \\
 2.0875 & 3.165 $\pm$ 0.083 & 2.9875 & 0.549 $\pm$ 0.030 & 3.8875 & 0.135 $\pm$ 0.014 & 4.7875 & 0.035 $\pm$ 0.006 \\
 2.1125 & 3.036 $\pm$ 0.080 & 3.0125 & 0.468 $\pm$ 0.028 & 3.9125 & 0.126 $\pm$ 0.013 & 4.8125 & 0.019 $\pm$ 0.006 \\
 2.1375 & 2.743 $\pm$ 0.077 & 3.0375 & 0.461 $\pm$ 0.027 & 3.9375 & 0.114 $\pm$ 0.013 & 4.8375 & 0.022 $\pm$ 0.006 \\
 2.1625 & 2.499 $\pm$ 0.073 & 3.0625 & 0.476 $\pm$ 0.028 & 3.9625 & 0.130 $\pm$ 0.013 & 4.8625 & 0.028 $\pm$ 0.006 \\
 2.1875 & 2.351 $\pm$ 0.070 & 3.0875 & 3.057 $\pm$ 0.065 & 3.9875 & 0.099 $\pm$ 0.012 & 4.8875 & 0.028 $\pm$ 0.005 \\
 2.2125 & 1.785 $\pm$ 0.062 & 3.1125 & 1.561 $\pm$ 0.048 & 4.0125 & 0.117 $\pm$ 0.013 & 4.9125 & 0.030 $\pm$ 0.005 \\
 2.2375 & 1.833 $\pm$ 0.061 & 3.1375 & 0.449 $\pm$ 0.028 & 4.0375 & 0.075 $\pm$ 0.011 & 4.9375 & 0.028 $\pm$ 0.005 \\
 2.2625 & 1.641 $\pm$ 0.059 & 3.1625 & 0.455 $\pm$ 0.027 & 4.0625 & 0.090 $\pm$ 0.011 & 4.9625 & 0.030 $\pm$ 0.005 \\
 2.2875 & 1.762 $\pm$ 0.059 & 3.1875 & 0.385 $\pm$ 0.025 & 4.0875 & 0.099 $\pm$ 0.012 & 4.9875 & 0.037 $\pm$ 0.005 \\

\end{tabular}
\end{ruledtabular}
\end{table*}

\begin{figure}[tbh]
\includegraphics[width=0.9\linewidth]{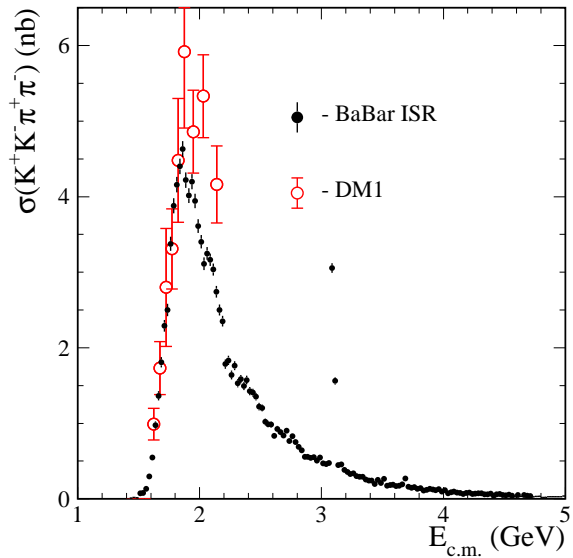}
\vspace{-0.4cm}
\caption{
   The $\epem \!\!\to\! \KKppch$ cross section as a function of  
   \epem c.m.\@ energy measured with ISR data at \babar\ (dots).
   The direct measurements from DM1~\cite{2k2pidm1} are shown as the
   open circles. 
   Only statistical errors are shown.
}
\label{2k2pi_ee_babar}
\end{figure} 
Our data sample contains about 3000 events in the $J/\psi$ peak.
Comparing this number 
with and without selection on \chiKKppch we find less than a 1\% 
difference between
data and MC simulation due to mis-modeling of the shape of 
the \chiKKppch distribution. This value is taken as an estimate of the 
systematic uncertainty associated with 
the \chiKKppch selection criterion.
To measure tracking efficiency, we consider data and simulated events
that contain a high-energy photon and exactly three charged-particle
tracks, which satisfy a set of kinematical criteria, 
including a good $\chi^2$ from a kinematic fit to the $\pipi\pipi$ hypothesis,
assuming one missing pion track in the event.
We find that the simulated track-finding efficiency is overestimated by
$(0.75\pm0.25)\%$ per track, so we apply a correction of $+(3\pm1)\%$ 
to the signal yield.

The kaon identification efficiency is studied in \babar~ using many different test
processes (e.g.\, $\epem \!\!\to\! \phi(1020)\gamma \!\to\! \Kp\Km\gamma$), and we
conservatively estimate a systematic uncertainty of  $\pm 1.0$\% per kaon due to
data-MC differences in our kaon momentum range.

The data-MC simulation correction due to ISR photon detection efficiency 
was studied with a sample of $\epem\to\mumu\gamma$ events and
was found to be $+(1.0\pm0.5)\%$.

\begin{table}[b]
\caption{
Summary of corrections and systematic uncertainties 
for the $\epem \!\!\to\! \KKppch$  cross section measurements.
The total correction is the linear sum of the contributions, and the
total uncertainty is obtained by summing the individual uncertainties in quadrature.
  }
\label{error_tab}
\begin{ruledtabular}
\begin{tabular}{l c r@{}l} 
     Source             & Correction & \multicolumn{2}{c}{Uncertainty}\\
\hline
                        &            &     &                \\[-0.2cm]
Rad. Corrections        &  --        & 1\% &              \\
Backgrounds             &  --        & 2\% &, $\Ecm <3.3~\gev$ \\
                        &            & 2-10 \%&, $\Ecm >3.3~\gev$ \\
Model Acceptance        &  --        & 2\% &             \\
\chiKKppch Distribution &  --        & 1\% &             \\ 
Tracking Efficiency     & +3\%       & 1\% &             \\
Kaon ID Efficiency      &  --        & 2\% &             \\
Photon Efficiency       & +1.0\%     & 0.5\% &            \\ 
ISR Luminosity          &  --        & 1\% &             \\[0.1cm]
\hline
                        &            &     &               \\[-0.2cm]
Total                   &   +4.0\%   & 4\% &, $\Ecm <3.3~\gev$ \\
                        &            & 4-11\% &, $\Ecm >3.3~\gev$ \\
\end{tabular}
\end{ruledtabular}
\end{table}

\subsection{\boldmath Cross Section for $\epem \!\to \KKppch$}
\label{sec:xs2k2pi}

We calculate the $\epem \!\!\to\! \KKppch$  cross section as a
function of the effective c.m.\ energy from
\begin{equation}
    \sigma_{\Kppch}(\Ecm)
  = \frac{{\it dN}_{\Kppch\gamma}(\Ecm)}
         {{\it d}{\cal L}(\Ecm) \cdot \epsilon_{\Kppch}(\Ecm)
  \cdot{\it R}(\Ecm)}\ ,
\label{xseqn}
\end{equation}
where 
$\Ecm \equiv {\it m}_{\Kppch}{\it c}^2$ with 
$m_{\Kppch}$ the measured invariant mass of the \KKppch system,
$dN_{\Kppch\gamma}$  the number of selected events after background 
subtraction in the interval $d\Ecm$, 
$\epsilon_{\Kppch}(\Ecm)$  the corrected detection efficiency,
and $R$  a radiative correction.

We calculate the differential luminosity $d{\cal L}(\Ecm)$
in each interval $d\Ecm$, with the photon in the same fiducial range
as that used for the simulation, using the simple leading order (LO) formula
described in Ref.~\cite{isr3pi}. From the mass spectra, obtained from the
MC simulation with and without extra-soft-photon (ISR and FSR) radiation, we
extract $R(\Ecm)$, which gives a correction less than 1\%.
Our data, calculated according to Eq.~\ref{xseqn}, include vacuum
polarization
(VP) and exclude any radiative effects, as is conventional for the reporting
of \epem cross sections.  Note that VP should be excluded and FSR included
for calculations of $a_\mu$.
From data-simulation comparisons for the $\epem\to\mumu\gamma$ events
  we estimate a systematic uncertainty on $d{\cal L}$ of 1\%~\cite{isr2pi}. 

We show the cross section as a function of $\Ecm$ in
Fig.~\ref{2k2pi_ee_babar} with statistical errors only
in comparison with the direct measurements from DM1~\cite{2k2pidm1}, 
and list our results in Table~\ref{2k2pi_tab}.  
The results are consistent
with our previous measurements for this 
reaction~\cite{isr4pi,isr2k2pi}, but have increased statistical precision.
Our data  lie systematically below the DM1 data
for \Ecm above 1.9~\gev.
The systematic uncertainties, summarized in Table~\ref{error_tab},
affect the normalization, but have little effect on the energy dependence.

The cross section rises from threshold to a peak value of about 4.6~nb near 
1.86~\gev, then generally decreases with increasing energy.
In addition to narrow peaks at the $J/\psi$ and $\psi(2S)$ mass values,
there are several possible wider structures in the 1.8--2.8~\gev
region.
Such structures might be due to thresholds for
intermediate resonant states, such as $\phi f_0(980)$ near 2~\gev.
Gaussian fits to the distributions of the mass difference between generated 
and reconstructed MC data yield \KKppch mass resolution values 
that vary from 4.2~\mevcc in the
1.5--2.5~\gevcc region to 5.5~\mevcc in the 2.5--3.5~\gevcc region. 
The resolution functions are not purely Gaussian due to soft-photon
radiation, 
but less than 10\% of the signal is outside the 0.025~\gevcc mass
interval used in Fig.~\ref{2k2pi_ee_babar}.
Since the cross section has no sharp structure other than the $J/\psi$
and $\psi(2S)$ peaks discussed in Sec.~\ref{sec:charmonium} below, 
we apply no correction for mass resolution.

\subsection{\boldmath Substructures in the \KKppch Final State}
\label{sec:kaons}

\begin{figure}[tbh]
\includegraphics[width=1.0 \linewidth]{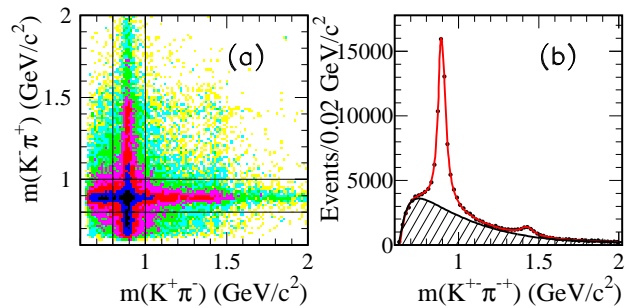}
\vspace{-0.2cm}
\caption{
  (a) Invariant mass of the $\Km\pip$ pair versus that of the $\Kp\pim$ pair;
(b)   The $K^\pm\pi^\mp$ mass distribution (two entries per event) for all
  selected \KKppch events:
  the solid line represents a fit including two resonances and a
  polynomial background function, which is
  shown as the hatched region.
  }
\label{kkstar}
\end{figure}

Our previous study~\cite{isr4pi,isr2k2pi} showed evidence for
 many intermediate resonances in the \KKppch final state. 
With the larger data sample used here, these can be seen more clearly
and, in some cases, studied in detail.
Figure~\ref{kkstar}(a) shows a  plot of the invariant mass of
the $\Km\pip$ pair versus that of the $\Kp\pim$ pair.
Signal for the $K^{*}(892)^{0}$  is clearly visible.
Figure~\ref{kkstar}(b) shows the $K^\pm\pi^\mp$ mass distribution 
(two entries per event) for all  selected \KKppch events.
As we show in our previous study~\cite{isr2k2pi}, the signal at about
1400~\gevcc has parameters consistent with $K_2^{*}(1430)^{0}$.  
 Therefore,
we perform a fit to this distribution using P- and D-wave Breit-Wigner (BW) 
functions for the $K^{*0}$ and $K_2^{*0}$ signals, respectively, 
and a third-order polynomial
function for the remainder of the distribution, taking into account the
$K\pi$ threshold.
The fit result is shown by the curves in Fig.~\ref{kkstar}(b).
The fit yields a $K^{*0}$ signal of $53997\pm526$ events with 
$m(K^{*0}) = 0.8932\pm0.0002$~\gevcc and $\Gamma(K^{*0}) = 0.0521\pm0.0007$~\gev, 
and a $K_2^{*0}$ signal of $4361\pm235$ events with  
$m(K_2^{*0}) = 1.4274\pm0.0019$~\gevcc and $\Gamma(K_2^{*0}) = 0.0902\pm0.0056$~\gev.
These values are consistent with current world averages for
$K^{*}(892)^0$ and $K_2^{*}(1430)^{0}$~\cite{PDG} ,
and the fit describes the data well, indicating that contributions
from other resonances decaying into $K^\pm\pi^\mp$, like 
$K^{*}(1410)^{0}$ and/or $K_0^{*}(1430)^{0}$, are small.

\begin{figure}[tbh]
\includegraphics[width=1.0\linewidth]{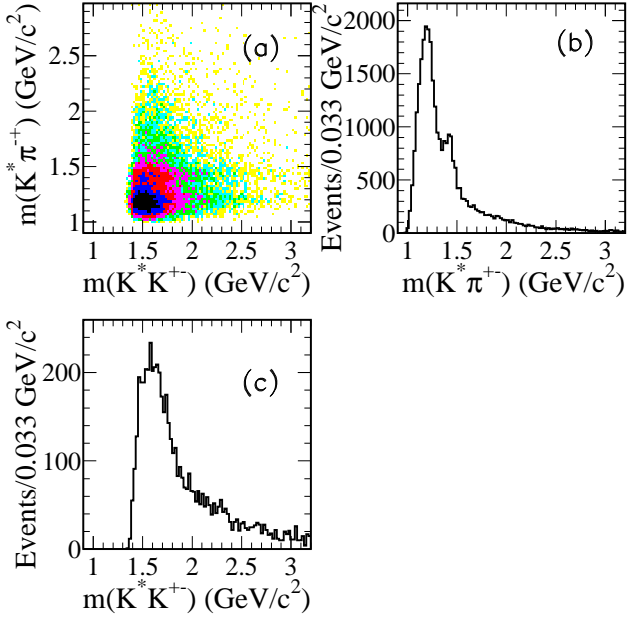}
\vspace{-0.2cm}
\caption{
   (a) 
Invariant mass of the $K^{*}(892)^{0}\pi^{\pm}$ system versus that of 
the $K^{*}(892)^{0} K^{\mp}$ system;
   (b) the $K^{*}(892)^{0}\pi^{\pm}$ projection plot of (a);
(c) the $K^{*}(892)^{0} K^{\mp}$ projection plot of (a) for $m(K^{*}(892)^{0}\pi^{\pm})>1.5$~\gevcc.
   }
\label{kstark}
\end{figure}

We combine $K^{*0}/\Kbar^{*0}$ candidates within the lines in 
Fig.~\ref{kkstar}(a) with the remaining pion and kaon to obtain the 
$K^{*}(892)^{0}\pi^{\pm}$ invariant mass distribution shown in Fig.~\ref{kstark}(b), 
and the $K^{*}(892)^{0}\pi^{\pm}$ versus\ $K^{*}(892)^{0} K^{\mp}$ mass  plot in 
Fig.~\ref{kstark}(a).
The bulk of Fig.~\ref{kstark}(a) shows a strong positive correlation,
characteristic of $K^{*0}K\pi$ final states with no higher resonances.
The horizontal bands in Fig.~\ref{kstark}(a) correspond to the peak
regions of the projection plot of Fig.~\ref{kstark}(b) and are
consistent with the contribution 
from the $K_1(1270)$ and $K_1(1400)$ resonances.
There is also an indication of a vertical band in Fig.~\ref{kstark}(a),
perhaps corresponding to a $K^{*}(892)^{0} K$ structure at
$\sim$1.5~\gevcc. The projection plot of Fig.~\ref{kstark}(c) for 
events with $m(K^{*}(892)^{0}\pi^{\pm})>1.5$~\gevcc shows the enhancement
not consistent with phase space behavior.

\begin{figure}[tbh]
\includegraphics[width=1.0\linewidth]{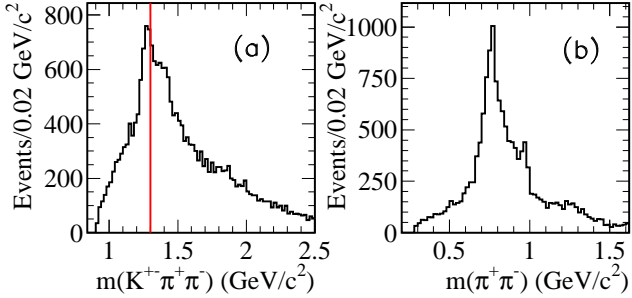}
\vspace{-0.2cm}
\caption{ 
   (a) The invariant mass of the $K^\pm\pipi$ combinations with
 $K^{*}(892)^{0}K\pi$  events excluded;
(b) the $\pipi$ invariant mass for events from (a) with the $K_1(1270)$ region
  suppressed by requiring $m(K^{\pm}\pipi)>1.3$~\gevcc as shown by vertical line in (a).
  }
\label{kpipi}
\end{figure}

We next suppress the $K^{*}(892)^{0}K\pi$ contribution by considering only events 
outside the lines in Fig.~\ref{kkstar}(a).
In Fig.~\ref{kpipi}(a) the $K^\pm\pipi$ invariant mass (two
entries per event) shows evidence of the $K_1(1270)$ and $K_1(1400)$ resonances, 
both of which decay into $K\rho(770)$, 
although the latter decay is very weak~\cite{PDG}.
In Fig.~\ref{kpipi}(b) we plot
the $\pipi$ invariant mass for events with $m(K^\pm\pipi)>1.3$~\gevcc.
There is a strong $\rho(770) \!\to\! \pipi$ signal, and
there are indications of additional structures in the 
$f_0(980)$ and $f_2(1270)$ regions.

The separation of all these, and any other, intermediate states
involving relatively broad resonances requires a partial wave analysis.
This is beyond the scope of this paper.
Instead we present the cross sections for the sum of all states that include
 $K^{*}(892)^{0}$,  $K_2^{*}(1430)^{0}$ or $\rho(770)$ signals,
and study intermediate states that include a narrow $\phi$ or $f_0$ resonance.

\subsection{\boldmath  The $\epem \to K^{*}(892)^{0} K \pi$,
  $K_2^{*}(1430)^{0} K \pi$ and $K^+ K^- \rho(770)$ Cross Sections}
\label{kstarxs}

\begin{figure}[tbh]
\includegraphics[width=1.0\linewidth]{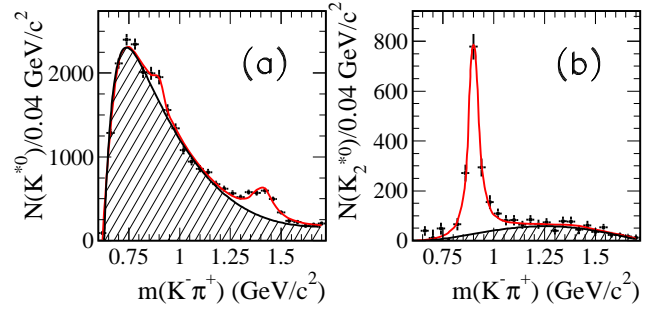}
\vspace{-0.2cm}
\caption{ 
The  $K^- \pi^+$  invariant mass distribution corresponding to the number of
$K^{*}(892)^{0}$ (a) and $K_{2}^{*}(1430)^{0}$  (b) events obtained from the fits to the
$K^+ \pi^-$ invariant mass distribution for each interval of $K^-\pi^+$ mass.  
 The curves result from the fits described in the text.
  }
\label{kstar_sel}
\end{figure}

Signals for the $K^{*}(892)^{0}$ and $K_2^{*}(1430)^{0}$ are clearly visible 
in the $K^\pm\pi^\mp$ mass distributions in Fig.~\ref{kkstar}(a,b).
To extract the number of events with correlated production of 
$K^{*}(892)^{0}\Kbar^{*}(892)^{0}$ 
and  $K^{*}(892)^{0}\Kbar_2^{*}(1430)^{0}+c.c.$, we perform the same fit as that 
shown in Fig.~\ref{kkstar}(b), but to the $K^+ \pi^-$ invariant mass 
distribution in each 0.04~\gevcc interval of $K^- \pi^+$ 
invariant mass. From each fit we obtain the number of 
 $K^{*}(892)^{0}$ and $K_2^{*}(1430)^{0}$ events and plot these values
as a function of $K^- \pi^+$ mass  in Fig.~\ref{kstar_sel}(a)
and Fig.~\ref{kstar_sel}(b), respectively. 
The fit to the data of Fig.~\ref{kstar_sel}(a)
indicates that only $548\pm 263$ events are associated with correlated
$\Kbar^{*}(892)^{0} K^{*}(892)^{0}$ production (about 1\% of the total  number of
$K^{*}(892)^{0}$ events), and that $1680\pm343$ events correspond to
 $\Kbar^{*}(892)^{0} K_2^{*}(1430)^{0}$ pairs, compared to  $4361\pm235$, the 
total number of events with a $K_2^{*}(1430)^{0}$ in the final state.
The distribution of the events from the $K_2^{*}(1430)^{0}$ peak shows a 
strong signal at the
 $\Kbar^{*}(892)^{0}$ mass in Fig.~\ref{kstar_sel}(b), which contains
$1648\pm32$ events, in  
agreement with the number of 
$K^{*}(892)^{0}\Kbar_2^{*}(1430)^{0}$ pairs obtained above.

\begin{figure}[tbh]
\includegraphics[width=1.0\linewidth]{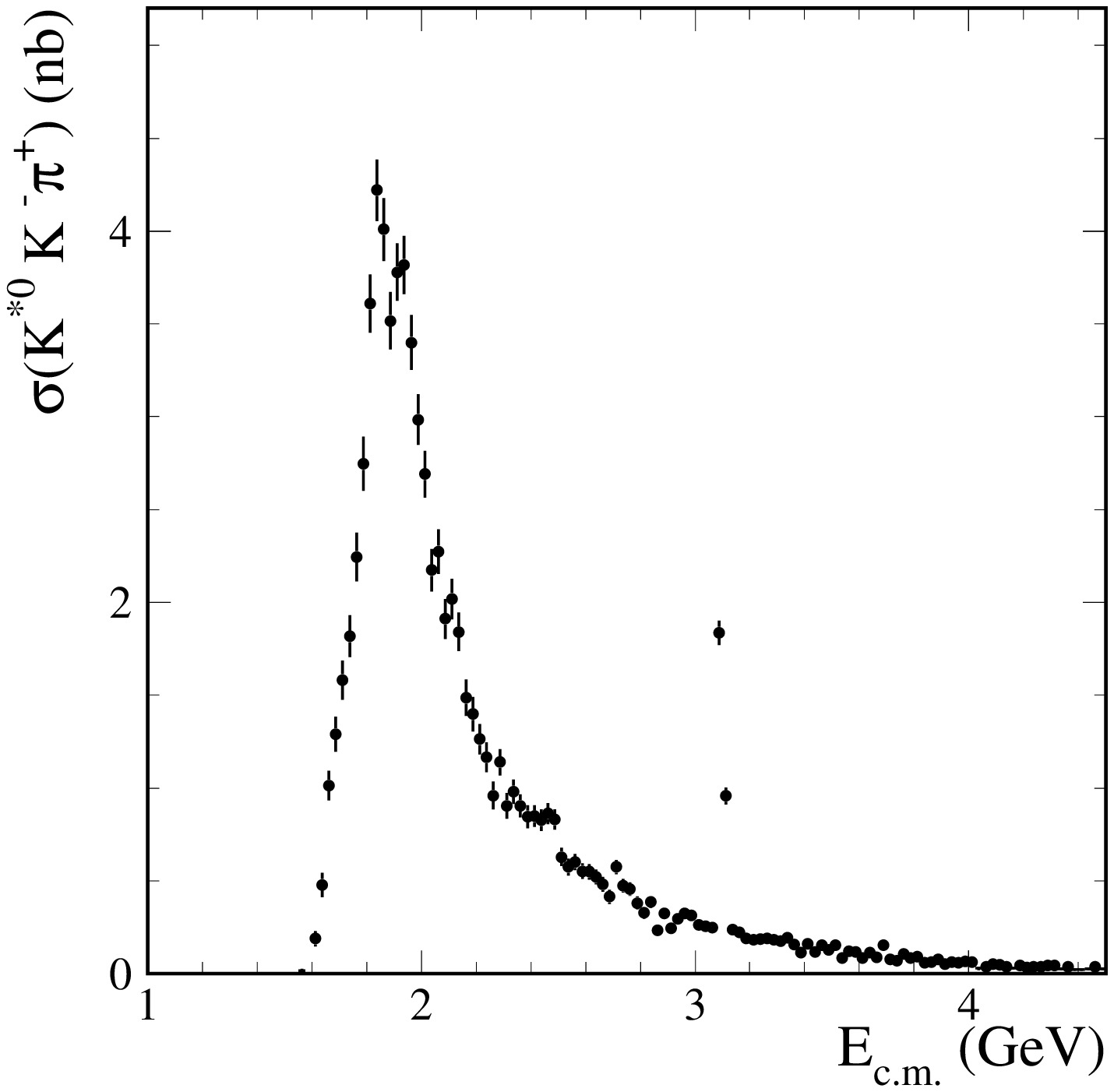}
\vspace{-0.2cm}
\caption{ 
 The $\epem\to K^{*}(892)^{0}K^- \pi^+ $ cross section,   
  obtained from the $K^{*}(892)^{0}$ signal of
Fig.~\ref{kkstar}(b). 
  }
\label{kstar_xs}
\end{figure}
\begin{figure}[tbh]
\includegraphics[width=1.0\linewidth]{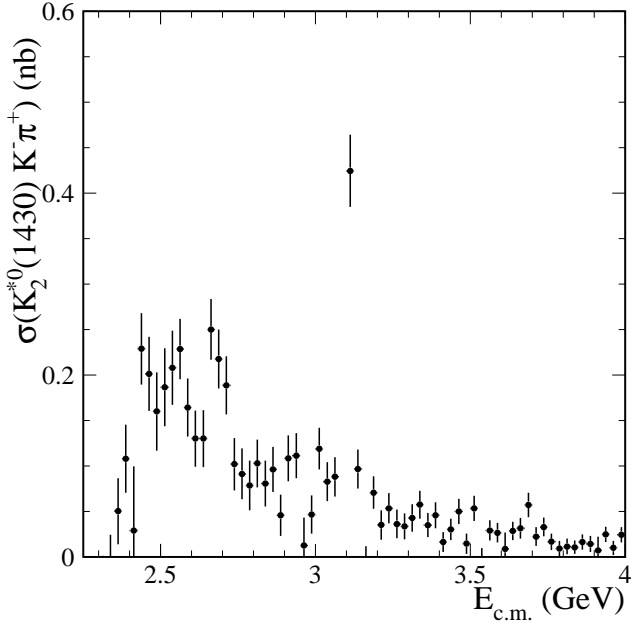}
\vspace{-0.2cm}
\caption{ 
 The  $K_2^{*}(1430)^{0}K^- \pi^+ $  cross section,
  obtained from the $K_2^{*}(1430)^{0}$ signal of
Fig.~\ref{kkstar}(b). 
  }
\label{kstar1_xs}
\end{figure}
\begin{figure}[tbh]
\vspace{-0.2cm}
\includegraphics[width=1.0\linewidth]{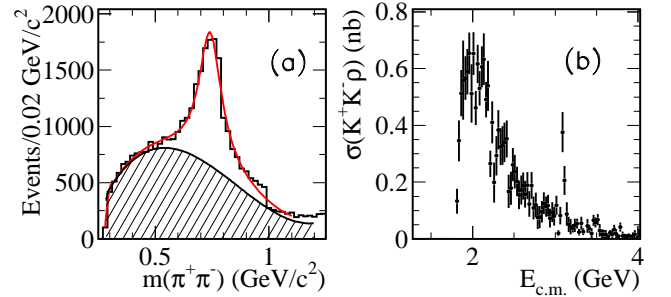}
\vspace{-0.2cm}
\caption{ 
  (a) The $\pi^+ \pi^-$ mass distribution for all
  selected \KKppch events with the $\phi$ and $K^{*0}$ regions excluded:
the solid curve represents a fit as described in the text, and the background
contribution is shown separately as the hatched region;
  (b) the $\epem \!\!\to\! K^+ K^- \rho(770)$  cross section 
  obtained from the $\rho$ signal from the fit in
  each 0.025~\gev c.m.~energy interval.
  }
\label{kkrho_sel}
\end{figure}

\begin{table*}
\caption{Summary of the cross section measurements for 
$\ep\en\to K^{*0}(892) K^- \pi^{+}$. 
Errors are statistical only.}
\label{kstar_tab}
\begin{ruledtabular}
\begin{tabular}{ c c c c c c c c }
\Ecm (GeV) & $\sigma$ (nb)  
& \Ecm (GeV) & $\sigma$ (nb) 
& \Ecm (GeV) & $\sigma$ (nb) 
& \Ecm (GeV) & $\sigma$ (nb)  
\\
\hline

 1.5875 &  0.00 $\pm$  0.00 & 2.1875 &  1.40 $\pm$  0.09 & 2.7875 &  0.38 $\pm$  0.03 & 3.3875 &  0.11 $\pm$  0.02 \\
 1.6125 &  0.19 $\pm$  0.04 & 2.2125 &  1.26 $\pm$  0.08 & 2.8125 &  0.33 $\pm$  0.03 & 3.4125 &  0.16 $\pm$  0.02 \\
 1.6375 &  0.48 $\pm$  0.07 & 2.2375 &  1.17 $\pm$  0.08 & 2.8375 &  0.39 $\pm$  0.03 & 3.4375 &  0.12 $\pm$  0.02 \\
 1.6625 &  1.01 $\pm$  0.08 & 2.2625 &  0.96 $\pm$  0.07 & 2.8625 &  0.24 $\pm$  0.03 & 3.4625 &  0.15 $\pm$  0.02 \\
 1.6875 &  1.29 $\pm$  0.10 & 2.2875 &  1.14 $\pm$  0.07 & 2.8875 &  0.32 $\pm$  0.03 & 3.4875 &  0.13 $\pm$  0.02 \\
 1.7125 &  1.58 $\pm$  0.11 & 2.3125 &  0.90 $\pm$  0.07 & 2.9125 &  0.24 $\pm$  0.03 & 3.5125 &  0.15 $\pm$  0.02 \\
 1.7375 &  1.82 $\pm$  0.11 & 2.3375 &  0.98 $\pm$  0.07 & 2.9375 &  0.30 $\pm$  0.03 & 3.5375 &  0.08 $\pm$  0.01 \\
 1.7625 &  2.24 $\pm$  0.13 & 2.3625 &  0.90 $\pm$  0.06 & 2.9625 &  0.33 $\pm$  0.03 & 3.5625 &  0.12 $\pm$  0.01 \\
 1.7875 &  2.75 $\pm$  0.15 & 2.3875 &  0.85 $\pm$  0.06 & 2.9875 &  0.31 $\pm$  0.03 & 3.5875 &  0.12 $\pm$  0.01 \\
 1.8125 &  3.61 $\pm$  0.16 & 2.4125 &  0.85 $\pm$  0.06 & 3.0125 &  0.26 $\pm$  0.03 & 3.6125 &  0.09 $\pm$  0.01 \\
 1.8375 &  4.22 $\pm$  0.17 & 2.4375 &  0.83 $\pm$  0.06 & 3.0375 &  0.26 $\pm$  0.03 & 3.6375 &  0.12 $\pm$  0.02 \\
 1.8625 &  4.01 $\pm$  0.17 & 2.4625 &  0.86 $\pm$  0.06 & 3.0625 &  0.25 $\pm$  0.02 & 3.6625 &  0.09 $\pm$  0.01 \\
 1.8875 &  3.52 $\pm$  0.15 & 2.4875 &  0.83 $\pm$  0.05 & 3.0875 &  1.84 $\pm$  0.06 & 3.6875 &  0.15 $\pm$  0.02 \\
 1.9125 &  3.78 $\pm$  0.15 & 2.5125 &  0.63 $\pm$  0.05 & 3.1125 &  0.96 $\pm$  0.05 & 3.7125 &  0.08 $\pm$  0.01 \\
 1.9375 &  3.82 $\pm$  0.16 & 2.5375 &  0.58 $\pm$  0.05 & 3.1375 &  0.24 $\pm$  0.02 & 3.7375 &  0.07 $\pm$  0.01 \\
 1.9625 &  3.40 $\pm$  0.15 & 2.5625 &  0.60 $\pm$  0.04 & 3.1625 &  0.22 $\pm$  0.02 & 3.7625 &  0.11 $\pm$  0.01 \\
 1.9875 &  2.98 $\pm$  0.14 & 2.5875 &  0.55 $\pm$  0.04 & 3.1875 &  0.19 $\pm$  0.02 & 3.7875 &  0.09 $\pm$  0.01 \\
 2.0125 &  2.69 $\pm$  0.13 & 2.6125 &  0.55 $\pm$  0.04 & 3.2125 &  0.18 $\pm$  0.02 & 3.8125 &  0.09 $\pm$  0.01 \\
 2.0375 &  2.17 $\pm$  0.11 & 2.6375 &  0.52 $\pm$  0.04 & 3.2375 &  0.19 $\pm$  0.02 & 3.8375 &  0.06 $\pm$  0.01 \\
 2.0625 &  2.27 $\pm$  0.12 & 2.6625 &  0.48 $\pm$  0.04 & 3.2625 &  0.19 $\pm$  0.02 & 3.8625 &  0.06 $\pm$  0.01 \\
 2.0875 &  1.91 $\pm$  0.11 & 2.6875 &  0.41 $\pm$  0.04 & 3.2875 &  0.18 $\pm$  0.02 & 3.8875 &  0.08 $\pm$  0.01 \\
 2.1125 &  2.02 $\pm$  0.11 & 2.7125 &  0.57 $\pm$  0.04 & 3.3125 &  0.17 $\pm$  0.02 & 3.9125 &  0.05 $\pm$  0.01 \\
 2.1375 &  1.84 $\pm$  0.10 & 2.7375 &  0.47 $\pm$  0.04 & 3.3375 &  0.19 $\pm$  0.02 & 3.9375 &  0.06 $\pm$  0.01 \\
 2.1625 &  1.49 $\pm$  0.10 & 2.7625 &  0.46 $\pm$  0.04 & 3.3625 &  0.16 $\pm$  0.02 & 3.9625 &  0.06 $\pm$  0.01 \\

\end{tabular}
\end{ruledtabular}
\end{table*}

We perform a fit similar to that shown in Fig.~\ref{kkstar}(b)
 to the data in intervals of \KKppch invariant
mass, with the resonance masses and widths fixed to the values
obtained from the overall fit.
Since correlated $K^{*}$ production is small, 
we convert the
resulting  $K^{*}$ yield in each interval into a
cross section value for 
$\epem \!\!\to\! K^{*}(892)^{0} K^- \pi^+$ or $K_2^{*}(1430)^{0} K^- \pi^+$, 
\footnote{The use of charge conjugate reactions 
is implied throughout the paper}
following the procedure described in Sec.~\ref{sec:xs2k2pi}. 
These cross section values take into account  only the $K\pi$ decay of
the  $K^{*}(892)^{0}$ and the $K_2^{*}(1430)^{0}$.

Note that the $\epem \!\!\to\! K^{*}(892)^{0} K \pi$ ($K_2^{*}(1430)^{0} K \pi$) 
cross section includes a small contribution from the
$K_2^{*}(1430)^{0} K \pi$ ($K^{*}(892)^{0} K \pi$) channel, because 
the  $K_2^{*}(1430)^{0} K^{*}(892)^{0}$ final state 
has not been taken into account.
 These cross sections are shown in Fig.~\ref{kstar_xs} and
 Fig.~\ref{kstar1_xs}, and 
the $\epem \!\!\to\! K^{*}(892)^{0} K^{-} \pi^{+} $ channel is listed in
Table~\ref{kstar_tab} for \Ecm energies from threshold up to 4.0~\gev.
At higher energies the signals are small and contain an unknown, but
possibly large, contribution from $\epem \!\!\to\! \qqbar$ events.
There is a rapid rise from threshold to a peak value of about 4~nb at 
1.84~\gev for the $\epem \!\!\to\! K^{*}(892)^{0} K^- \pi^+$ cross section, 
followed by a very rapid decrease with increasing energy.
There are suggestions of narrow structures in the peak region, 
but the only statistically significant structure is the $J/\psi$ peak, 
which is discussed below. 
There are some structures in the $\epem \!\!\to\! K_2^{*}(1430)^{0} K^-
\pi^+$ cross section, but the signal size is too small to make any 
definite statement.

The $\epem \!\!\to\! K^{*}(892)^{0} K^- \pi^+$ contribution is a large fraction of
the total \KKppch cross 
section at all energies above its threshold,
and dominates in the 1.8--2.0~\gev region.
The $\Kp\Km\rho^0(770)$ intermediate state makes up the majority of the
remainder of the cross section. We exclude a small $\phi$ contribution by
requiring
$|m(K^+ K^- )-m(\phi)| >0.01$~\gevcc, and suppress the large
$K^{*}(892)^{0}$ contribution by means of the anti-selection
$|m([K^\pm \pi^\mp )-0.892| >0.035$~\gevcc. 
Figure~\ref{kkrho_sel}(a) shows the $\pipi$ mass distribution for
the remaining events. 
The combinatorial background is relatively large, and includes 
a small contribution from $f_0(980)\to\pipi$ decays.
We fit the $\rho(770)$ signal with a single BW (mass and
width are fixed to 0.77~\gevcc and 0.15~\gev, respectively) and a
polynomial background (contribution shown by the hatched area) in each
0.025~\gev c.m.~energy interval. The cross section obtained
is shown in Fig.~\ref{kkrho_sel}(b),  and has no significant
structures except the $J/\psi$ signal.
The uncertainty in the $\rho(770)$ shape, and also in the background shape,
   provides the largest contribution to the systematic error, estimated
   to be 20-30\%. A small contribution to the background from $f_0(980)\to\pipi$
   is ignored in the fit, which does not result in a significant
uncertainty.

\subsection{\boldmath The $\phi(1020)\pipi$ Intermediate State}
\label{sec:phipipi}

\begin{figure}[tbh]
\includegraphics[width=1.0\linewidth]{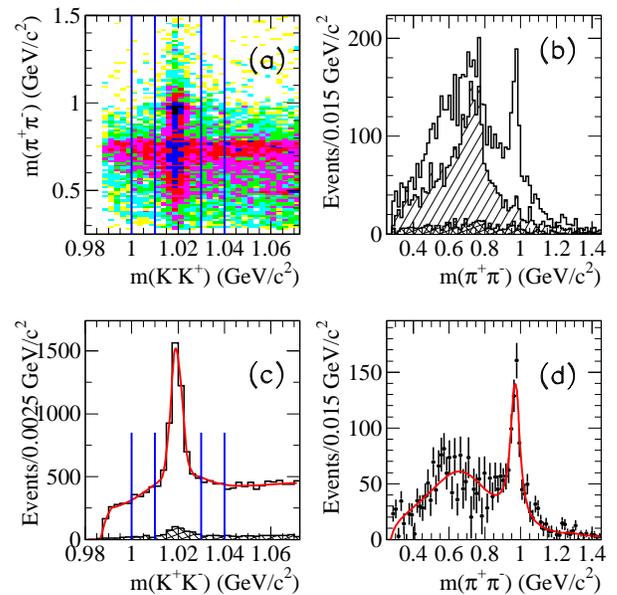}
\vspace{-0.4cm}
\caption{
  (a) $m(\pipi)$ versus $m(\Kp\Km)$ 
  for all selected \KKppch events;
  (b) the $\pipi$ invariant mass projections for events in the $\phi$ peak
  (open histogram), sidebands (hatched), and background control region
  (cross-hatched);
  (c) the $\Kp\Km$ mass projections for all events (open) and control
  region (cross-hatched);
  (d) the difference between the open histogram and the sum of the
  other contributions to (b).  
  }
\label{phif0_sel}
\end{figure}
\begin{figure}[tbh]
\includegraphics[width=0.9\linewidth]{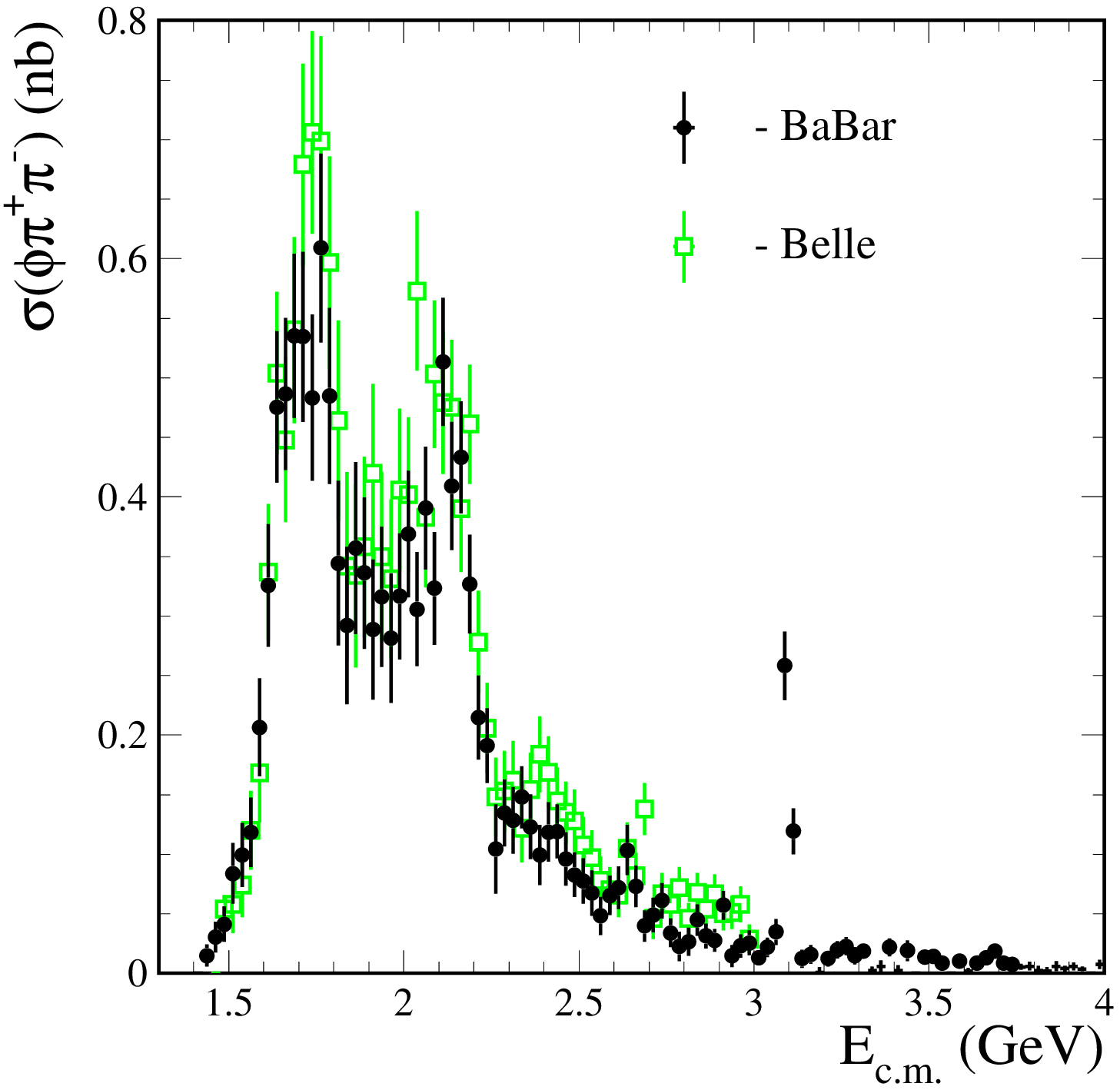}
\vspace{-0.4cm}
\caption{
  The $\epem \!\!\to\! \phi\pipi$  cross section 
as a  function of  \epem c.m.\ energy obtained by \babar~
(dots) and Belle (squares) ~\cite{belle_phif0}.
  }
\label{phipipixs}
\end{figure}
Intermediate states containing narrow resonances can be
studied more easily. For the \Ecm energy range below 3.0~\gev, 
Fig.~\ref{phif0_sel}(a) shows a plot of the 
invariant mass of the $\pipi$ pair versus that of the $\Kp\Km$ pair.
Horizontal and vertical bands corresponding to the $\rho^0(770)$ and
$\phi$, respectively, are visible,
and there is a concentration of entries in the $\phi$ band corresponding
to the correlated production of $\phi$ and $f_{0}(980)$,
as demonstrated by the open histogram of Fig.~\ref{phif0_sel}(b).
The $\phi$ signal is clearly visible in the $\Kp\Km$ mass projection of
Fig.~\ref{phif0_sel}(c).
The large contribution from the $\rho(770)$ 
is nearly uniform in $\Kp\Km$ mass,
and the cross-hatched histogram shows the non-\KKppch background
estimated from the control region in \chiKKppch.
The cross-hatched histogram 
also shows a $\phi$ peak, but this is a small fraction of the events.
When we subtract this background and fit the remaining data with a double-Gaussian 
function for the $\phi$ signal, and a first-order polynomial function for the non-$\phi$
background (with a cut-off at the KK threshold), we obtain
3951$\pm$91 events corresponding to the $\phi\pipi$ intermediate state.

To study the $\phi\pipi$ channel, we select candidate events with a
$\Kp\Km$ invariant mass    
within 10~\mevcc of the $\phi$ mass, 
indicated by the inner vertical lines in Figs.~\ref{phif0_sel}(a,c),
and estimate the non-$\phi$ contribution from the mass sidebands between
the inner and outer vertical lines.
In Fig.~\ref{phif0_sel}(b) we show the \pipi invariant mass distributions
for $\phi$ candidate events, sideband events, and \chisq control region events
as the open, hatched and cross-hatched histograms, respectively,
and in Fig.~\ref{phif0_sel}(d) we show the $\pipi$ distribution after
subtracting the non-$\phi$ background contributions. We observe a
clear, narrow peak in the $f_0(980)$ mass region, together with a broad
enhancement that reaches a maximum at about 0.6~\gevcc, which could
indicate $f_0(600)$ production. We defer a
detailed analysis of this distribution to Secs.~\ref{sec:phif01},
~\ref{phipipistudy}, and ~\ref{phif0bump}. 

\begin{table*}[tbh]
\caption{Summary of the cross section measurements for
  $\ep\en\to\phi(1020)\pipi$. Errors are statistical only.}
\label{phi2pi_tab}
\begin{ruledtabular}
\begin{tabular}{ c c c c c c c c }
\Ecm (GeV) & $\sigma$ (nb)  
& \Ecm (GeV) & $\sigma$ (nb) 
& \Ecm (GeV) & $\sigma$ (nb) 
& \Ecm (GeV) & $\sigma$ (nb)  
\\
\hline

 1.4875 &  0.04 $\pm$  0.01 & 1.8375 &  0.29 $\pm$  0.07 & 2.1875 &  0.33 $\pm$  0.04 & 2.5375 &  0.07 $\pm$  0.02 \\
 1.5125 &  0.08 $\pm$  0.03 & 1.8625 &  0.36 $\pm$  0.07 & 2.2125 &  0.21 $\pm$  0.04 & 2.5625 &  0.05 $\pm$  0.02 \\
 1.5375 &  0.10 $\pm$  0.03 & 1.8875 &  0.34 $\pm$  0.06 & 2.2375 &  0.19 $\pm$  0.03 & 2.5875 &  0.07 $\pm$  0.02 \\
 1.5625 &  0.12 $\pm$  0.03 & 1.9125 &  0.29 $\pm$  0.06 & 2.2625 &  0.10 $\pm$  0.04 & 2.6125 &  0.07 $\pm$  0.02 \\
 1.5875 &  0.21 $\pm$  0.04 & 1.9375 &  0.32 $\pm$  0.06 & 2.2875 &  0.13 $\pm$  0.03 & 2.6375 &  0.10 $\pm$  0.02 \\
 1.6125 &  0.33 $\pm$  0.05 & 1.9625 &  0.28 $\pm$  0.05 & 2.3125 &  0.13 $\pm$  0.03 & 2.6625 &  0.07 $\pm$  0.02 \\
 1.6375 &  0.48 $\pm$  0.06 & 1.9875 &  0.32 $\pm$  0.05 & 2.3375 &  0.15 $\pm$  0.03 & 2.6875 &  0.04 $\pm$  0.01 \\
 1.6625 &  0.49 $\pm$  0.06 & 2.0125 &  0.37 $\pm$  0.05 & 2.3625 &  0.12 $\pm$  0.03 & 2.7125 &  0.05 $\pm$  0.01 \\
 1.6875 &  0.54 $\pm$  0.07 & 2.0375 &  0.31 $\pm$  0.05 & 2.3875 &  0.10 $\pm$  0.03 & 2.7375 &  0.06 $\pm$  0.01 \\
 1.7125 &  0.53 $\pm$  0.07 & 2.0625 &  0.39 $\pm$  0.05 & 2.4125 &  0.12 $\pm$  0.02 & 2.7625 &  0.03 $\pm$  0.01 \\
 1.7375 &  0.48 $\pm$  0.07 & 2.0875 &  0.32 $\pm$  0.05 & 2.4375 &  0.12 $\pm$  0.02 & 2.7875 &  0.02 $\pm$  0.01 \\
 1.7625 &  0.61 $\pm$  0.08 & 2.1125 &  0.51 $\pm$  0.05 & 2.4625 &  0.10 $\pm$  0.02 & 2.8125 &  0.03 $\pm$  0.01 \\
 1.7875 &  0.48 $\pm$  0.07 & 2.1375 &  0.41 $\pm$  0.05 & 2.4875 &  0.08 $\pm$  0.02 & 2.8375 &  0.04 $\pm$  0.01 \\
 1.8125 &  0.34 $\pm$  0.07 & 2.1625 &  0.43 $\pm$  0.05 & 2.5125 &  0.08 $\pm$  0.02 & 2.8625 &  0.03 $\pm$  0.01 \\

\end{tabular}
\end{ruledtabular}
\end{table*}

We obtain the number of $\epem \!\!\to\! \phi\pipi$ events in
0.025~\gevcc intervals of the
$\phi\pipi$ invariant mass by fitting the $K^+ K^-$ invariant mass
projection in that interval after subtracting non-\KKppch background.
Each projection is a subset of Fig.~\ref{phif0_sel}(c), 
where the curve represents the fit to the full sample. 
In each mass interval, all parameters other than number of events in the 
  $\phi$ peak and the normalization of the background distribution are fixed
  to the values obtained from the overall fit.  As a check, we also describe
  the background as a linear function, with all parameters free in each mass
  interval; the alternative fit yields consistent results with the nominal
fit  to within 5\%, which is taken as a systematic uncertainty.

The reconstruction efficiency may depend on the details of the production mechanism.
Using the two-pion mass distribution in Fig.~\ref{phif0_sel}(d) 
as input, 
we simulate the \pipi system as an S-wave composition of two 
structures both described by the BW amplitudes,
with parameters set to the values obtained in Sec.~\ref{phipipistudy}.
The BW amplitudes represent the $f_0(980)$ and the bump at 0.6~\gevcc,
which we call $f_0(600)$ (see Sec.~\ref{phipipistudy}).
We describe the $\phi\pipi$ mass distribution using a simple model
with one resonance of mass 1.68~\gevcc and
width 0.3~\gev, which decays to $\phi\pipi$ or $\phi f_0(980)$ when
phase space allows. 
The reconstructed spectrum that results then has 
a sharp increase at about 2~\gevcc due to the 
$\phi f_0(980)$ threshold.

We obtain the efficiency as a function of $\phi\pipi$ mass by
dividing the number of reconstructed events in each interval by the number 
generated; the result is
 shown in Fig.~\ref{mc_acc1} by the dashed curve. 
Comparison with the solid curve
 in the same figure shows that the model dependence is weak,
 giving confidence in the efficiency calculation.
We calculate the $\epem \!\!\to\! \phi\pipi$ cross section as described
in Sec.~\ref{sec:xs2k2pi}, and divide by the $\phi \!\to\! K^+ K^-$
branching fraction (0.489~\cite{PDG}).
We show our results as a function of c.m.~energy in
Fig.~\ref{phipipixs},  and list them
in Table~\ref{phi2pi_tab}.
The cross section has a peak value of about 0.6~nb at about 1.7~\gev, 
then decreases with increasing energy until the $\phi(1020) f_0(980)$ 
threshold, around 2.0~\gev.
From this point it rises, falls sharply at about 2.2~\gev,
and then decreases slowly.
Except in the charmonium region, the results
at energies above 3~\gev are not meaningful due to 
small signals and potentially large backgrounds, and are omitted from
Table~\ref{phi2pi_tab}.
Figure~\ref{phipipixs}
displays the cross section up to 4.0~\gev in order to show 
the $J/\psi$ and $\psi(2S)$ signals, which
are discussed in Sec.~\ref{sec:charmonium}.

The cross section obtained is in agreement with our previous
measurement~\cite{isr2k2pi}.
The cross section measured by
the Belle Collaboration~\cite{belle_phif0}, also shown in Fig.~\ref{phipipixs}, presents
very similar features, and a general consistency with
our data, although a small systematic
difference at higher c.m. energies is visible.

\begin{table*}
\caption{
Summary of the $\ep\en\to\phi(1020)\pi\pi$ cross section, dominated by 
  $\phi(1020) f_0(980)$,  $f_0(980)\to\pi\pi$, obtained from 
$\phi(1020)\pipi$  events with   $0.85<m(\pipi)<1.1$~\gevcc.
Errors are statistical only.}
\label{phif0_tab}
\begin{ruledtabular}
\begin{tabular}{ c c c c c c c c }
\Ecm (GeV) & $\sigma$ (nb)  
& \Ecm (GeV) & $\sigma$ (nb) 
& \Ecm (GeV) & $\sigma$ (nb) 
& \Ecm (GeV) & $\sigma$ (nb)  
\\
\hline

 1.8875 &  0.00 $\pm$  0.01 & 2.1625 &  0.54 $\pm$  0.06 & 2.4375 &  0.11 $\pm$  0.02 & 2.7125 &  0.04 $\pm$  0.03 \\
 1.9125 &  0.01 $\pm$  0.02 & 2.1875 &  0.38 $\pm$  0.05 & 2.4625 &  0.11 $\pm$  0.03 & 2.7375 &  0.04 $\pm$  0.02 \\
 1.9375 &  0.16 $\pm$  0.04 & 2.2125 &  0.19 $\pm$  0.04 & 2.4875 &  0.08 $\pm$  0.02 & 2.7625 &  0.03 $\pm$  0.02 \\
 1.9625 &  0.15 $\pm$  0.04 & 2.2375 &  0.19 $\pm$  0.04 & 2.5125 &  0.07 $\pm$  0.02 & 2.7875 &  0.03 $\pm$  0.02 \\
 1.9875 &  0.19 $\pm$  0.04 & 2.2625 &  0.10 $\pm$  0.04 & 2.5375 &  0.06 $\pm$  0.02 & 2.8125 &  0.02 $\pm$  0.02 \\
 2.0125 &  0.32 $\pm$  0.05 & 2.2875 &  0.15 $\pm$  0.03 & 2.5625 &  0.05 $\pm$  0.02 & 2.8375 &  0.05 $\pm$  0.02 \\
 2.0375 &  0.28 $\pm$  0.05 & 2.3125 &  0.14 $\pm$  0.03 & 2.5875 &  0.07 $\pm$  0.02 & 2.8625 &  0.03 $\pm$  0.02 \\
 2.0625 &  0.38 $\pm$  0.06 & 2.3375 &  0.16 $\pm$  0.03 & 2.6125 &  0.07 $\pm$  0.02 & 2.8875 &  0.02 $\pm$  0.02 \\
 2.0875 &  0.35 $\pm$  0.05 & 2.3625 &  0.14 $\pm$  0.03 & 2.6375 &  0.07 $\pm$  0.02 & 2.9125 &  0.04 $\pm$  0.02 \\
 2.1125 &  0.60 $\pm$  0.06 & 2.3875 &  0.07 $\pm$  0.03 & 2.6625 &  0.07 $\pm$  0.02 & 2.9375 &  0.01 $\pm$  0.02 \\
 2.1375 &  0.50 $\pm$  0.07 & 2.4125 &  0.11 $\pm$  0.03 & 2.6875 &  0.03 $\pm$  0.02 & 2.9625 &  0.01 $\pm$  0.01 \\

\end{tabular}
\end{ruledtabular}
\end{table*}
\begin{figure*}[tbh]
\includegraphics[width=0.32\linewidth]{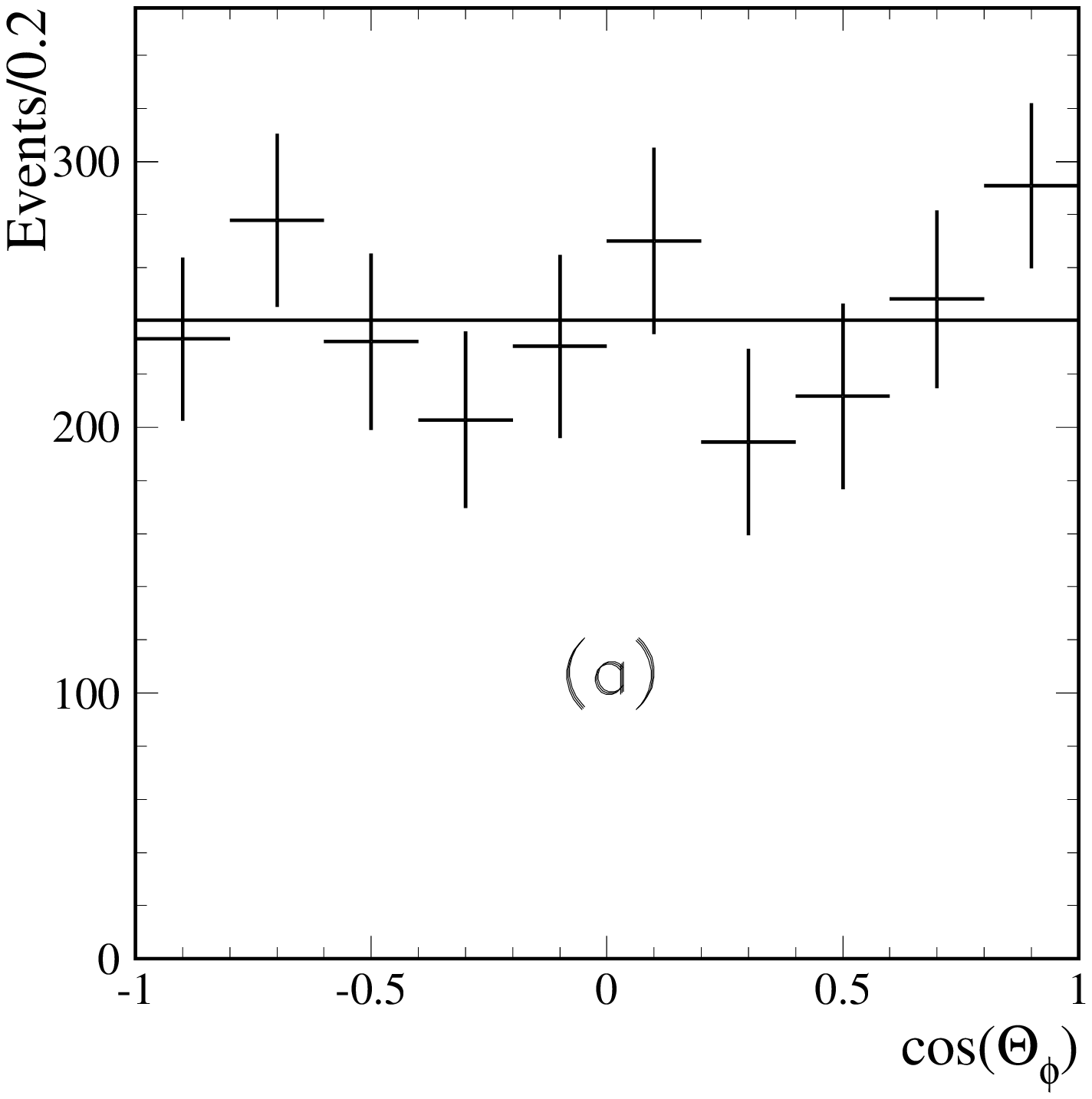}
\includegraphics[width=0.32\linewidth]{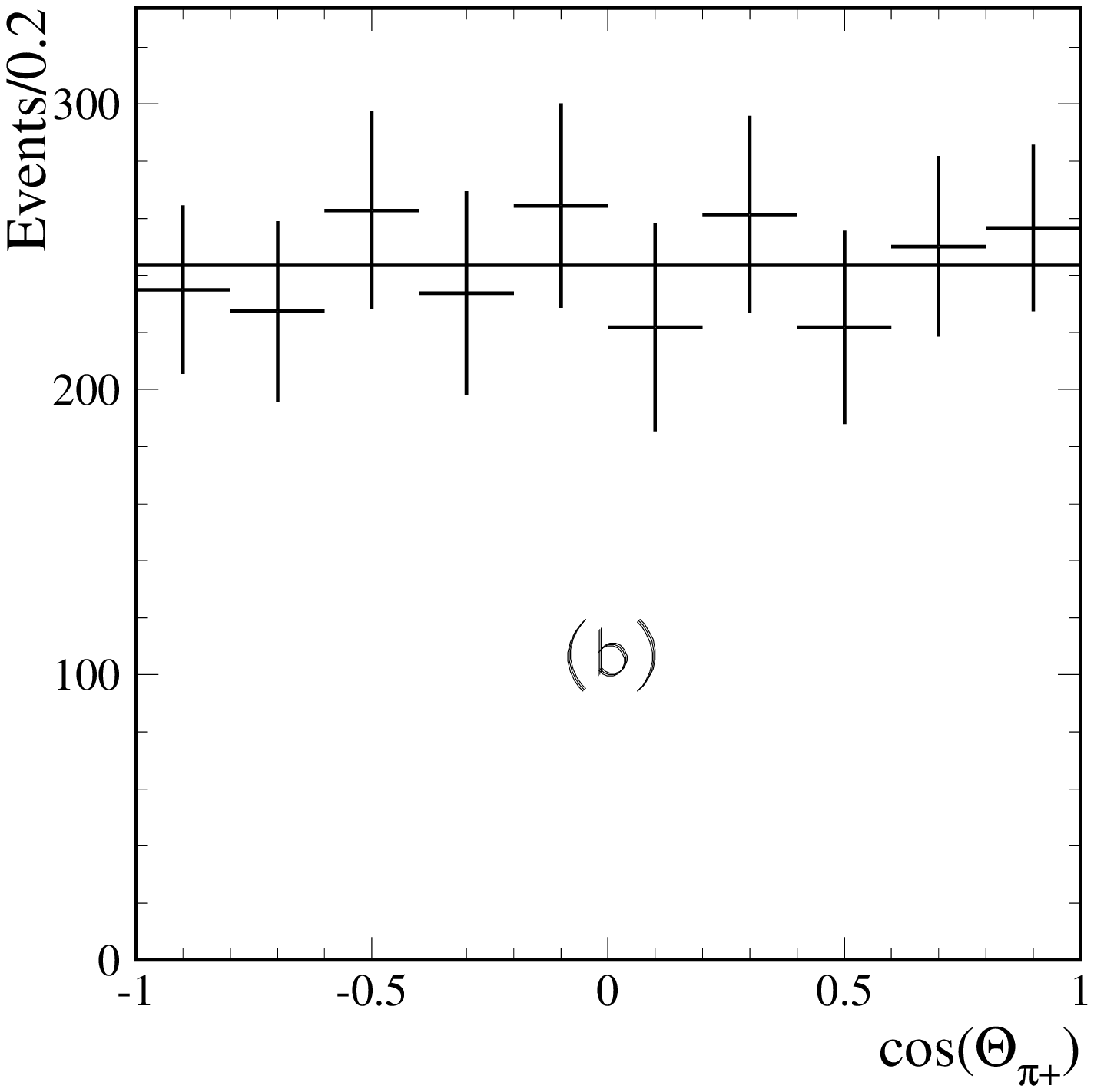}
\includegraphics[width=0.32\linewidth]{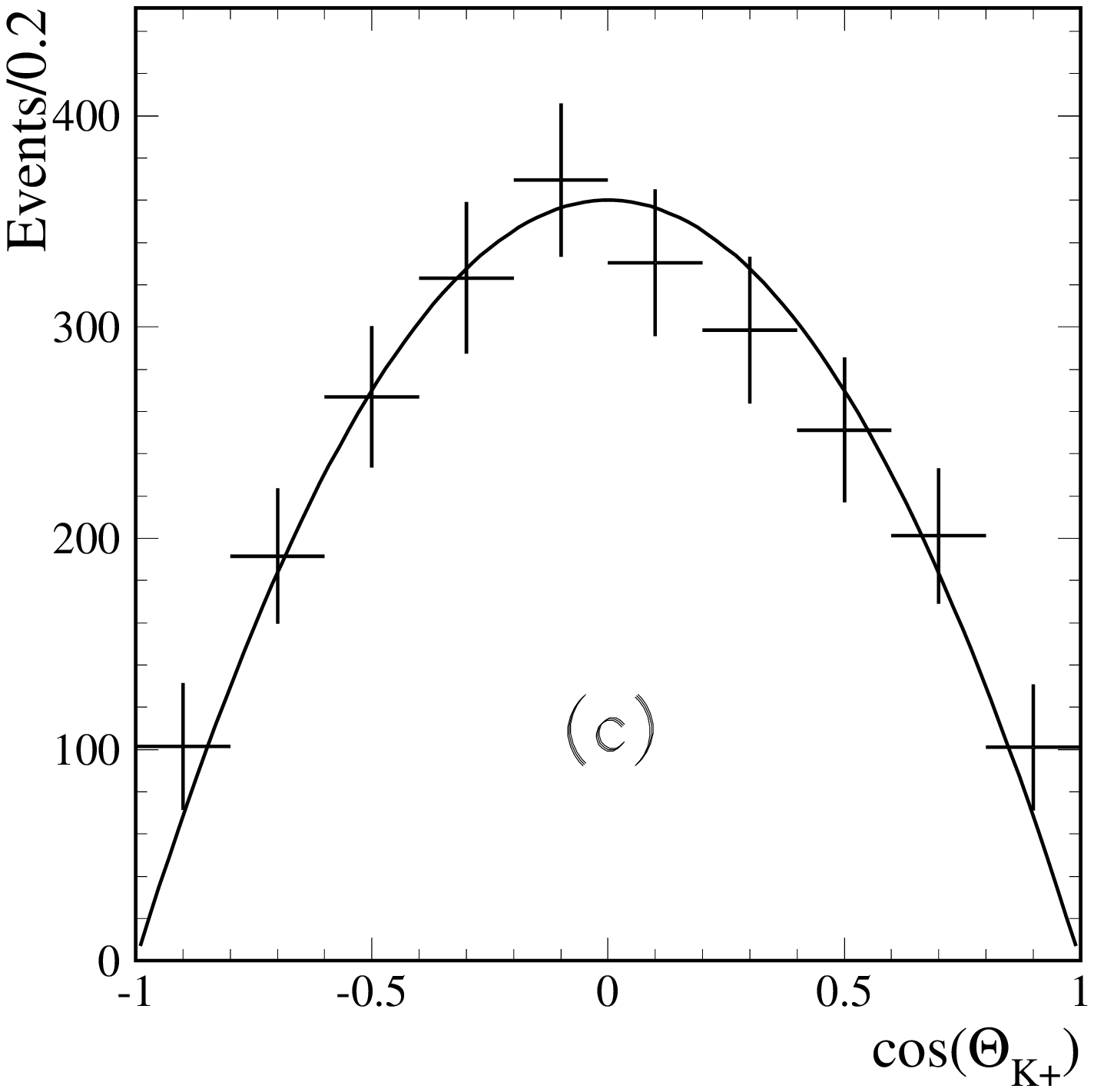}
\vspace{-0.2cm}
\caption{
  Distributions of the cosine of
  (a) the $\phi$ production angle,
  (b) the pion helicity angle, and
  (c) the kaon helicity angle (see text) for $\epem \!\!\to\! \phi\pipi$
  events:
  the curves (normalized to the data) represent the distributions
  expected if the \pipi system 
  recoiling against the vector $\phi$ meson is an S-wave system
produced in an S-wave orbital angular momentum state. 
  }
\label{phi_angle}
\end{figure*}
\begin{figure}[tbh]
\includegraphics[width=1.0\linewidth]{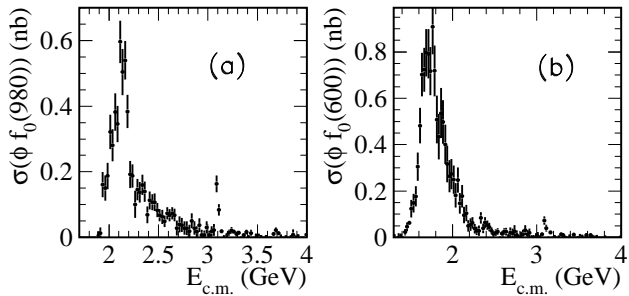}
\vspace{-0.4cm}
\caption{
The  $\epem \!\!\to\! \phi\pipi$ cross section derived from the \KKppch final
  state as a function of c.m. energy, for (a) the $0.85<m(\pipi)<1.1$~\gevcc
  region, dominated by the $\phi(1020) f_0(980)$, and (b) $m(\pipi)<0.85$~\gevcc.
  }
\label{phif0xs}
\end{figure}

We perform a study of the angular distributions in the $\phi(1020)\pipi$ 
final state by considering all \KKppch candidate events with mass
below 3~\gevcc in intervals 
of the cosine of each angle defined below,
and fitting the background-subtracted $\Kp\Km$ mass projection in each
interval.
The efficiency is nearly uniform in the cosine of each angle, and
so we study the number of events in each interval.
We define the $\phi$ production angle, $\Theta_\phi$, as the angle
between the $\phi$ direction and the ISR photon direction in the rest  
frame of the $\phi\pipi$ system (i.e., the effective $\epem$ collision axis). 
The distribution of $\cos\Theta_\phi$, shown in Fig.~\ref{phi_angle}(a),
is consistent with the uniform distribution expected if 
the quasi-two-body final state $\phi X$, $X \!\to\! \pipi$, is produced
in an S-wave angular-momentum state.
We define the pion helicity angle, $\Theta_{\pip}$, as that between 
the \pip and the recoil $\phi$ direction in the \pipi rest frame.
The kaon helicity angle, $\Theta_{\Kp}$ is defined as that between the \Kp 
direction and the ISR photon direction in the $\phi$ rest frame. 
The distributions of $\cos\Theta_{\pip}$ and $\cos\Theta_{\Kp}$, 
shown in Figs.~\ref{phi_angle}(b) and~\ref{phi_angle}(c), respectively,
are consistent with those expected from scalar (uniform) and vector
($\cos^2 \Theta_{\Kp}$) meson decays, 
where for the latter the $\phi$ retains the helicity of the virtual photon
to which the $\phi X$ system couples.
\begin{table*}
\caption{Summary of the cross section measurements, dominated by 
  $\ep\en\to\phi(1020) f_{0}(600)$, $f_0(600)\to\pi\pi$ process, obtained for $m(\pipi)<0.85$~\gevcc.
Errors are statistical only.}
\label{phis_tab}
\begin{ruledtabular}
\begin{tabular}{ c c c c c c c c }
\Ecm (GeV) & $\sigma$ (nb)  
& \Ecm (GeV) & $\sigma$ (nb) 
& \Ecm (GeV) & $\sigma$ (nb) 
& \Ecm (GeV) & $\sigma$ (nb)  
\\
\hline

 1.2875 &  0.00 $\pm$  0.01 & 1.7125 &  0.79 $\pm$  0.11 & 2.1375 &  0.10 $\pm$  0.04 & 2.5625 &  0.00 $\pm$  0.01 \\
 1.3125 &  0.01 $\pm$  0.01 & 1.7375 &  0.72 $\pm$  0.10 & 2.1625 &  0.10 $\pm$  0.04 & 2.5875 &  0.02 $\pm$  0.01 \\
 1.3375 &  0.00 $\pm$  0.01 & 1.7625 &  0.91 $\pm$  0.12 & 2.1875 &  0.05 $\pm$  0.03 & 2.6125 &  0.03 $\pm$  0.01 \\
 1.3625 &  0.01 $\pm$  0.01 & 1.7875 &  0.72 $\pm$  0.11 & 2.2125 &  0.05 $\pm$  0.03 & 2.6375 &  0.03 $\pm$  0.02 \\
 1.3875 &  0.01 $\pm$  0.01 & 1.8125 &  0.51 $\pm$  0.10 & 2.2375 &  0.06 $\pm$  0.03 & 2.6625 &  0.01 $\pm$  0.01 \\
 1.4125 &  0.00 $\pm$  0.01 & 1.8375 &  0.43 $\pm$  0.10 & 2.2625 &  0.04 $\pm$  0.02 & 2.6875 &  0.02 $\pm$  0.02 \\
 1.4375 &  0.02 $\pm$  0.01 & 1.8625 &  0.54 $\pm$  0.11 & 2.2875 &  0.03 $\pm$  0.02 & 2.7125 &  0.02 $\pm$  0.02 \\
 1.4625 &  0.05 $\pm$  0.02 & 1.8875 &  0.50 $\pm$  0.09 & 2.3125 &  0.03 $\pm$  0.02 & 2.7375 &  0.03 $\pm$  0.03 \\
 1.4875 &  0.06 $\pm$  0.02 & 1.9125 &  0.40 $\pm$  0.09 & 2.3375 &  0.08 $\pm$  0.02 & 2.7625 &  0.01 $\pm$  0.02 \\
 1.5125 &  0.12 $\pm$  0.04 & 1.9375 &  0.32 $\pm$  0.08 & 2.3625 &  0.04 $\pm$  0.02 & 2.7875 &  0.00 $\pm$  0.01 \\
 1.5375 &  0.15 $\pm$  0.04 & 1.9625 &  0.26 $\pm$  0.07 & 2.3875 &  0.06 $\pm$  0.02 & 2.8125 &  0.01 $\pm$  0.02 \\
 1.5625 &  0.18 $\pm$  0.04 & 1.9875 &  0.27 $\pm$  0.07 & 2.4125 &  0.05 $\pm$  0.02 & 2.8375 &  0.01 $\pm$  0.02 \\
 1.5875 &  0.31 $\pm$  0.06 & 2.0125 &  0.25 $\pm$  0.06 & 2.4375 &  0.04 $\pm$  0.02 & 2.8625 &  0.03 $\pm$  0.02 \\
 1.6125 &  0.48 $\pm$  0.08 & 2.0375 &  0.18 $\pm$  0.05 & 2.4625 &  0.03 $\pm$  0.01 & 2.8875 &  0.01 $\pm$  0.02 \\
 1.6375 &  0.70 $\pm$  0.09 & 2.0625 &  0.25 $\pm$  0.05 & 2.4875 &  0.01 $\pm$  0.01 & 2.9125 &  0.02 $\pm$  0.02 \\
 1.6625 &  0.72 $\pm$  0.09 & 2.0875 &  0.15 $\pm$  0.05 & 2.5125 &  0.02 $\pm$  0.01 & 2.9375 &  0.00 $\pm$  0.01 \\
 1.6875 &  0.80 $\pm$  0.10 & 2.1125 &  0.18 $\pm$  0.05 & 2.5375 &  0.03 $\pm$  0.01 & 2.9625 &  0.00 $\pm$  0.01 \\

\end{tabular}
\end{ruledtabular}
\end{table*}

\subsection{\boldmath The $\phi(1020) f_{0}(980)$ and 
$\phi(1020) f_{0}(600)$ Intermediate States}
\label{sec:phif01}

The narrow $f_0(980)$ peak seen in Fig.~\ref{phif0_sel}(d) allows the
selection of a fairly clean sample of $\phi f_0(980)$ events.
We repeat the analysis just described with the additional requirement
that the \pipi invariant mass be in the range 0.85--1.10~\gevcc.
A fit to the $\Kp\Km$ mass spectrum
for this sample, analogous to that
shown in Fig.~\ref{phif0_sel}(c),
yields about 1350 events;
all of these contain a true $\phi$, with a small fraction of events 
   with the pion pair not produced through the $f_0(980)$,
but the latter contribution is relatively small (see discussion
in Sec.~\ref{phipipistudy}).
By selecting events with the \pipi invariant mass below
0.85~\gevcc, we similarly obtain a sample composed mostly of $\phi f_0(600)$ events. 

We convert the above two samples of  $f_0(980)$ and  $f_0(600)$ events 
in each mass interval 
into measurements of the $\epem \!\!\to\! \phi(1020) f_{0}(980)$
and $\epem \!\!\to\! \phi(1020) f_{0}(600)$
cross sections as described above, dividing by the 
$f_0 \!\!\to\! \pipi$ branching fraction of 2/3
to account for $f_0\to\ppz$ decays.
The cross sections are shown in Fig.~\ref{phif0xs} as functions of
c.m.~energy
and are listed in Table~\ref{phif0_tab} and  Table~\ref{phis_tab}.
The $\phi(1020) f_{0}(980)$ cross section
behavior near threshold does not appear to be smooth, 
but is more consistent with a steep rise to a value of about 0.3~nb
at 2.0~\gev followed by a slow decrease that is interrupted
by a structure around 2.175~\gev.
In contrast, the $\phi(1020) f_{0}(600)$ cross section has a smooth
threshold increase to about 0.8 nb, followed by a smooth decrease
thereafter, and can be interpreted as the $\phi(1680)$ resonance.
It is 
important to note that all structures above 2.0~\gev seen in
Fig.~\ref{phipipixs} relate only to the $f_0(980)$ resonance.
 Possible interpretations of these structures are discussed 
in Sec.~\ref{phif0bump}.
Again, the cross section values are not meaningful for c.m.~energy
above about 3~\gev,  
except for the $J/\psi$ and $\psi(2S)$ signals,
discussed in Sec.~\ref{sec:charmonium}.

\section{The {\boldmath $K^+ K^-\ppz$} Final State}
\subsection{Final Selection and Backgrounds}

The \KKppnt sample contains background from the ISR processes
$\epem \!\!\to\! \Kp\Km\piz\gamma$ and $\Kp\Km\eta\gamma$,
in which two soft photon candidates from machine- or detector-related
backgrounds combine with the relatively energetic photons from
the \piz or $\eta$ to form two fake \piz candidates.
We reduce this background using the angle between each 
reconstructed \piz direction and the direction of its higher-energy photon
daughter calculated in the \piz rest frame.
If the cosines of both angles are larger than 0.85,
we remove the event.

\begin{figure}[tbh]
\includegraphics[width=0.9\linewidth]{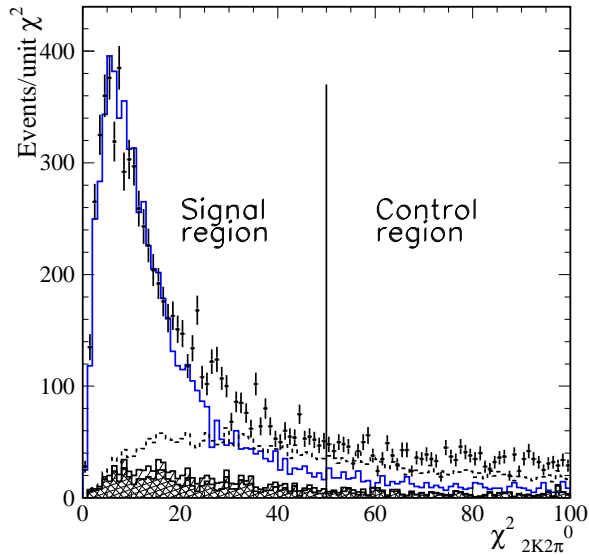}
\vspace{-0.4cm}
\caption{
  Distribution of \chisq from the six-constraint fits to \KKppnt candidates
  in the data (points).
  The open histogram is the distribution for simulated signal events, 
  normalized as described in the text.
  The cross-hatched, hatched, and dashed regions represent, respectively, 
  the backgrounds from non-ISR \qqbar events, ISR-produced $\pipi\ppz$ and
 $K_S K\pi$ events, 
  and ISR-produced $\Kp\Km\piz$, $\Kp\Km\eta$ and $\KpKm\ppz\piz$ events.
}
\label{2k2pi0_chi2_all}
\end{figure}

Figure~\ref{2k2pi0_chi2_all} shows the distribution of \chiKKppnt for
the remaining candidates together with the
simulated \KKppnt events.
Again, the distributions are broader than those for a typical 6C \chisq
distribution due to higher order ISR, 
and we normalize the histogram to the data in the region 
$\chiKKppnt\! <\! 15$.
The cross-hatched histogram in Fig.~\ref{2k2pi0_chi2_all} represents 
 background from $\epem \!\!\to\! \qqbar$ events, 
evaluated in the same way as for the \KKppch final state.
The hatched region represents the
ISR backgrounds from final states with similar kinematics.
The first of these is
 $\pipi\ppz$, which yields events with 
both charged pions misidentified
as kaons, and the second is the $K_S K\pi$, 
which yields $K_S\to\ppz$ and a misidentified pion.
Each contribution is small. 

The dominant background in this case is from residual ISR $\Kp\Km\piz$
and $\Kp\Km\eta$ events, 
as well as ISR-produced $\KpKm\ppz\piz$ events.
Their net simulated contribution, 
indicated by the dashed contour in Fig.~\ref{2k2pi0_chi2_all}, 
is consistent with the data in the high \chiKKppnt region.
All other backgrounds are either negligible or distributed uniformly 
in \chiKKppnt.
We define the signal region by $\chiKKppnt\! <\! 50$, which  
contains 7967 data and 7402 simulated events, 
and a control region by $50\! <\! \chiKKppnt\! <\! 100$, which
contains 2007 data and 704 simulated signal events.

\begin{figure}[tbh]
\includegraphics[width=0.9\linewidth]{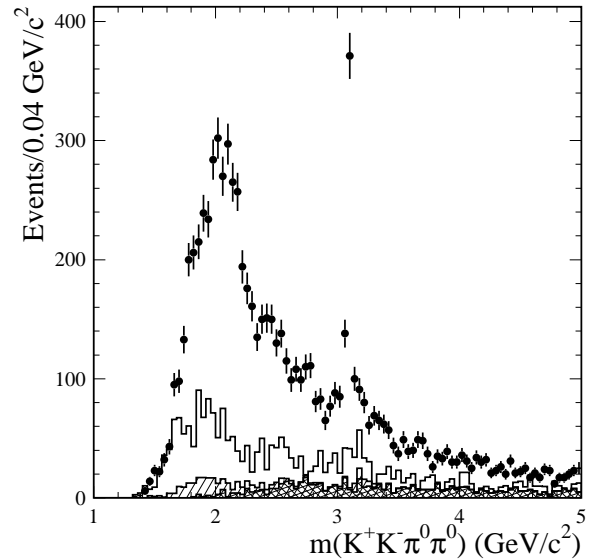}
\vspace{-0.4cm}
\caption{
  Invariant mass distribution for \KKppnt candidates in the signal region for
 data (points).
  The cross-hatched, hatched, and open regions represent, respectively, 
  the non-ISR \qqbar background, the contribution from ISR-produced
 $\pipi\ppz$ and $K_SK\pi$ events,
  and the contribution from the other ISR processes described in the text.
  }
\label{2k2pi0_babar}
\end{figure}

Figure~\ref{2k2pi0_babar} shows the \KKppnt invariant mass
distribution from threshold up to 5~\gevcc for events in the signal region.
The \qqbar background (cross-hatched histogram) is negligible at low 
masses but yields a significant fraction of the selected events above about 4~\gevcc.
The ISR $\pipi\ppz$ contribution (hatched region) is negligible except
in the 1.5--2.5~\gevcc region.
The sum of all other backgrounds, estimated from the control region,
is the dominant contribution below 2.5~\gevcc and is non-negligible everywhere.
The total background varies
from ~100\% below 1.6~\gevcc to ~25\% at higher masses. 

We subtract the sum of the estimated background contributions
 from the number of selected events 
in each mass interval to obtain the number of signal events.
Considering uncertainties in the cross sections for the background processes,
the normalization of events in the control region and the simulation
statistics, we estimate a systematic uncertainty on the signal yield
after background subtraction of
about 5\% in the 1.6--3.0~\gevcc region; 
this increases linearly from 5\% to 15\% in the region above 3~\gevcc.

\subsection{Selection Efficiency}

\begin{figure}[tbh]
\includegraphics[width=0.9\linewidth]{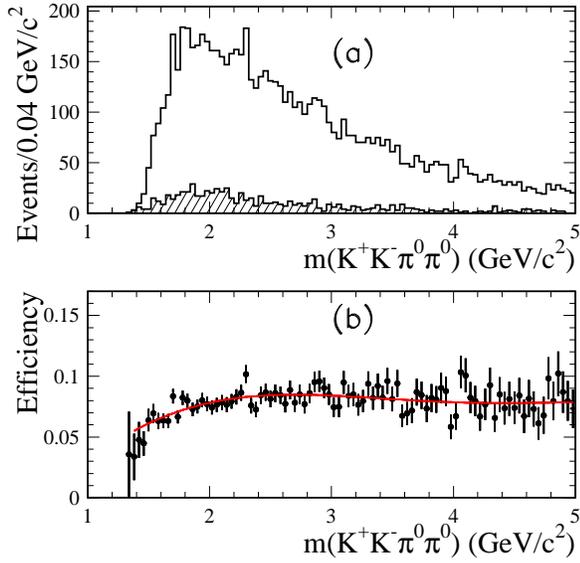}
\vspace{-0.4cm}
\caption{
  (a) Invariant mass distribution for simulated \KKppnt events in the
  signal (open) and control (hatched) regions (see Fig.~\ref{2k2pi0_chi2_all});
  (b) net reconstruction and selection efficiency as a function of mass
  obtained from this simulation
  (the curve represents the result of a third-order-polynomial fit).
  }
\label{mc_acc3}
\end{figure}

The detection efficiency is determined in the same manner as in 
Sec.~\ref{sec:eff1}.
Figure~\ref{mc_acc3}(a) shows the simulated \KKppnt invariant mass
distributions in the signal and control regions obtained from the phase space
model.
We divide the number of reconstructed events in each
0.04~\gevcc mass interval by the number generated in that interval to
obtain the efficiency estimate shown by the points in Fig.~\ref{mc_acc3}(b);
a third-order-polynomial fit to the efficiency
is used in calculating the cross section.
Again, the simulation of the ISR photon
covers a limited angular range, which is 
about 30\% wider than the EMC acceptance.
Simulations assuming dominance of the $\phi \!\to\! \Kp\Km$ and/or
the $f_0 \!\!\to\! \ppz$ channels give results consistent with those of 
Fig.~\ref{mc_acc3}(b),  and we apply
a 3\% systematic uncertainty for possible model dependence, 
as in Sec.~\ref{sec:eff1}.

We correct for mis-modeling of the track-finding and kaon identification 
efficiencies as in Sec.~\ref{sec:eff1} (corrections of $(+1.9\pm0.6)$\% and
$(0\pm2.0)$\%, respectively).
We do not observe any large discrepancy in the shape of the \chiKKppnt
distribution, and so apply no correction for the $\chiKKppnt<50$ selection,
 but introduce 3\% as an associated systematic uncertainty.
We correct the \piz-finding efficiency using the procedure described in
detail in Ref.~\cite{isr6pi}. 
From ISR $\epem \!\!\to\! \omega\piz\gamma \!\!\to\! \pipi\ppz\gamma$ events
selected with and without the \piz from the $\omega$ decay,
we find that the simulated efficiency for one \piz is too large by (3.0$\pm$1.0)\%,
and we apply a correction of $(+6.0\pm 2.0)$\% 
because of the two $\piz$s in each event.
\begin{table*}
\caption{Summary of the cross section measurements for $\ep\en\to K^+ K^- \ppz$. 
Errors are statistical only.}
\label{2k2pi0_tab}
\begin{ruledtabular}
\begin{tabular}{ c c c c c c c c }
\Ecm (GeV) & $\sigma$ (nb)  
& \Ecm (GeV) & $\sigma$ (nb) 
& \Ecm (GeV) & $\sigma$ (nb) 
& \Ecm (GeV) & $\sigma$ (nb)  
\\
\hline

 1.5000 &  0.00 $\pm$  0.04 & 2.1400 &  0.65 $\pm$  0.07 & 2.7800 &  0.19 $\pm$  0.03 & 3.4200 &  0.05 $\pm$  0.02 \\
 1.5400 &  0.01 $\pm$  0.05 & 2.1800 &  0.65 $\pm$  0.06 & 2.8200 &  0.11 $\pm$  0.03 & 3.4600 &  0.05 $\pm$  0.02 \\
 1.5800 &  0.00 $\pm$  0.05 & 2.2200 &  0.47 $\pm$  0.05 & 2.8600 &  0.09 $\pm$  0.03 & 3.5000 &  0.02 $\pm$  0.02 \\
 1.6200 &  0.01 $\pm$  0.06 & 2.2600 &  0.37 $\pm$  0.05 & 2.9000 &  0.09 $\pm$  0.02 & 3.5400 &  0.06 $\pm$  0.02 \\
 1.6600 &  0.14 $\pm$  0.08 & 2.3000 &  0.38 $\pm$  0.05 & 2.9400 &  0.09 $\pm$  0.03 & 3.5800 &  0.04 $\pm$  0.01 \\
 1.7000 &  0.14 $\pm$  0.07 & 2.3400 &  0.26 $\pm$  0.04 & 2.9800 &  0.10 $\pm$  0.03 & 3.6200 &  0.03 $\pm$  0.02 \\
 1.7400 &  0.35 $\pm$  0.07 & 2.3800 &  0.26 $\pm$  0.05 & 3.0200 &  0.12 $\pm$  0.02 & 3.6600 &  0.07 $\pm$  0.02 \\
 1.7800 &  0.59 $\pm$  0.08 & 2.4200 &  0.32 $\pm$  0.04 & 3.0600 &  0.18 $\pm$  0.03 & 3.7000 &  0.05 $\pm$  0.02 \\
 1.8200 &  0.66 $\pm$  0.08 & 2.4600 &  0.26 $\pm$  0.04 & 3.1000 &  0.71 $\pm$  0.04 & 3.7400 &  0.03 $\pm$  0.01 \\
 1.8600 &  0.48 $\pm$  0.08 & 2.5000 &  0.21 $\pm$  0.04 & 3.1400 &  0.12 $\pm$  0.03 & 3.7800 &  0.01 $\pm$  0.01 \\
 1.9000 &  0.64 $\pm$  0.08 & 2.5400 &  0.21 $\pm$  0.04 & 3.1800 &  0.06 $\pm$  0.03 & 3.8200 &  0.03 $\pm$  0.01 \\
 1.9400 &  0.54 $\pm$  0.08 & 2.5800 &  0.17 $\pm$  0.04 & 3.2200 &  0.08 $\pm$  0.02 & 3.8600 &  0.04 $\pm$  0.01 \\
 1.9800 &  0.74 $\pm$  0.08 & 2.6200 &  0.15 $\pm$  0.03 & 3.2600 &  0.05 $\pm$  0.02 & 3.9000 &  0.04 $\pm$  0.01 \\
 2.0200 &  0.84 $\pm$  0.08 & 2.6600 &  0.19 $\pm$  0.03 & 3.3000 &  0.10 $\pm$  0.02 & 3.9400 &  0.02 $\pm$  0.01 \\
 2.0600 &  0.63 $\pm$  0.08 & 2.7000 &  0.14 $\pm$  0.03 & 3.3400 &  0.08 $\pm$  0.02 & 3.9800 &  0.03 $\pm$  0.01 \\
 2.1000 &  0.78 $\pm$  0.07 & 2.7400 &  0.20 $\pm$  0.03 & 3.3800 &  0.07 $\pm$  0.02 & 4.0200 &  0.02 $\pm$  0.01 \\

\end{tabular}
\end{ruledtabular}
\end{table*}

\begin{table}[tbh]
\caption{
Summary of corrections and systematic uncertainties 
for the $\epem \!\!\to\! \KKppnt$  cross  section measurements.
The total correction is the linear sum of the contributions, and the
total uncertainty is obtained by summing the individual contributions
in quadrature.
  }
\label{error2_tab}
\begin{ruledtabular}
\begin{tabular}{l c r@{}l} 
     Source             & Correction & \multicolumn{2}{c}{Uncertainty}   \\
\hline
                        &            &       &              \\[-0.2cm]
Rad. Corrections        &  --        &  1\%  &              \\
Backgrounds             &  --        &  5\%  &, $\Ecm <3~\gev$ \\
                        &            & 5-15\%&, $\Ecm >3~\gev$ \\
Model Dependence        &  --        &  3\%  &              \\
\chiKKppnt Distribution &  --        &  3\%  &              \\ 
Tracking Efficiency     & +1.9\%     &  0.6\%&              \\
Kaon ID Efficiency      &   --       &  2\%  &              \\
\piz Efficiency         &  +6\%      &  2\%  &              \\
ISR-photon Efficiency   &  +1.0\%    &  0.5\%&            \\ 
ISR Luminosity          &  --        &  1\%  &              \\[0.1cm]
\hline
                        &            &       &             \\[-0.2cm]
Total                   &  +8.9\%    &  7\%  &, $\Ecm <3~\gev$ \\
                        &            & 7-16\%&, $\Ecm >3~\gev$ \\
\end{tabular}
\end{ruledtabular}
\end{table}

\subsection{\boldmath Cross Section for $\epem \to \KKppnt$}
\label{sec:2k2pi0xs}

\begin{figure}[tbh]
\includegraphics[width=0.9\linewidth]{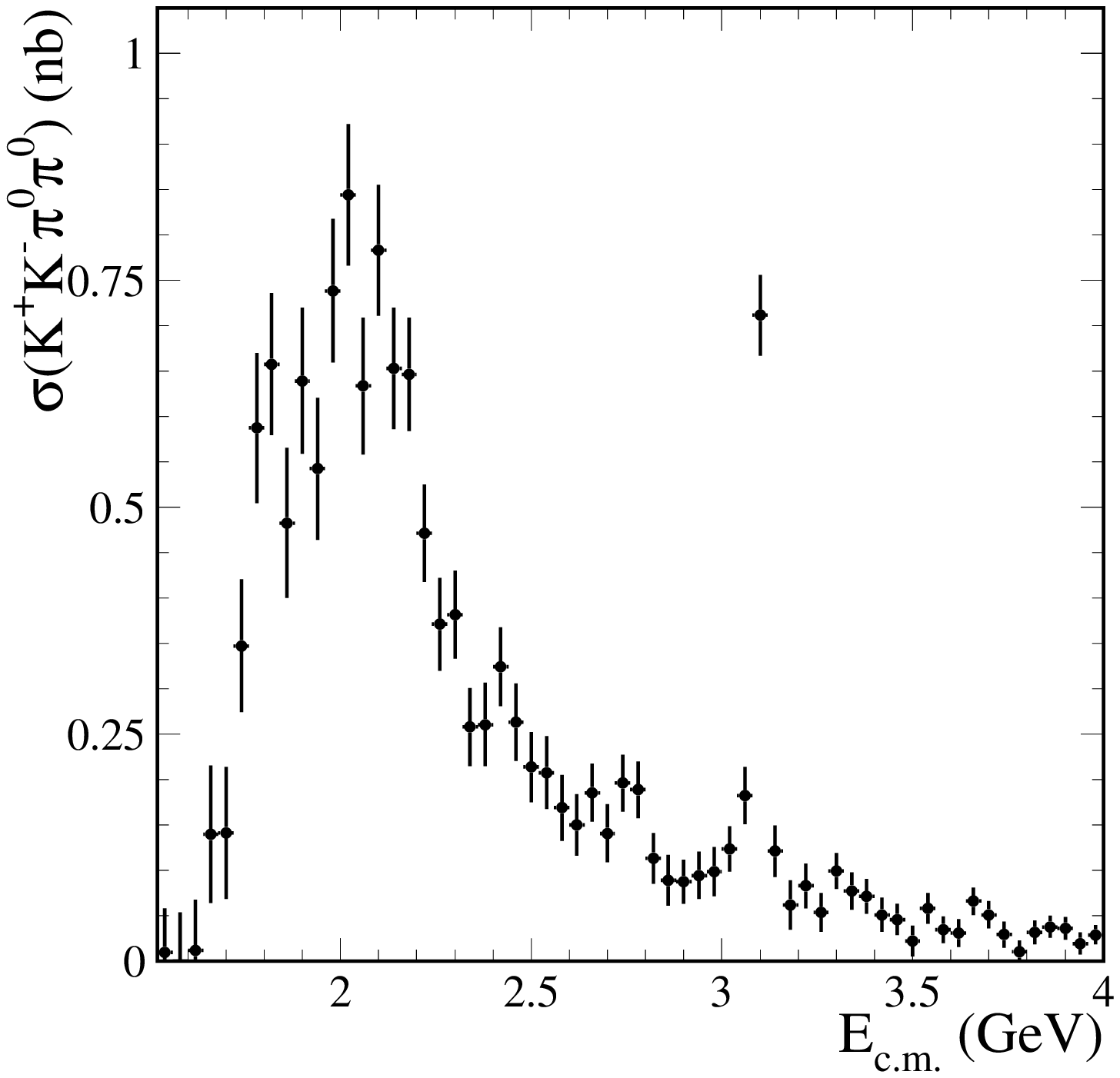}
\vspace{-0.5cm}
\caption{
  The $\epem \!\!\to\! \KKppnt$ cross section as a function of 
  \epem c.m.\ energy measured with ISR data at \babar.
  The errors are statistical only.
  }
\label{2k2pi0_ee_babar}
\end{figure}

We calculate the cross section for $\epem \to \KKppnt$ in 0.04~\gev \Ecm
intervals from the analog of Eq.(\ref{xseqn}),
using the invariant mass of the \KKppnt system to determine the
c.m.\ energy.
We show the results
in Fig.~\ref{2k2pi0_ee_babar} and list the values and statistical errors
 in Table~\ref{2k2pi0_tab}.
The cross section rises to a peak value near 0.8~nb at 2~\gev then
shows a rapid decrease, which is
interrupted by a large $J/\psi$ signal; the charmonium region is
discussed in Sec.~\ref{sec:charmonium} below.
The drop at 2.2~\gev is similar to that seen for the \KKppch final state.
Again, the differential luminosity includes corrections for vacuum
polarization that should be omitted for calculations of $a_\mu$.

The simulated \KKppnt invariant mass resolution is 8.8~\mevcc in the 
1.5--2.5~\gevcc mass range,
and increases with mass to 11.2~\mevcc in the 2.5--3.5~\gevcc range. 
Since less than 20\% of the events in a 0.04~\gev interval are
reconstructed outside that interval,
and the cross section has no sharp structure other than the  $J/\psi$ peak,
we again make no correction for resolution.
The point-to-point systematic uncertainties are much smaller than the statistical
uncertainties, and
the errors on the normalization are summarized in Table~\ref{error2_tab},
along with the corrections that were applied to the measurements.
The total correction is $+8.9$\%, and the total systematic uncertainty
is 7\% at low mass, increasing linearly from 7\% to 16\% above 3~\gevcc.

\subsection{\boldmath Substructure in the \KKppnt Final State}
\label{sec:kaons2}

\begin{figure}[tbh]
\includegraphics[width=1.0\linewidth]{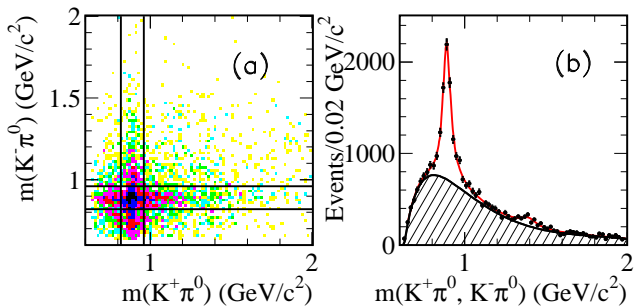}
\vspace{-0.4cm}
\caption{
  (a) Invariant mass of the $\Km\piz$ pair versus that of the
  $\Kp\piz$ pair in selected \KKppnt events (two entries per event);
  (b) sum of the projections of (a) (dots, four entries per event).
  The curves represent the result of the fit described in the text.
  }
\label{kkstarpi0}
\end{figure}
\begin{figure}[tbh]
\includegraphics[width=1.0\linewidth]{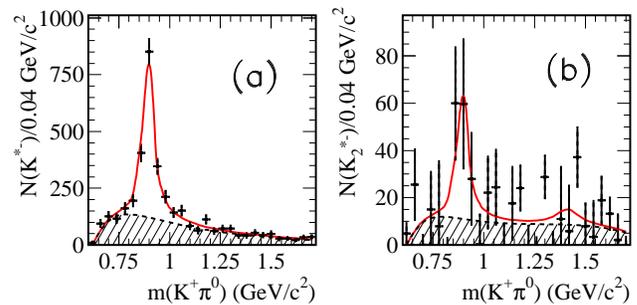}
\vspace{-0.4cm}
\caption{ 
The number of
$K^{*}(892)^{-}$ (a) and $K_{2}^{*}(1430)^{-}$  (b) events obtained from the fits to the
$K^- \pi^0$ invariant mass distributions for each 0.04~\gevcc interval of $K^+\pi^0$ mass.  
 The curves result from the fits described in the text.
  }
\label{k2stars0}
\end{figure}

A plot of the invariant mass 
of the $K^-\pi^0$ pair versus that of the $K^+\pi^0$ pair is shown in
Fig.~\ref{kkstarpi0}(a)  
(two entries per event) for the \chisq signal region after removing the
$\phi(1020)$ contribution by $|m(K^+ K^-)-m(\phi)|>0.01$~\gevcc. 
Horizontal and vertical bands corresponding to the $K^{*}(892)^{-}$ and
$K^{*}(892)^{+}$, respectively, are visible.
Figure~\ref{kkstarpi0}(b) shows as points the sum of the two projections of 
Fig.~\ref{kkstarpi0}(a);
a large $K^{*}(892)^{\pm}$ signal is evident.
Fitting this distribution with the function used in
Sec.~\ref{kstarxs}, we obtain the number of events corresponding to 
 $K^{*}(892)^{\pm}$ ($7734\pm320$) and $K^{*}(1430)^{\pm}$ ($793\pm137$) production. 
The $K^{*}(1430)^{\pm}$:$K^{*}(892)^{\pm}$ ratio is consistent with that
obtained for neutral $K^*$ production in the \KKppch channel,
but the number of $K^{*}(892)^{\pm}$ combinations in the peak 
is larger  than the total number of  \KKppnt events (5522). 
This indicates the presence of 
some number of correlated $K^{*}(892)^{+} K^{*}(892)^{-}$ pairs.
Fitting the $K^-\pi^0$ mass distribution in each 0.04~\gevcc bin of
$K^+\pi^0$ invariant mass, 
 we obtain the number of $K^{*}(892)^{-}$ and $K^{*}(1430)^{-}$ events 
shown in Fig.~\ref{k2stars0}(a,b).  The correlated production of
$K^{*}(892)^{+} K^{*}(892)^{-}$  and $K^{*}(892)^{+}
K_2^{*}(1430)^{-}$ is clearly seen, and the  fits yield $1750\pm60$ and
$140\pm49$ events, respectively.
Note that $K^{*}(892)^{+} K^{*}(892)^{-}$ accounts for about 30\% of all
\KKppnt events, in contrast with the \KKppch channel, where only
$548\pm263$ events (less than 1\% of the total) are found to result from the
$\Kbar^{*}(892)^{0} K^{*}(892)^{0}$ pair production.  

We find no evidence for resonance production 
in the $\Kp\Km\piz$ or $K^\pm\ppz$ subsystems.
Since the
statistics are low in any given mass interval, we do not attempt to extract a
separate $K^{*}(892)^{+} K^{-}\pi^0 + c.c.$ cross section.
The total \KKppnt cross section is roughly a factor of four lower than
the $K^{*}(892)^{0} K^{-}\pi^+$ cross section observed in the \KKppch
final state.
This is consistent with what is expected from isospin considerations and the
charged versus neutral $K^*$ branching fractions involving charged kaons.
\begin{table*}
\caption{
Summary of the $\ep\en\to\phi(1020)\pi\pi$ cross section, dominated by
  $\phi(1020) f_0(980)$,  $f_0(980)\to\pi\pi$, obtained from 
$\phi(1020)\ppz$  events with  $0.85<m(\ppz)<1.1$~\gevcc.
Errors are statistical only.}
\label{phif0_tab2}
\begin{ruledtabular}
\begin{tabular}{ c c c c c c c c }
\Ecm (GeV) & $\sigma$ (nb)  
& \Ecm (GeV) & $\sigma$ (nb) 
& \Ecm (GeV) & $\sigma$ (nb) 
& \Ecm (GeV) & $\sigma$ (nb)  
\\
\hline

 1.9000 &  0.15 $\pm$  0.07 & 2.1000 &  0.45 $\pm$  0.11 & 2.3200 &  0.12 $\pm$  0.06 & 2.9200 &  0.02 $\pm$  0.01 \\
 1.9400 &  0.14 $\pm$  0.06 & 2.1400 &  0.47 $\pm$  0.11 & 2.4000 &  0.14 $\pm$  0.03 & 3.0800 &  0.05 $\pm$  0.01 \\
 1.9800 &  0.19 $\pm$  0.09 & 2.1800 &  0.55 $\pm$  0.10 & 2.4800 &  0.12 $\pm$  0.03 & 3.2400 &  0.01 $\pm$  0.01 \\
 2.0200 &  0.47 $\pm$  0.11 & 2.2200 &  0.11 $\pm$  0.05 & 2.6000 &  0.04 $\pm$  0.01 & 3.4000 &  0.01 $\pm$  0.00 \\
 2.0600 &  0.22 $\pm$  0.08 & 2.2600 &  0.13 $\pm$  0.05 & 2.7600 &  0.04 $\pm$  0.01 &        &                    \\

\end{tabular}
\end{ruledtabular}
\end{table*}

\subsection{\boldmath The $\phi(1020)\ppz$ Intermediate State}
\label{sec:phipipi2}

\begin{figure}[tbh]
\includegraphics[width=1.0\linewidth]{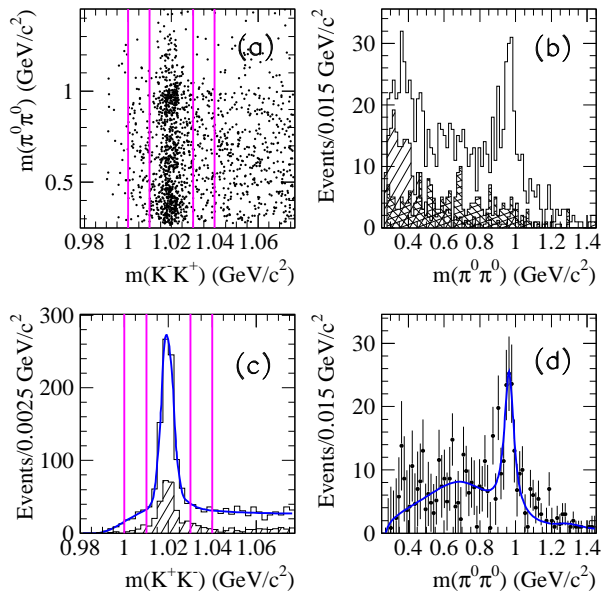}
\vspace{-0.4cm}
\caption{
  (a) Plot of the $\ppz$ invariant mass versus the $\Kp\Km$ invariant
  mass for all selected \KKppnt events;
  (b) the $\ppz$ invariant mass projections for events in the $\phi$ peak
  (open histogram), sidebands (cross-hatched), and control region
  (hatched);
  (c) the $\Kp\Km$ mass projection for events in the signal (open) and
  control (hatched) regions;
  (d) the difference between the open histogram  and sum of the 
other contributions to (b).
  }
\label{phif0_sel2}
\end{figure}

The selection of events containing  $\phi(1020) \!\to\! \KpKm$ decays
follows that in Sec.~\ref{sec:phipipi}. 
Figure~\ref{phif0_sel2}(a) shows the
plot of the invariant mass of the $\ppz$ pair versus that of the 
\KpKm pair.
The $\phi$ resonance is visible as a vertical band,
whose intensity decreases with increasing \ppz mass except for an enhancement
in the $f_{0}(980)$ region (Fig.~\ref{phif0_sel2}(b)).
The $\phi$ signal is also visible in the $\Kp\Km$ invariant mass
projection for events in the control region,
shown in Fig.~\ref{phif0_sel2}(c).
The relative non-$\phi$ background is smaller than in the \KKppch
mode, 
but there is a large background from ISR $\phi\pi^0$, $\phi\eta$
and/or $\phi\ppz\piz$ events, 
as indicated by the control region histogram (hatched) in 
Fig.~\ref{phif0_sel2}(c).
The contributions from non-ISR and ISR $\pipi\ppz$ events are negligible.
Selecting $\phi$ candidate and side band events as for the \KKppch mode
(vertical lines in Figs.~\ref{phif0_sel2}(a,c)),
we obtain the \ppz mass projections shown as the open and cross-hatched
histograms, respectively, in Fig.~\ref{phif0_sel2}(b).
Control region events (hatched histogram) are concentrated at low mass values
in Fig.~\ref{phif0_sel2}(b), and
a peak corresponding to the $f_{0}(980)$ is visible over a relatively
low background. 

In Fig.~\ref{phif0_sel2}(d) we show the $\ppz$ mass distribution
associated with $\phi$ production after subtraction of all background
contributions. The distribution is consistent in shape with that of
Fig.~\ref{phif0_sel}(d), but with a data sample which is about six times
smaller. 

\begin{figure}[tbh]
\includegraphics[width=0.9\linewidth]{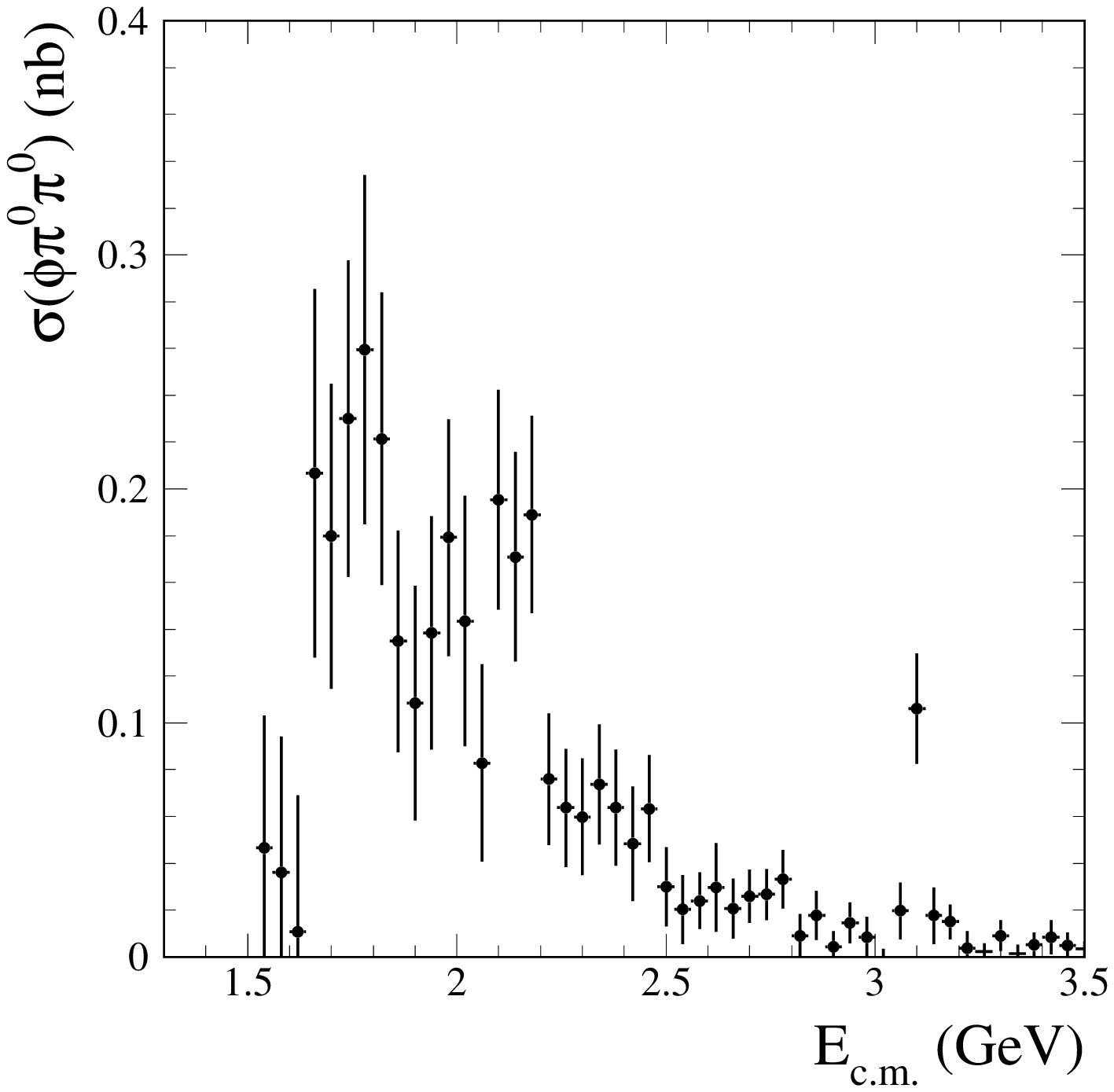}
\vspace{-0.4cm}
\caption{
  Cross section for the reaction $\epem \!\!\to\! \phi(1020)
  \ppz$ 
  as a function of \epem c.m.\ energy obtained from the
  \KKppnt final state.
}
\label{phi2pi0xs}
\end{figure}

We obtain the number of $\epem \!\!\to\! \phi\ppz$ events in 0.04~\gevcc
intervals of
$\phi\ppz$ invariant mass by fitting the $K^+ K^-$ invariant mass
projection in that interval to the $\phi$ signal,
after subtracting the non-\KKppnt background, the same way as described in
Sec.~\ref{sec:phipipi}. The obtained cross section is shown in
Fig.~\ref{phi2pi0xs} and is very similar to that obtained from the
\KKppch final state shown in Fig.~\ref{phipipixs}. The errors shown
reflect not only that there are six times fewer events, but also
a much larger background level.

As before, we defer discussion to Secs.~\ref{phipipistudy} and ~\ref{phif0bump}.

\subsection{\boldmath The $\phi(1020) f_0(980)$ Intermediate State}
\label{sec:phif02}

Since the background under the $f_0(980)$ peak in Figs.~\ref{phif0_sel2}(b,d)
is 25\% or less, we are able to extract the $\phi(1020) f_0(980)$ 
contribution. 
As in Sec.~\ref{sec:phif01}, we require the dipion mass to be in the
range 0.85--1.10~\gevcc and fit the background-subtracted $\Kp\Km$ 
mass projection in each 0.04~\gevcc interval of \KKppnt mass to obtain the number of   
$\phi f_0$ events.
Again, some $\phi\ppz$ events are present in which the \ppz
pair is not produced through the $f_0$.

\begin{figure}[tbh]
\includegraphics[width=0.9\linewidth]{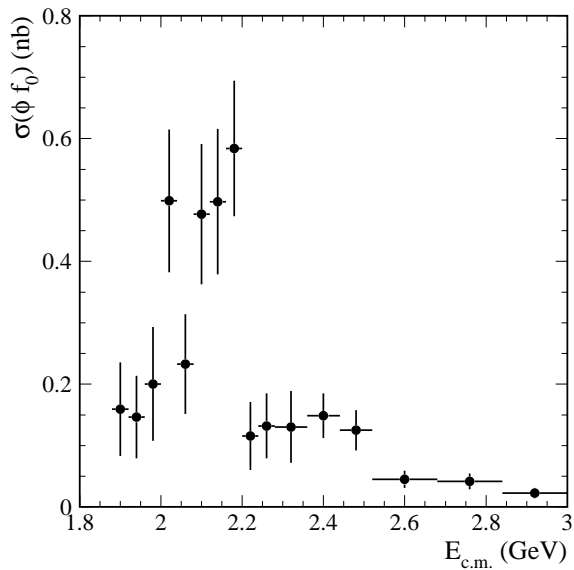}
\vspace{-0.4cm}
\caption{
  Cross section for the reaction $\epem \!\!\to\! \phi(1020)
  f_{0}(980)$, $f_0\to\pi\pi$ 
  as a function of \epem c.m.\ energy obtained from the
  \KKppnt final state.
}
\label{phif0xs2}
\end{figure}

We convert the number of $f_0(980)$ events in each mass interval into a
measurement of the $\epem \!\!\to\! \phi(1020) f_{0}(980)$ cross
section as described previously, and divide by the
$f_{0}(980) \!\!\to\! \ppz$ branching fraction of 1/3
to obtain the $f_{0}(980)\to\pi\pi$ value.
The cross section, corrected for the $\phi(1020)\to\KpKm$ decay rate,
 is shown in Fig.~\ref{phif0xs2} as a function of
\Ecm and is listed in Table~\ref{phif0_tab2}.
Due to the smaller number of events, we have used larger 
intervals at higher energies.
The overall shape is consistent with that obtained from the \KKppch final state 
(see Fig.~\ref{phif0xs}),
and there seems to be a sharp drop near 2.2~\gev;
however, the statistical errors are large and no conclusion can be drawn
from this mode alone.
Possible interpretations are discussed in Section~\ref{phif0bump}.

\section{\boldmath The \KKKK Final State}
\subsection{Final Selection and Background}

\begin{figure}[tbh]
\includegraphics[width=0.9\linewidth]{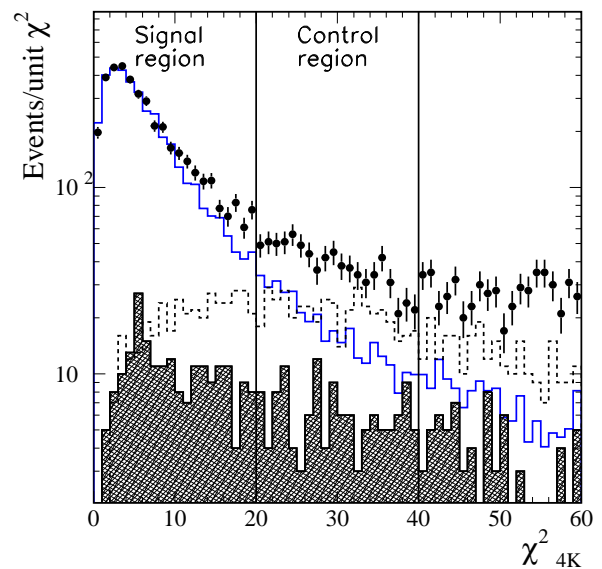}
\vspace{-0.4cm}
\caption{
  Distribution of \chisq from the three-constraint fit for \KKKK
  candidates in the data (points).
  The open histogram is the distribution for simulated signal events,
  normalized as described in the text.
  The shaded histogram represents the background from
  non-ISR events, estimated
  as described in the text. The region defined by the dashed contour
is for  simulated ISR $\Kp\Km\pipi$ events with at least one pion
  misidentified as a kaon.  
  }
\label{4k_chi2_all}
\end{figure}

Figure~\ref{4k_chi2_all} shows the distribution of \chifourK for the
\KKKK candidates as points. The open histogram is the distribution for
simulated \KKKK events,
normalized to the data in the region $\chifourK \! <\! 5$ where
the relative contributions of the backgrounds and radiative corrections are small.
The shaded histogram represents the background from non-ISR 
$\epem \!\!\to\! \qqbar$ events, evaluated as for the other modes.
The region defined by the dashed contour represents the background
from simulated ISR \KKppch events with at least one charged pion 
misidentified as a kaon. 

\begin{figure}[tbh]
\includegraphics[width=0.9\linewidth]{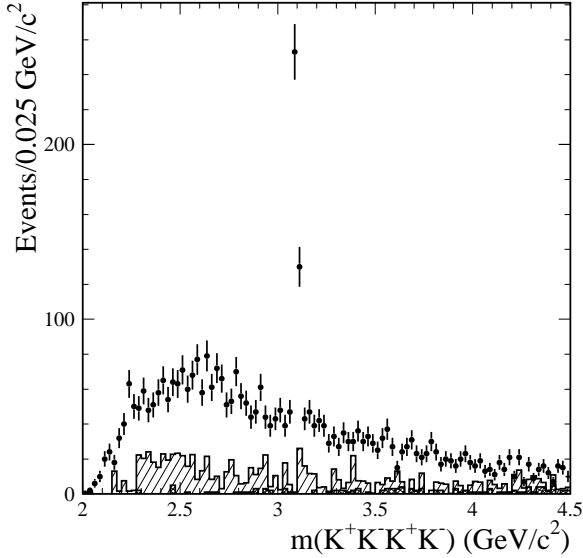}
\vspace{-0.4cm}
\caption{
  Invariant mass distribution for \KKKK candidates in the data (points).
  The shaded histogram represents the non-ISR background, and  
the hatched region is for the ISR background from the control region,
  which is dominated by the contribution from misidentified ISR \KKppch events.
  }
\label{4k_babar}
\end{figure}

We define signal and control regions by $\chifourK\! <\! 20$ and 
$20\! <\! \chifourK\! <\! 40$, respectively.
The signal region contains 4190 data and 14904 simulated events, 
and the control region contains 877 data and 1437 simulated events.
Figure~\ref{4k_babar} shows the \KKKK invariant mass distribution from
threshold up to 4.5~\gevcc for events in the signal region as points
with errors.
The \qqbar background (shaded histogram) is small at all masses.
Since the ISR \KKppch background does not peak at low \chifourK values, 
we include it in the background evaluated from the control region,
according to the method explained in Sec.~\ref{sec:selection1}.
It dominates this background, which is about 20\% for 2.3-2.6~\gevcc and
10\% or lower at all other mass values.
The total background is shown as the hatched histogram in Fig.~\ref{4k_babar}.

We subtract the sum of backgrounds from the number of selected events
in each mass interval to obtain the number of signal events.
Considering the uncertainties in the cross sections for the background processes,
the normalization of events in the control region, 
and the simulation statistics,
we estimate that the systematic uncertainty on the signal yield is less than
5\% in the 2--3~\gevcc region, but it
 increases to about 10\% above 3~\gevcc.

\begin{figure}[tbh]
\includegraphics[width=0.9\linewidth]{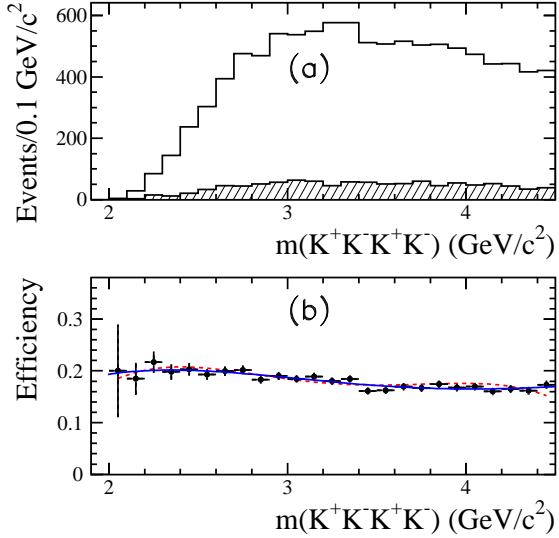}
\vspace{-0.4cm}
\caption{
  (a) Invariant mass distributions for simulated \KKKK events in the
  signal (open) and control (hatched) regions (see Fig.~\ref{4k_chi2_all});
  (b) net reconstruction and selection efficiency as a function of mass
  obtained from this simulation; 
  the curves represent third-order polynomial fits for
  the phase space model (solid) and the $\phi K^+ K^-$ model (dashed).
  }
\label{mc_acc4}
\end{figure}

\subsection{Selection Efficiency}

The detection efficiency is determined as for the other two final states.
Figure~\ref{mc_acc4}(a) shows the simulated \KKKK invariant-mass
distributions in the signal and control regions from the phase space
model.
We divide the number of reconstructed  events in each
mass interval by the number generated in that interval
to obtain the efficiency shown by the points in Fig.~\ref{mc_acc4}(b).
It is quite uniform, and we fit the measurements using a third-order
polynomial, which we then use to obtain the cross section.
As discussed previously, this efficiency includes
the difference between the EMC acceptance and
the region of ISR photon simulation.
A simulation assuming dominance of the $\phi\Kp\Km$ channel, 
with the \KpKm pair in an angular-momentum S-wave state, gives
consistent results, as shown by the
dashed curve in Fig.~\ref{mc_acc4}(b),
and we estimate a  5\% systematic  uncertainty associated with the difference.
We correct only for mis-modeling of the track-finding and 
ISR-photon-detection efficiency
as in Sec.~\ref{sec:eff1}.

\subsection{\boldmath Cross Section for $\epem\to K^+ K^- K^+ K^-$}
\label{sec:4kxs}

\begin{figure}[tbh]
\includegraphics[width=0.9\linewidth]{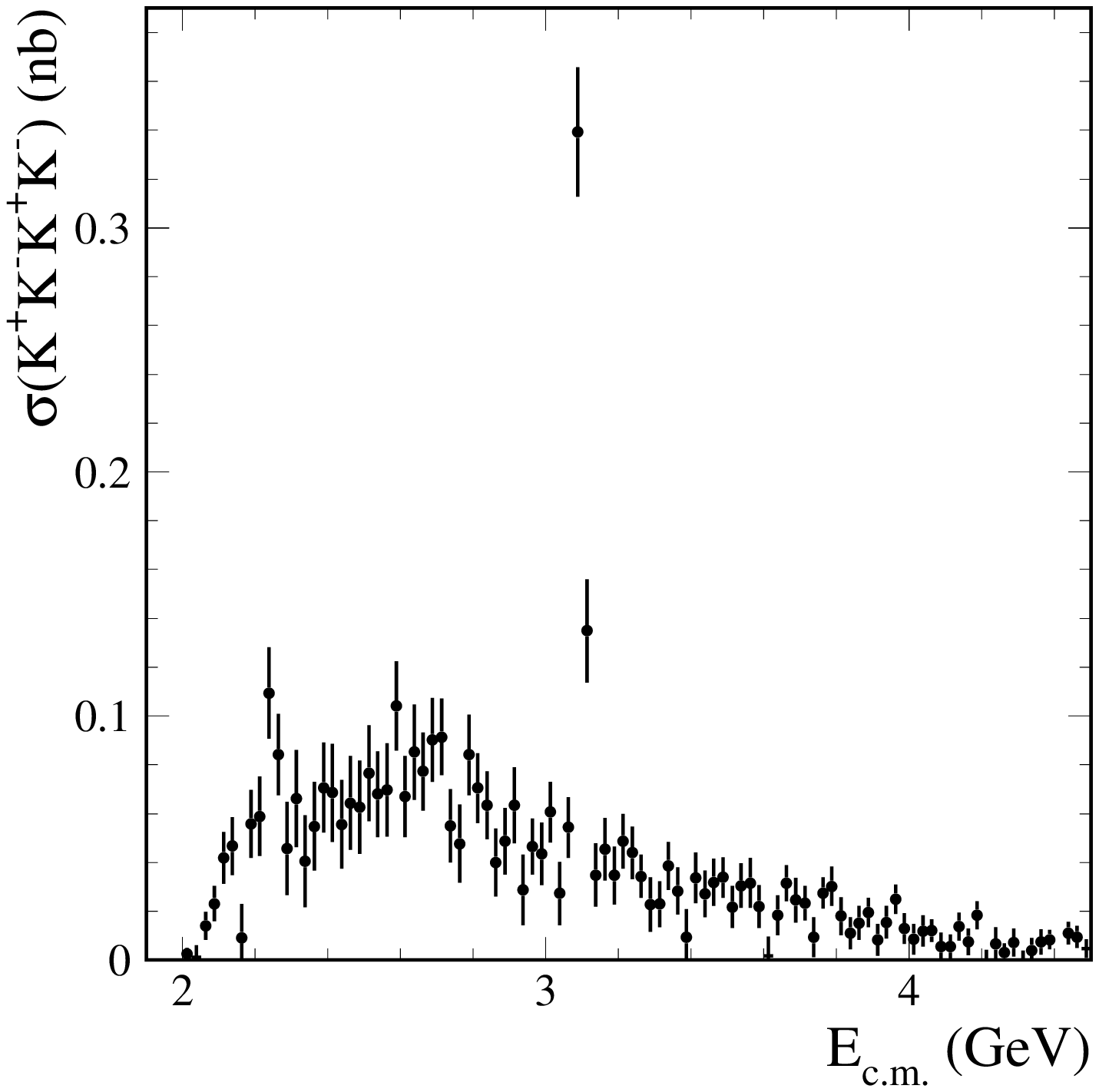}
\vspace{-0.5cm}
\caption{
  The $\epem \!\!\to\! \KKKK$ cross section as a function of \epem
  c.m.\ energy measured with ISR data at \babar.
  The errors are statistical only.
  }
\label{4k_ee_babar}
\end{figure}

We calculate the $\epem \!\!\to\! \KKKK$ cross section in
0.025~\gev intervals of \Ecm from the analog of Eq.(\ref{xseqn}),
using the invariant mass of the \KKKK system to determine the
c.m.\ energy.
We show the cross section in Fig.~\ref{4k_ee_babar}, and list the
measured values in
Table~\ref{4k_tab}.  The cross section increases from threshold
to a peak value of about 0.1~nb near 2.7~\gev, 
then decreases slowly with increasing energy.
The only statistically significant narrow structures are the large $J/\psi$ peak and 
a possible narrow structure near ~2.3~\gev, which will be discussed in
Sec.~\ref{sec:phif03}.
Again, the differential luminosity contribution in each \Ecm interval
includes corrections for vacuum polarization that
should be omitted for the calculations of $a_\mu$.
This measurement supersedes our previous result~\cite{isr4pi}.

\begin{table*}
\caption{Summary of the cross section measurements for $\ep\en\to K^+ K^- K^+ K^-$. 
Errors are statistical only.}
\label{4k_tab}
\begin{ruledtabular}
\begin{tabular}{ c c c c c c c c }
\Ecm (GeV) & $\sigma$ (nb)  
& \Ecm (GeV) & $\sigma$ (nb) 
& \Ecm (GeV) & $\sigma$ (nb) 
& \Ecm (GeV) & $\sigma$ (nb)  
\\
\hline
 2.0125 & 0.002 $\pm$ 0.002 & 2.6375 & 0.100 $\pm$ 0.016 & 3.2625 & 0.035 $\pm$ 0.008 & 3.8875 & 0.020 $\pm$ 0.005 \\
 2.0375 & 0.003 $\pm$ 0.004 & 2.6625 & 0.083 $\pm$ 0.013 & 3.2875 & 0.030 $\pm$ 0.009 & 3.9125 & 0.011 $\pm$ 0.005 \\
 2.0625 & 0.013 $\pm$ 0.005 & 2.6875 & 0.097 $\pm$ 0.014 & 3.3125 & 0.027 $\pm$ 0.008 & 3.9375 & 0.017 $\pm$ 0.005 \\
 2.0875 & 0.021 $\pm$ 0.007 & 2.7125 & 0.094 $\pm$ 0.013 & 3.3375 & 0.040 $\pm$ 0.008 & 3.9625 & 0.023 $\pm$ 0.006 \\
 2.1125 & 0.040 $\pm$ 0.010 & 2.7375 & 0.064 $\pm$ 0.012 & 3.3625 & 0.032 $\pm$ 0.008 & 3.9875 & 0.015 $\pm$ 0.005 \\
 2.1375 & 0.046 $\pm$ 0.010 & 2.7625 & 0.061 $\pm$ 0.012 & 3.3875 & 0.021 $\pm$ 0.009 & 4.0125 & 0.012 $\pm$ 0.005 \\
 2.1625 & 0.021 $\pm$ 0.010 & 2.7875 & 0.091 $\pm$ 0.014 & 3.4125 & 0.037 $\pm$ 0.009 & 4.0375 & 0.015 $\pm$ 0.005 \\
 2.1875 & 0.057 $\pm$ 0.012 & 2.8125 & 0.074 $\pm$ 0.012 & 3.4375 & 0.031 $\pm$ 0.008 & 4.0625 & 0.012 $\pm$ 0.004 \\
 2.2125 & 0.066 $\pm$ 0.013 & 2.8375 & 0.067 $\pm$ 0.012 & 3.4625 & 0.035 $\pm$ 0.008 & 4.0875 & 0.008 $\pm$ 0.005 \\
 2.2375 & 0.112 $\pm$ 0.016 & 2.8625 & 0.050 $\pm$ 0.011 & 3.4875 & 0.034 $\pm$ 0.007 & 4.1125 & 0.008 $\pm$ 0.004 \\
 2.2625 & 0.086 $\pm$ 0.014 & 2.8875 & 0.054 $\pm$ 0.011 & 3.5125 & 0.025 $\pm$ 0.007 & 4.1375 & 0.015 $\pm$ 0.005 \\
 2.2875 & 0.063 $\pm$ 0.015 & 2.9125 & 0.073 $\pm$ 0.013 & 3.5375 & 0.033 $\pm$ 0.008 & 4.1625 & 0.010 $\pm$ 0.004 \\
 2.3125 & 0.083 $\pm$ 0.016 & 2.9375 & 0.042 $\pm$ 0.011 & 3.5625 & 0.035 $\pm$ 0.008 & 4.1875 & 0.018 $\pm$ 0.005 \\
 2.3375 & 0.060 $\pm$ 0.014 & 2.9625 & 0.048 $\pm$ 0.010 & 3.5875 & 0.025 $\pm$ 0.007 & 4.2125 & 0.003 $\pm$ 0.004 \\
 2.3625 & 0.070 $\pm$ 0.014 & 2.9875 & 0.050 $\pm$ 0.010 & 3.6125 & 0.008 $\pm$ 0.006 & 4.2375 & 0.012 $\pm$ 0.005 \\
 2.3875 & 0.083 $\pm$ 0.015 & 3.0125 & 0.062 $\pm$ 0.010 & 3.6375 & 0.020 $\pm$ 0.007 & 4.2625 & 0.004 $\pm$ 0.003 \\
 2.4125 & 0.087 $\pm$ 0.016 & 3.0375 & 0.037 $\pm$ 0.010 & 3.6625 & 0.031 $\pm$ 0.007 & 4.2875 & 0.009 $\pm$ 0.005 \\
 2.4375 & 0.071 $\pm$ 0.014 & 3.0625 & 0.057 $\pm$ 0.010 & 3.6875 & 0.028 $\pm$ 0.008 & 4.3125 & 0.003 $\pm$ 0.004 \\
 2.4625 & 0.079 $\pm$ 0.016 & 3.0875 & 0.334 $\pm$ 0.023 & 3.7125 & 0.023 $\pm$ 0.006 & 4.3375 & 0.006 $\pm$ 0.004 \\
 2.4875 & 0.080 $\pm$ 0.015 & 3.1125 & 0.151 $\pm$ 0.017 & 3.7375 & 0.014 $\pm$ 0.006 & 4.3625 & 0.009 $\pm$ 0.004 \\
 2.5125 & 0.093 $\pm$ 0.016 & 3.1375 & 0.045 $\pm$ 0.010 & 3.7625 & 0.026 $\pm$ 0.006 & 4.3875 & 0.008 $\pm$ 0.004 \\
 2.5375 & 0.079 $\pm$ 0.014 & 3.1625 & 0.053 $\pm$ 0.010 & 3.7875 & 0.031 $\pm$ 0.007 & 4.4125 & 0.001 $\pm$ 0.004 \\
 2.5625 & 0.086 $\pm$ 0.015 & 3.1875 & 0.041 $\pm$ 0.010 & 3.8125 & 0.021 $\pm$ 0.006 & 4.4375 & 0.012 $\pm$ 0.004 \\
 2.5875 & 0.110 $\pm$ 0.015 & 3.2125 & 0.051 $\pm$ 0.009 & 3.8375 & 0.013 $\pm$ 0.005 & 4.4625 & 0.010 $\pm$ 0.004 \\
 2.6125 & 0.077 $\pm$ 0.013 & 3.2375 & 0.046 $\pm$ 0.009 & 3.8625 & 0.018 $\pm$ 0.006 & 4.4875 & 0.006 $\pm$ 0.003 \\

\end{tabular}
\end{ruledtabular}
\end{table*}

The simulated \KKKK invariant mass resolution is 3.0~\mevcc in the 
2.0--2.5~\gevcc range, increasing with mass to 4.7~\mevcc in the 
2.5--3.5~\gevcc range, and to about 6.5~\mevcc at higher masses.
Since the cross section has no sharp structure except for the  $J/\psi$
peak, we again make no correction for resolution.
The errors shown in Fig.~\ref{4k_ee_babar} and listed in Table~\ref{4k_tab} are
statistical only.
The point-to-point systematic uncertainties are much smaller, and
the errors on the normalization are summarized in Table~\ref{error3_tab},
along with the corrections applied to the measurements.
The total correction is +4.0\%, 
and the total systematic uncertainty is 9\% at low mass, linearly
increasing to 13\% above 3~\gevcc.
\begin{figure}[tbh]
\includegraphics[width=0.9\linewidth]{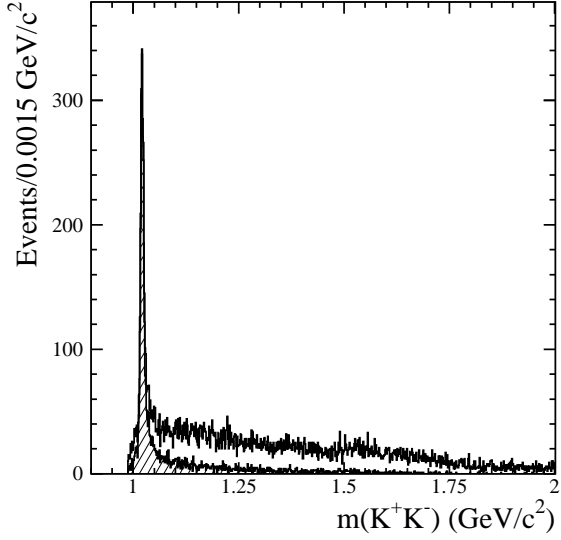}
\vspace{-0.6cm}
\caption{
  Invariant mass distribution for all \KpKm pairs in selected
  $\epem \!\!\to\! \KKKK$ events (open histogram), and for the 
  combination in each event closest to the $\phi$-meson mass (hatched).
  }
\label{mkk_phi}
\end{figure}
\begin{table}[tbh]
\caption{
Summary of corrections and systematic uncertainties 
for the $\epem \!\!\to\! \KKKK$    cross section measurements.
The total correction is the linear sum of the individual corrections, and the
total uncertainty is the sum in quadrature of the separate uncertainties.
  }
\label{error3_tab}
\begin{ruledtabular}
\begin{tabular}{l c r@{}l} 
     Source             & Correction & \multicolumn{2}{c}{Uncertainty}   \\
\hline
                        &            &        &              \\[-0.2cm]
Rad. Corrections        &  --        &  1\%   &              \\
Backgrounds             &  --        &  5\%   &, $\Ecm <3~\gev$ \\
                        &            &  5-10\%&, $\Ecm >3~\gev$ \\
Model Dependence        &  --        &  5\%   &              \\
\chifourK Distribution  &  --        &  3\%   &              \\ 
Tracking Efficiency     & +3.0\%     &  2\%   &              \\
Kaon ID Efficiency      &  --        &  4\%   &              \\
ISR-photon Efficiency   & +1.0\%     &  0.5\% &            \\ 
ISR Luminosity          &  --        &  3\%   &              \\[0.1cm]
\hline
                        &            &        &             \\[-0.2cm]
Total                   & +4.0\%     &  9\%   &, $\Ecm <3~\gev$ \\
                        &            &  9-13\%&, $\Ecm >3~\gev$ \\
\end{tabular}
\end{ruledtabular}
\end{table}
\begin{figure*}[t]
\includegraphics[width=0.25\linewidth]{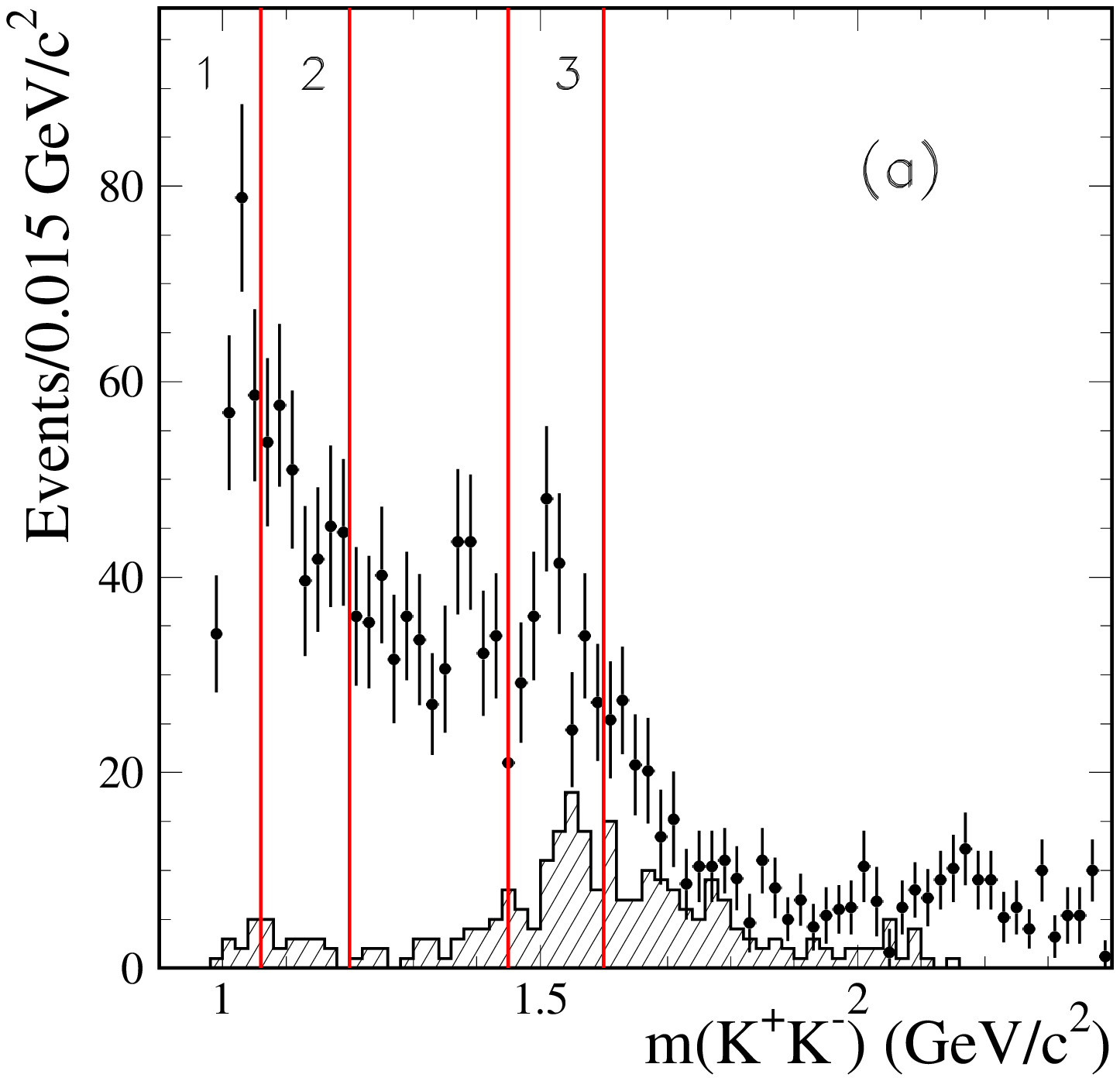}
\includegraphics[width=0.25\linewidth]{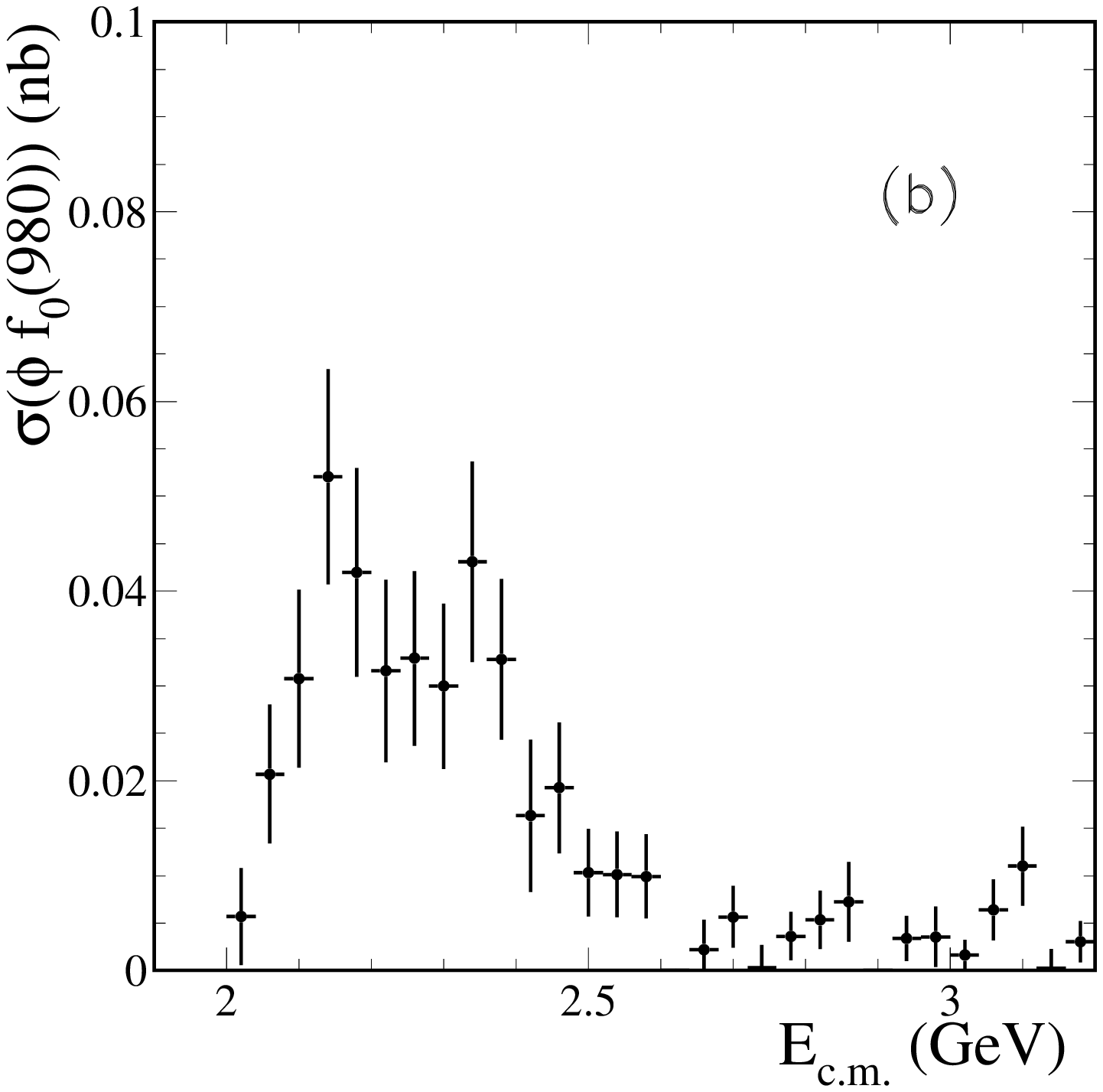}
\includegraphics[width=0.25\linewidth]{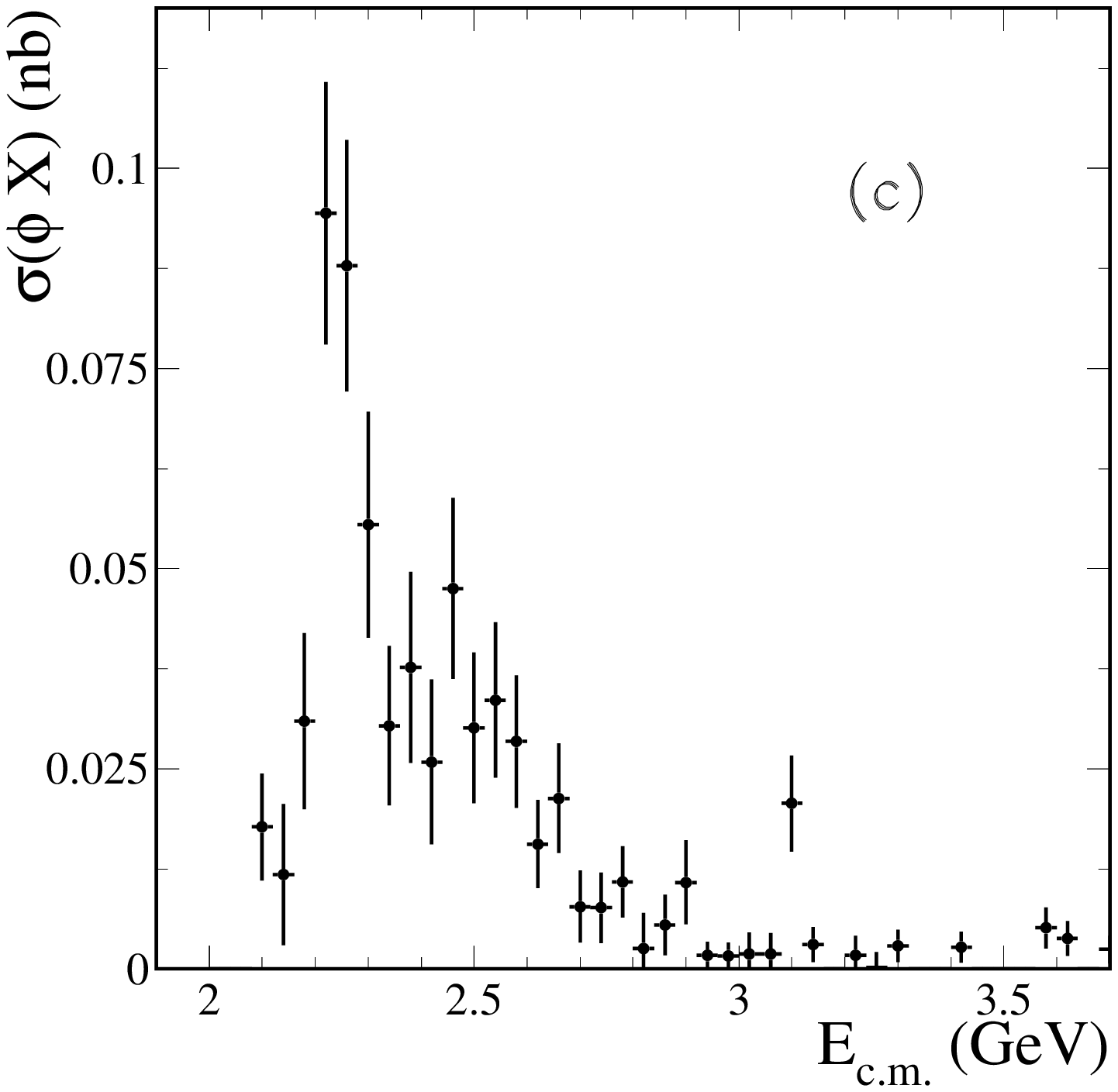}
\includegraphics[width=0.25\linewidth]{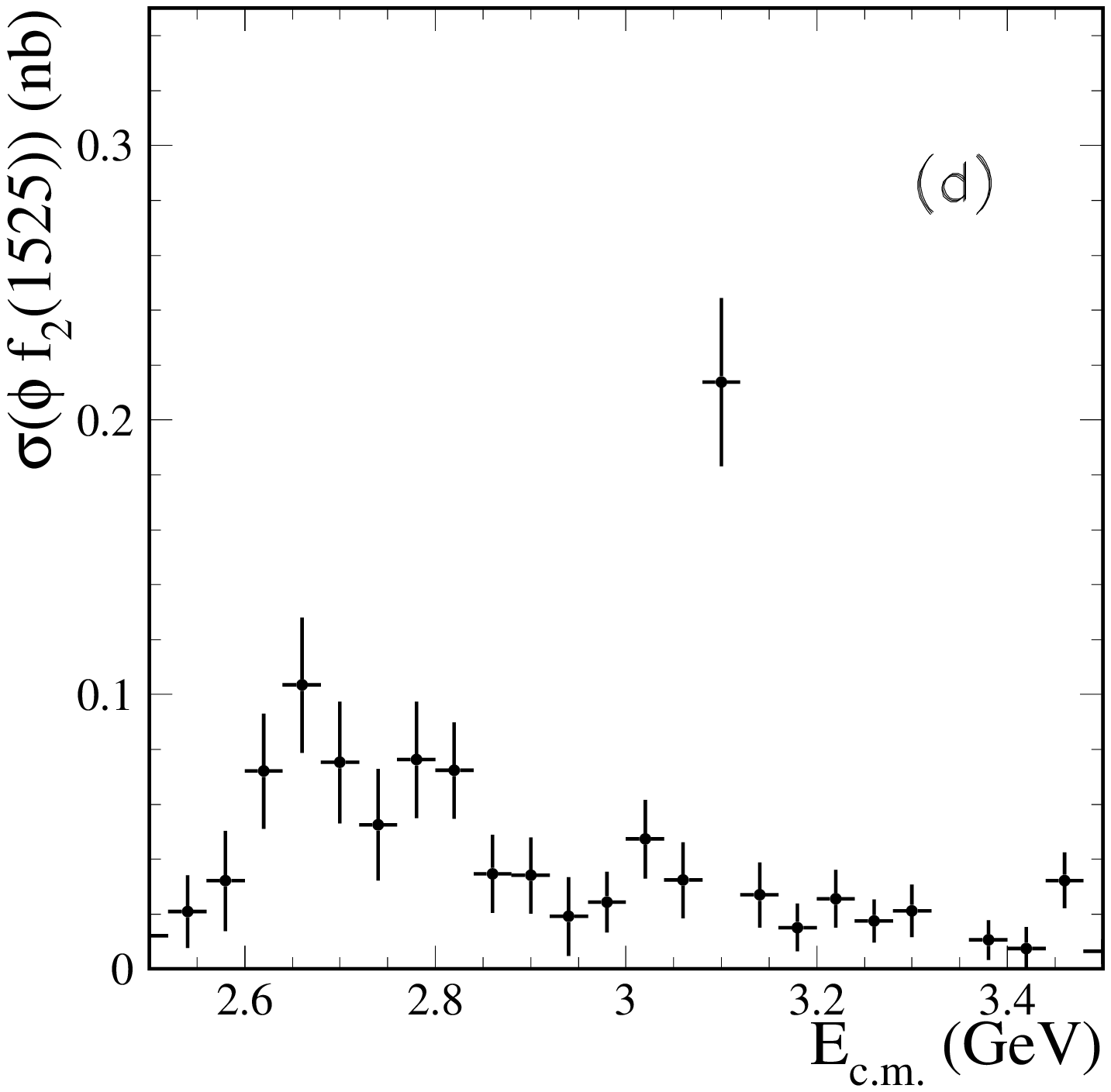}
\vspace{-0.4cm}
\caption{ (a)
The  invariant mass distribution for \KpKm pairs in events in which
  the other \KpKm pair has mass closest to, and within 10~\mevcc of, 
  the nominal $\phi$ mass (open histogram); events within $\pm 50\mevcc$
of the  $J/\psi$ mass have been excluded. 
The hatched histogram corresponds to events with \KKKK invariant mass in the
  $J/\psi$ peak. The numbered regions of the combined histograms from (a) are
 used to calculate the cross sections shown in Figs. (b), (c) and (d)
 for regions 1, 2 and 3 respectively.
  }
\label{mkk_notphi}
\end{figure*}

\subsection{\boldmath The $\phi(1020) \KpKm$  Intermediate State}
\label{sec:phif03}

Figure~\ref{mkk_phi} shows the invariant mass distribution for all
\KpKm pairs in the selected \KKKK events (4 entries per event) as the
open histogram.
A prominent $\phi$ peak is visible along with a possible excess near ~1.5~\gevcc.
The hatched histogram is for the pair in each event with mass closest
to the nominal $\phi$ mass,
and indicates that the $\phi \KpKm$ channel dominates the \KKKK final state; 
we do not see any other significant contribution. 
If the invariant mass of the $\KpKm$ pair that
is closest to the $\phi$ mass is within $\pm$10~\mevcc of the $\phi$
peak, then we include the invariant mass of the other $\KpKm$ combination
in Fig.~\ref{mkk_notphi}(a).
Events with \KKKK mass within $\pm 50\mevcc$ of
  the $J/\psi$ mass are excluded.
Events within $\pm 50$~\mevcc of the $J/\psi$ mass 
are shown as the
hatched histogram. The latter is in agreement with results from the BES
experiment~\cite{bes4k}, for  which 
the structures around 1.5, 1.7, and 2.0~\gevcc were studied in detail.
For the dots with error bars
there is an enhancement at threshold that can be
interpreted as being due to $f_0(980)\to\KpKm$ decay.
This is expected in light of the $\phi f_0$ cross sections measured
above in the \KKppch and \KKppnt final states, but a contribution from
the $a_0(980)\to\KpKm$ can not be excluded. 
For the combined histograms of Fig.~\ref{mkk_notphi}(a), we
select events with $m(\KpKm)<1.06$~\gevcc (shown as region 1) and
calculate a cross
section enriched in the $\epem \to \phi f_0(980)$ reaction 
(Fig.~\ref{mkk_notphi}(b)).
A bump at $\Ecm = 2.175\gev$  is seen;
however, the small number of events and uncertainties in the $f_0(980) \!\to\! \KpKm$
line-shape do not allow a meaningful extraction of the cross section
for this $f_0(980)$ decay mode.

A clear signal corresponding to the $f'_2(1525)$ is seen in both histograms shown in 
Fig.~\ref{mkk_notphi}(a). The $f'_2(1525)$ region is defined by
$1.45<m(\KpKm)<1.6$~\gevcc, and is indicated as region 3 in Fig.~\ref{mkk_notphi}(a).
  The corresponding cross section is shown
in  Fig.~\ref{mkk_notphi}(d) and exhibits a broad (about
0.10-0.15~\gev) structure at 2.7~\gev and a strong  $J/\psi$ signal. 
In Fig.~\ref{mkk_notphi}(a)(open histogram)  there is an indication
of structure for \KpKm invariant mass in the 1.3-1.4~\gevcc region; this
may correspond to production of the $\phi f_0(1370)$ final state.

Finally, we tried to find a region of  \KpKm invariant mass
corresponding to the spike seen at about 2.3~\gev in the total  $\epem\to\KKKK$
cross section shown in Fig.~\ref{4k_ee_babar}. This spike is much more
significant if we require $1.06<m(\KpKm)<1.2$~\gevcc,
shown as region 2 in  Fig.~\ref{mkk_notphi}(a),
 with corresponding cross section shown in Fig.~\ref{mkk_notphi}(c).
We have no explanation of this structure.

We observe no significant structure in the $K^+ K^- K^{\pm}$ mass distribution.

We use the  $\phi\KpKm$ events to investigate the possibility that part of our
$\phi\pipi$ signal is due to $\phi\KpKm$ events with the two kaons interpreted
as pions.
No structure is present in the resulting \KKppch invariant mass
distribution.

\section{\boldmath The  $\epem \!\to \phi\pi\pi$ cross section}
\label{phipipistudy}
\begin{figure}[tbh]
\includegraphics[width=0.9\linewidth]{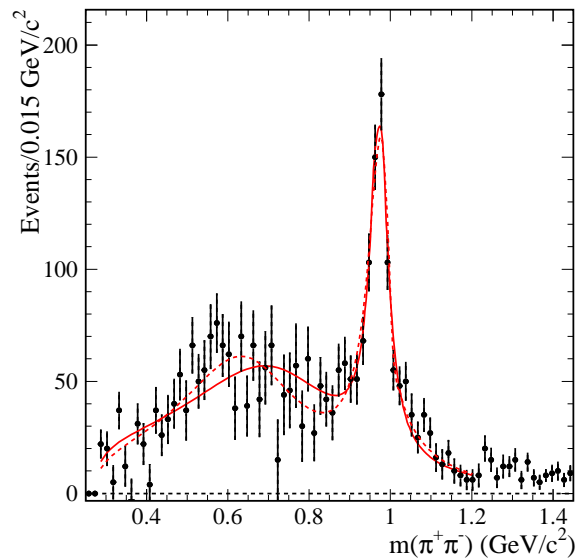}
\vspace{-0.3cm}
\caption{
The two-Breit-Wigner fit to the $\pipi$ invariant mass distribution
of Fig.~\ref{phif0_sel}(d).
The dashed curve corresponds to the inclusion of the partial width to
$K\Kbar$ in the propagator of the $f_0(980)$ BW.
 }
\label{pi2fit2bw}
\end{figure}
We next perform a more detailed study in the \Ecm region  from
threshold to 3.0~\gev of the 
$\epem\to\phi(1020)\pi\pi$ cross section.
For this study we use the cross section for the $\phi\pipi$
final state shown in Figs.~\ref{phipipixs} and 
\ref{phif0xs}, after scaling by a factor of 1.5 to take into account the
$\phi\ppz$ contribution.
The cross section for the $\phi\ppz$ (see Fig.~\ref{phi2pi0xs}) 
final state does not help much due to large statistical errors. 
There are at least two candidate resonant 
structures in Fig.~\ref{phipipixs}. These are associated with the
peaks observed at 1.7~\gev and at 2.1~\gev. 
As shown in Sec.~\ref{sec:phif01}, the latter is related to 
$\phi(1020) f_0(980)$ production, while the best candidate for
 the former may be 
the $\phi(1680)$, which is a radial excitation of the $s \bar s$ state
decaying predominantly to $K^*(892) \bar K$~\cite{isrkkpi}. 
This would be another confirmation of the decay of this state to
$\phi(1020)\pi\pi$, previously reported in Refs.~\cite{isr2k2pi,belle_phif0}. 

As discussed in Sec.~\ref{sec:phipipi} we associate 
the narrow peak in the $\pipi$ invariant mass distribution, shown on a
larger scale in Fig.~\ref{pi2fit2bw},  with the $f_0(980)$ (denoted as the
$f_0$ meson), and observe a broad enhancement at about 0.6~\gevcc; the angular
distributions of Fig.~\ref{phi_angle} justify that these structures are in
an S-wave state. 
 This low mass bump 
 can not be formed by pure three-body phase space. Indeed, the
$\phi(1020)\pi\pi$ threshold is 1.3~\gev, but the observed cross
section has a slow rise starting at 1.4~\gev. This indicates that the
observed structure could  be a result of 
$f_0(600)$ resonance decay. 

The observed two-pion-mass shape of the $f_0(600)$ (denoted as the
$\sigma$ meson) is distorted by
the $\phi(1020)\pi\pi$ final state. This is less of an issue for the
narrower $f_0(980)$. Nevertheless, 
to obtain mass and width parameter values
for these states, we fit the data of Fig.~\ref{pi2fit2bw} using a
function consisting of an incoherent sum of two S-wave relativistic
BW intensity
distributions, modified to account
for the two pion phase space.
The fit values obtained are
\begin{equation}
 m_{\sigma}=(0.692\pm0.030)~\gevcc, 
\Gamma_{\sigma}=(0.538\pm0.075)~\gev,
\label{sigpar}
\end{equation}
and
\begin{equation}
m_{f_0}=(0.972\pm0.002)~\gevcc, \Gamma_{f_0}=(0.056\pm0.011)~\gev,
\label{f0par}
\end{equation}
and the fit result is represented by the solid curve in
Fig.~\ref{pi2fit2bw}. Note that the $f_0(980)$ parameters are consistent
with the PDG values~\cite{PDG}, indicating that interference
with  the $f_0(600)$ (or $\pi\pi$ coherent continuum) is minimal. This is expected
because events with $m(\pi\pi)<0.85$~\gevcc are associated with
the resonance at 1.7~\gevcc in the $\phi(1020)\pi\pi$ mass, while the $f_0(980)$
contributes only to a structure above 2~\gevcc (see Fig.~\ref{phif0xs}). 
To confirm this, we examine two $m(\pipi)$ distributions using the selections shown
in Fig.~\ref{phif0_sel}, but  for events  with either $m(\phi\pipi)<1.95$~\gevcc
or   $1.95<m(\phi\pipi)<3.0$~\gevcc.
For the first case we observe only the bump at  0.6~\gevcc of
Fig.~\ref{pi2fit2bw}, with no evidence for the $f_0(980)$. 
For the second case we see a clear $f_0(980)$ signal but no
evidence for the $f_0(600)$. 
We fit each distribution
the same way as the data in Fig.~\ref{pi2fit2bw}. The resulting parameters for the $f_0(600)$
and $f_0(980)$ are in agreement with those presented above.

The dashed curve of Fig.~\ref{pi2fit2bw} is obtained when the $f_0(980)\to\Kbar K$ partial
width is incorporated into the BW propagator (the so called Flatt\'e
approximation used in Ref.~\cite{wa76} with parameters $c1/c2$ and
$m\cdot c1$,  which correspond to the ratio of the coupling constants
$g^2_{KK}/g^2_{\pi\pi}$ and effective $f_0(980)$ width).  It differs only slightly at
the top of the $f_0(980)$, but the wider shape of the Flatt\'e function
leaves less room for the remaining events and we obtain:
\begin{equation}
 m_{\sigma}=(0.631\pm0.020)~\gevcc, 
\Gamma_{\sigma}=(0.472\pm0.075)~\gev.
\label{sigpar1}
\end{equation}
The obtained Flatt\'e function parameters are in agreement with those
obtained in Ref.~\cite{wa76}: $c2/c1=2.20\pm0.67$, $m\cdot
c1=0.131\pm0.033$.

The Flatt\'e approximation gives a little better description of the
observed $\pi\pi$ mass spectrum, and so we
use it in the analysis of the structures observed
in the $\phi\pi\pi$ cross section.

It appears that the structure at $\Ecm\approx 2.1\gev$ in the
$\phi\pipi$ cross section (Figs.~\ref{phipipixs} and~\ref{phif0xs}) couples to the
$f_0(980)$ but not to the $f_0(600)$. This is very similar to the
behavior observed for the $\pipi$ system in  $J/\psi\to\phi\pipi$
decay~\cite{bes4k} (and demonstrated with our data in
Fig.~\ref{jpsi_phipipi} of
Sec.~\ref{sec:charmonium}), and in $D_s\to\pipi\pi^+$ decay~\cite{D+s}. In 
both instances a clear $f_0(980)$ signal is observed, while the broad
$f_0(600)$ enhancement of Fig.~\ref{pi2fit2bw} is absent. In contrast
we note that in $J/\psi\to\omega\pipi$ decay~\cite{besomega} exactly
the opposite behavior is observed; the $\pipi$ system exhibits a broad
low-mass enhancement, and there is no evidence of an $f_0(980)$ signal.

In contrast with the ``clean'' $m(\phi\pipi)$ distribution, obtained from the fit on the
$\phi$ peak, the $m(\pipi)$ distribution is obtained by the selection 
of $\phi$ signal in the $K^+ K^-$ invariant mass distribution, with 
background subtraction performed using the $\phi$ side bands and control
region of the \chisq distribution (see Fig.~\ref{phif0_sel}).
To minimize these uncertainties, we use a BW description for $\sigma$
and the Flatt\'e approximation for $f_0$ to
incorporate these two states in a simple model
describing the structures in the $\phi\pi\pi$ cross section data of
Fig.~\ref{phipipixs} (after scaling by a factor of 1.5 to take into account the
$\phi\ppz$ contribution). The model consists of the incoherent
addition of two contributions at each value of \Ecm. The first
represents the decay process $\phi(1680)\to\phi f_0(600)$, with the
parameters of the $\sigma$ given by Eq.~(\ref{sigpar1}); the second
results from the coherent superposition of amplitudes describing the
processes $\phi(1680)\to\phi f_0(980)$ and   $Y(2175)\to\phi
f_0(980)$, where the  $Y(2175)$ BW amplitude describes the peak
observed at $\approx$2.2~\gev in Fig.~\ref{phipipixs}. We note that
in Ref.~\cite{belle_phif0} the contribution from  $\phi(1680)\to\phi
f_0(980)$ decay was not taken into account. We see no physical evidence to justify
doing this, and so allow the presence of this amplitude in our model.
The angular distributions of Fig.~\ref{phi_angle} are consistent with
the $\phi(1020)$ and the S-wave $\pi\pi$ system being in an S-wave
orbital angular momentum state, and so our model includes no
centrifugal barrier factor in the amplitude representations.
\begin{figure*}[tbh]
\includegraphics[width=0.32\linewidth]{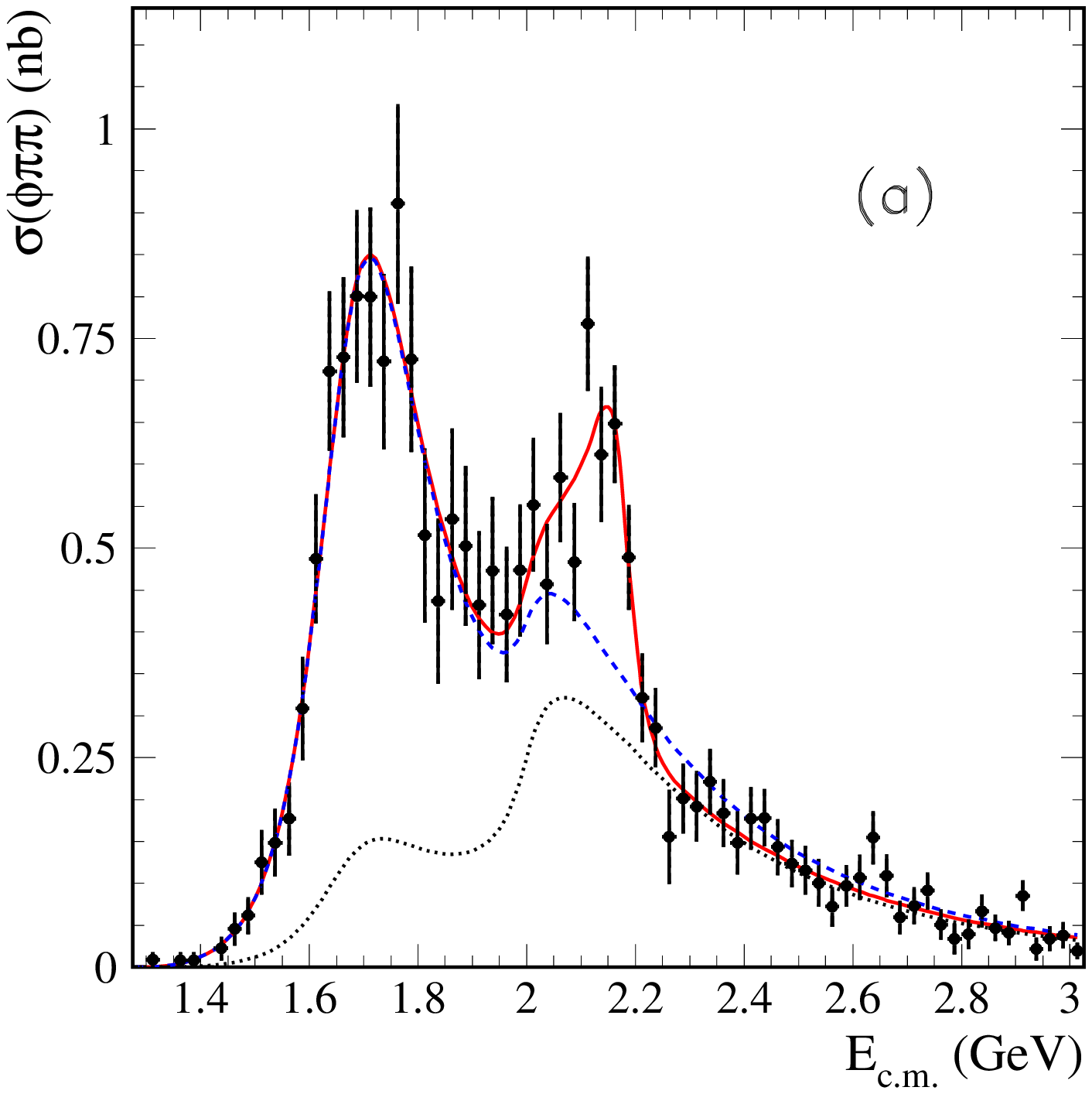}
\includegraphics[width=0.32\linewidth]{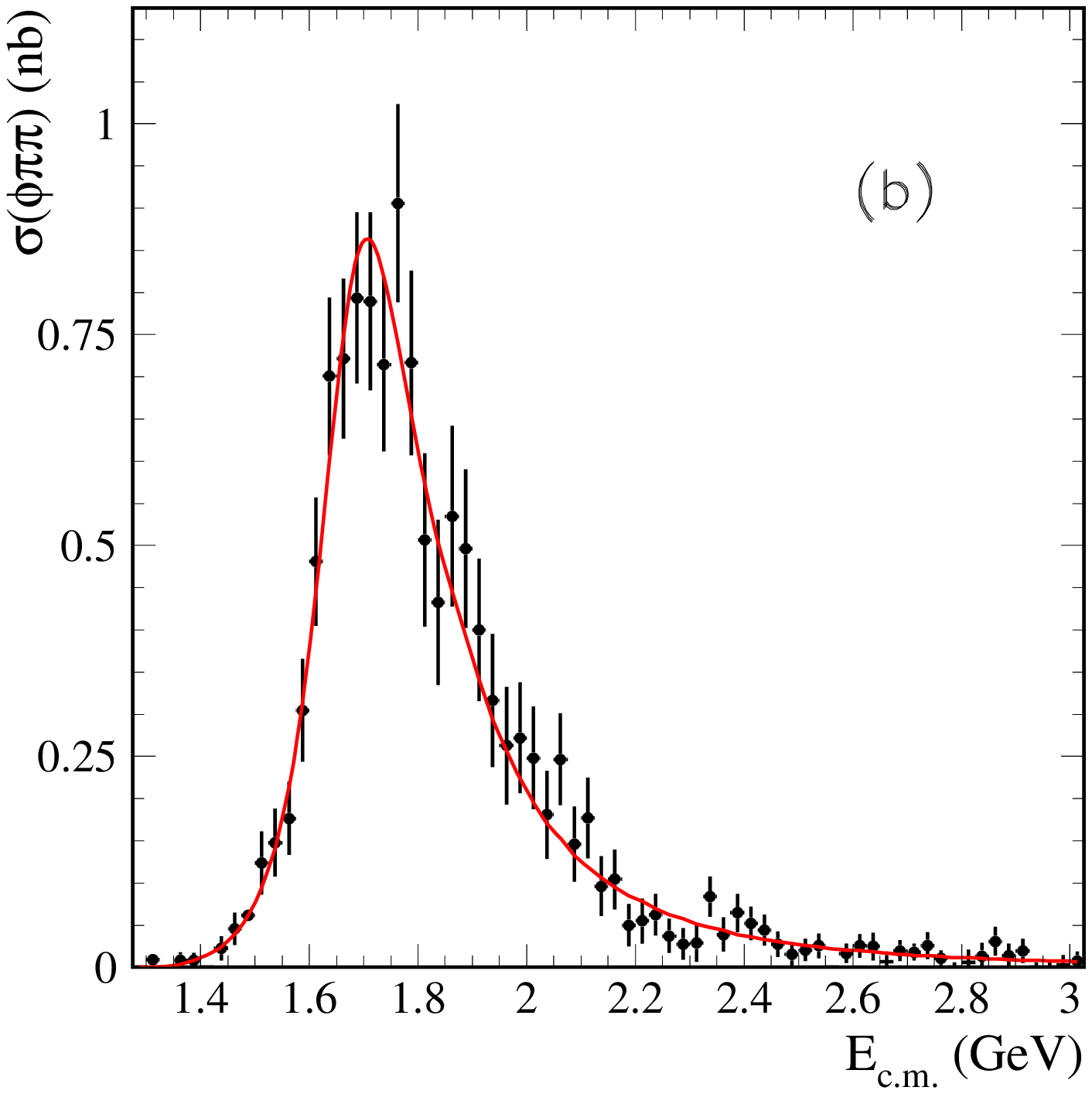}
\includegraphics[width=0.32\linewidth]{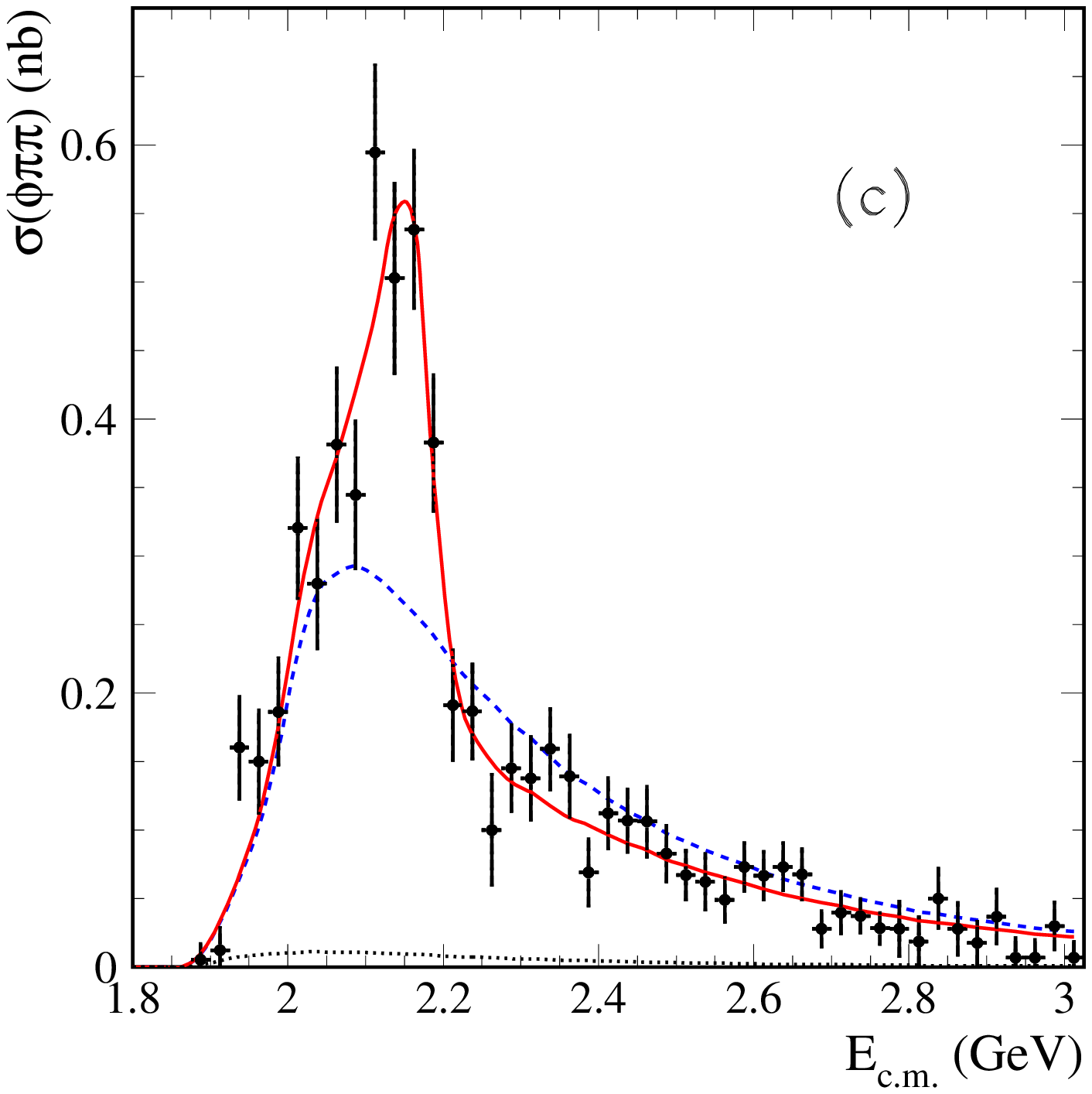}
\vspace{-0.3cm}
\caption{
(a) The fit to the $\epem\to\phi\pi\pi$ cross section
using the model described in the text; the entire contribution due to
the $\phi(1680)$ is shown by the dashed curve. The dotted curve shows
the contribution for only $\phi f_0(980)$ decay. (b) Comparison of the
data and the curve obtained from the overall fit, with the restriction
$m(\pi\pi)<0.85$~\gevcc. (c) The $\epem \!\!\to\! \phi(1020)
f_{0}(980)$ cross section with the requirement
$0.85<m(\pi\pi)<1.1$~\gevcc; the dashed and dotted curves represent
the contributions from $\phi(1680)\to\phi(1020) f_{0}(980)$, and
$\phi(1680)\to\phi(1020) f_{0}(600)$ calculated using the parameter
values from the overall fit to the cross section data.
 }
\label{phipipifit3}
\end{figure*}
\begin{table*}[tbh]
\caption{
  Summary of parameter values obtained from the fits with
Eq.~(\ref{2bwysig}) described in the text.
  An asterisk denotes a value that was fixed in that fit.
  }
\label{fittab3}
\begin{ruledtabular}
\begin{tabular}{ c c c c } 
Fit &  All $m(\pi\pi)$ & $m(\pi\pi)<0.85$~\gevcc & $0.85<m(\pi\pi)<1.1$~\gevcc \\
\hline
$\sigma_{11}$ (nb)    & 0.655$\pm$0.039$\pm$0.040     & 0.678$\pm$0.047$\pm$0.040  &  
0.655*  \\
$m_1$(\gevcc)         & 1.742$\pm$0.013$\pm$0.012     & 1.733$\pm$0.010$\pm$0.010  &
1.742*  \\
$\Gamma_1$(\gev)  & 0.337$\pm$0.043$\pm$0.061     & 0.300$\pm$0.015$\pm$0.037  &  
0.337* \\
$\sigma_{22}$ (nb)    & 0.082$\pm$0.024$\pm$0.010     &       0.082*     &  
0.094$\pm$0.023$\pm$0.010 \\
$m_2$(\gevcc)         & 2.176$\pm$0.014$\pm$0.004     &       2.176*     &  
2.172$\pm$0.010$\pm$0.008  \\
$\Gamma_2$(\gev)  & 0.090$\pm$0.022$\pm$0.010     &       0.090*     &  
0.096$\pm$0.019$\pm$0.012  \\
$\sigma_{12}$(nb)     & 0.152$\pm$0.034$\pm$0.040     &       0.152*     &  
0.132$\pm$0.010$\pm$0.010  \\
$\psi$ (rad)              & -1.94$\pm$0.34$\pm$0.10      &        -1.94*     &  
-1.92$\pm$0.24$\pm$0.12     \\
\chisq /n.d.f.           &    48/(67-9)                &     46/(66-4)&  
38/(46-6)    \\
P(\chisq)                  &     0.74                       &       0.96       &
  0.40   \\
\end{tabular}
\end{ruledtabular}
\end{table*}

We fit the observed $\epem\to\phi\pi\pi$ cross section
using the function
 \begin{eqnarray}
  \sigma(s) &  =  &  
\frac{P_{\phi\sigma}(s)}{s^{3/2}}  \cdot
  \left| \frac{A_{11}(s)}{\sqrt{P_{\phi\sigma}(m_1)}}\right|^2 \label{2bwysig}  \\
 & + & \frac{P_{\phi f_0}(s)}{s^{3/2}}  \cdot
 \left| \frac{A_{12}(s)e^{i\psi}}{\sqrt{P_{\phi f_0}(m_1)}} + 
         \frac{A_{22}(s)}{\sqrt{P_{\phi f_0}(m_2)}} \right|^2,\nonumber
\end{eqnarray}
where
\begin{eqnarray}
    A_{ij}(s)    &  =  &   \frac{\sqrt{\sigma_{ij}} m^{3/2}_{i} m_{i} \Gamma_i}
                     {m_{i}^2- s -i \sqrt{s} \Gamma_{i}(s)} ,    \nonumber
\end{eqnarray}
with \\ 
\\
$i=1$ for the $\phi(1680)$, $i=2$ for the $Y(2175)$,\\
$j=1$ for the $f_0(600)$, $j=2$ for the $f_0(980)$, \\
\\
so that \\
\\
$A_{11}(s)$ describes $\phi(1680)\to\phi(1020) f_0(600)$ decay, \\
$A_{12}(s)$ describes $\phi(1680)\to\phi(1020) f_0(980)$ decay, \\
$A_{22}(s)$ describes $Y(2175)\to\phi(1020) f_0(980)$ decay;\\
\\
$s=\Ecm^2$, $m_1$ and $\Gamma_1$ are the mass and width of the
$\phi(1680)$, $m_2$ and $\Gamma_2$  are the mass and width of the
$Y(2175)$, and the $\sigma_{ij}$ represent the peak cross section values. 

The factors
$P_{\phi\sigma}(s)$ and $P_{\phi f_0}(s)$ represent quasi-two body
phase space integrated over the range of $\pi\pi$ invariant mass
available at $\Ecm=\sqrt{s}$, and are obtained from
\begin{equation}
  P_{\phi\pi\pi}(s)=\int_{2m_\pi}^{\sqrt{s}-m_{\phi}} 
BW_{\pi\pi}(m) q(s,m,m_{\phi}) dm,
\label{ps}
\end{equation}
where $BW_{\pi\pi}(m)$ is a BW function with  
$f_0(980)$ parameters ($BW_{f_0}(m)$) to define $P_{\phi f_0}(s)$, or with
$f_0(600)$ parameters ($BW_{\sigma}(m)$)  to define
$P_{\phi\sigma}(s)$~\cite{achasov_koz}, and  $q$ is the
momentum of the particles with masses $m$ and $m_{\phi}$ in the
two-body reaction at $\Ecm=\sqrt{s}$.

Since the decay $\phi(1680)\to\phi(1020) f_0(980)$ is suppressed by
phase space near $\sqrt{s}=m_1$, the value of $\sigma_{12}$ is much
smaller than that of $\sigma_{11}$, but its contribution to $\sigma(s)$
increases rapidly beyond the $\phi(1020) f_0(980)$ threshold.

The $\phi(1680)$ resonance 
decays mainly to $\Kbar K^*(892)$ and $\phi(1020)\eta$~\cite{PDG,isrkkpi}.
We find that it has a branching fraction of about
10\% to $\phi\pi\pi$, which together with  other modes
listed in  PDG, leads to an energy-dependent width 
that can be written as 
\begin{eqnarray}
\Gamma_{1}(s) & = & \Gamma_{1}\Bigl[0.7\frac{m_1^3 P_{2K}(s)}{s^{3/2} P_{2K}(m_1^2)} \nonumber \\
& + &  0.2\frac{m_1 P_{\phi\eta}(s)}{s^{1/2} P_{\phi\eta}(m_1^2)} 
+ 0.1\frac{m_1 P_{\phi\pi\pi}(s)}{s^{1/2} P_{\phi\pi\pi}(m_1^2)}\Bigr], 
\end{eqnarray}
with $P_{2K}(s)=q^{3}(\sqrt{s},m_{K},m_{K^*})$, and $P_{\phi\eta}(s)=q(\sqrt{s},m_{\phi},m_{\eta})$.

For the second resonance candidate, which decays mostly to $\phi\pi\pi$, 
the energy dependence of  the width is written as
\begin{eqnarray}
\Gamma_{2}(s) & = & \Gamma_{2}\frac{m_2 P_{\phi\pi\pi}(s)}
{s^{1/2} P_{\phi\pi\pi}(m_2^2)}. 
\end{eqnarray}
We note that the introduction of an energy dependence for each width
significantly increases the values of the resonance mass and width,
especially for broad structures. 

The results of the fits are shown in Fig.~\ref{phipipifit3} and 
 summarized in Table~\ref{fittab3}. The first error is statistical, and the second
 error represents the systematic uncertainty estimated as a difference in fitted
 values for two different descriptions of the two-pion spectrum as shown in
 Fig.~\ref{pi2fit2bw}.  
In Fig.~\ref{phipipifit3}(a) we show the contribution from the $\phi(1680)$ for 
both modes (dashed curves), and for  $\phi(1680)\to\phi(1020)
 f_0(980)$ only (dotted curve). 
The increase of the cross section at about 2~\gev is explained by
the opening of the $\phi f_0(980)$ decay channel of the $\phi(1680)$
resonance. However the fit shows that an additional relatively narrow
 state is needed in order to provide a
better description of the observed data. 

It is important to note that this model describes the observed data very well 
independently of the $m(\pi\pi)$ region selected. 
Figure~\ref{phipipifit3}(b) shows the $\phi\pi\pi$ cross section for
$m(\pipi)<0.85$~\gevcc for the data; the curve is obtained by using the
parameter values from the overall fit and yields
\chisq/n.d.f. = 63/(66-1) (P(\chisq) = 0.54). 
If we fit this distribution, slightly better parameter values can
be obtained (see Table~\ref{fittab3}), but these still agree well with
those from the overall fit. We consider them as our measurement of the
$\phi(1680)$ resonance parameters. They correspond to
the product of the electronic
width, $\Gamma_{ee}$, and branching fraction to $\phi\pi\pi$, $\BR_{\phi\pi\pi}$, 
$$
  \BR_{\phi\pi\pi} \cdot \Gamma_{ee} = 
  \frac{\Gamma_1 \sigma_{11} m_1^2}{12\pi C } = (42\pm 2\pm 3)~\ev\ ,
$$
where we fit the product $\Gamma_1 \sigma_{11}$ to reduce correlations, 
and $C$, the conversion constant, is $0.389$~mb(\gevcc)$^2$~\cite{PDG}.
The second error is systematic, and corresponds to the normalization
uncertainty on the cross section, and to the uncertainty in the $m(\pi\pi)$
distribution description. 

If we require $0.85 < m(\pi\pi) < 1.1$~\gevcc   
(Fig.~\ref{phipipifit3}(c)), then
without additional fitting the model yields \chisq/n.d.f. = 48/(46-1)
(P(\chisq) = 0.31), and improves 
to  \chisq/n.d.f. = 38/(46-6) (P(\chisq) = 0.40 )
by refitting using the parameter values listed
in Table~\ref{fittab3}. If we try to explain the observed cross
section only in terms of the $\phi(1680)$ without any narrow state (dashed
curve in Fig.~\ref{phipipifit3}(c)), the fit
gives \chisq/n.d.f. = 123/(46-2) (P(\chisq) = $10^{-7}$) and so this hypothesis
is not compatible with the data.
Note, that the contribution of $\phi f_0(600)$, shown by
dotted curve in Fig.~\ref{phipipifit3}(c), is very small.

The model described above provides an excellent description
of the observed cross section behavior, and suggests that 
the $Y(2175)$ may not be a radially excited $s\bar s$  state,
since such a state would be expected to be
much wider (300-400~\gevcc) and also should decay to $\phi f_0(600)$,
like the $\phi(1680)$.

\begin{figure}[tbh]
\includegraphics[width=0.95\linewidth]{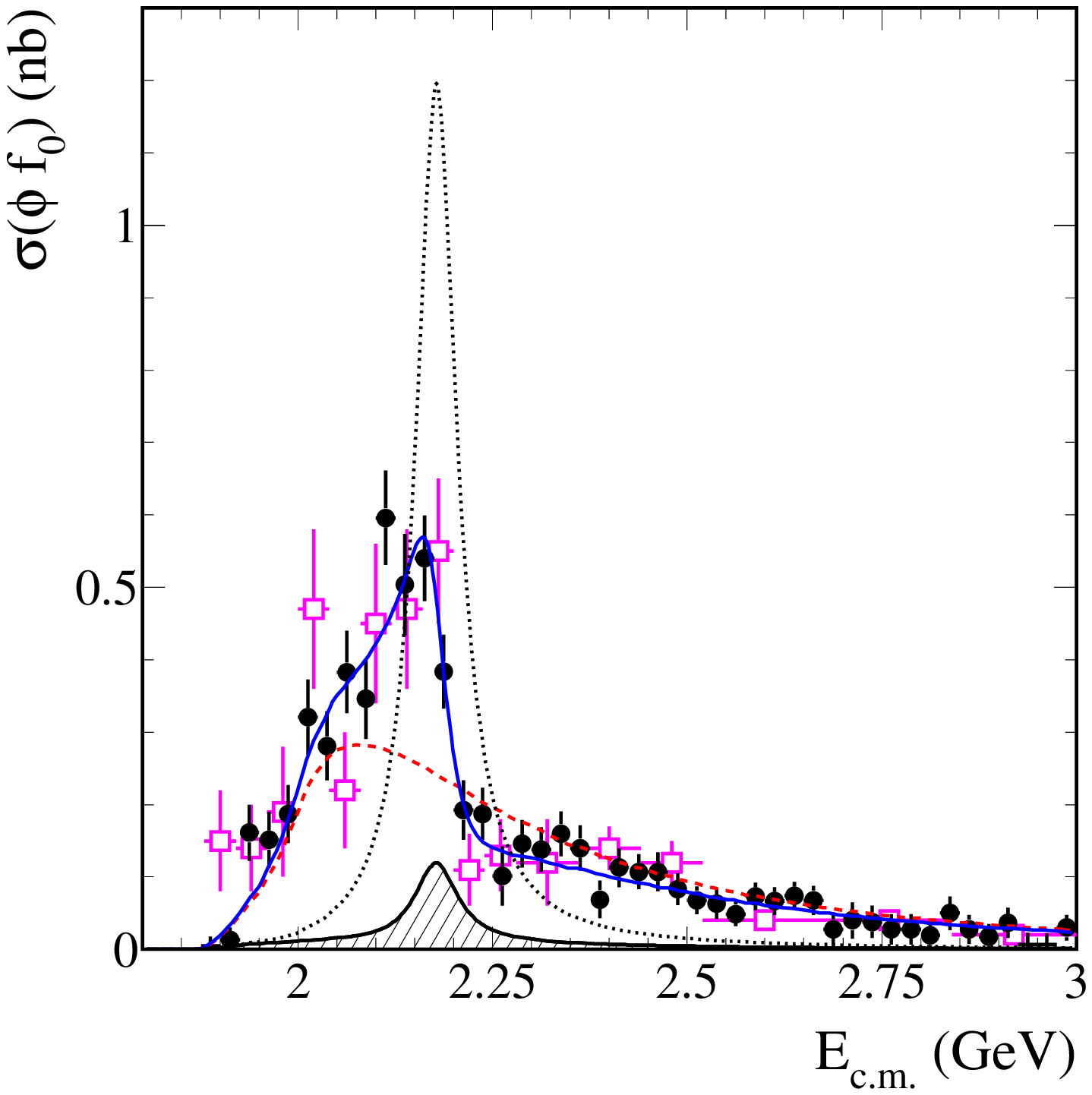}
\vspace{-0.3cm}
\caption{
The $\epem \!\!\to\! \phi(1020) f_{0}(980)$ cross section measured 
  in the \KKppch (solid dots) and \KKppnt (open squares) final states.
  The solid (dashed) curve represents the result of the two-resonance
  (one-resonance -  $\phi(1680)\to\phi(1020) f_{0}(980)$) 
fit using Eq.(\ref{2bwysig}), as described in the text. 
The hatched area and dotted curve show the $Y(2175)$ contribution for two solutions. 
}
\label{phif0xsall}
\includegraphics[width=0.95\linewidth]{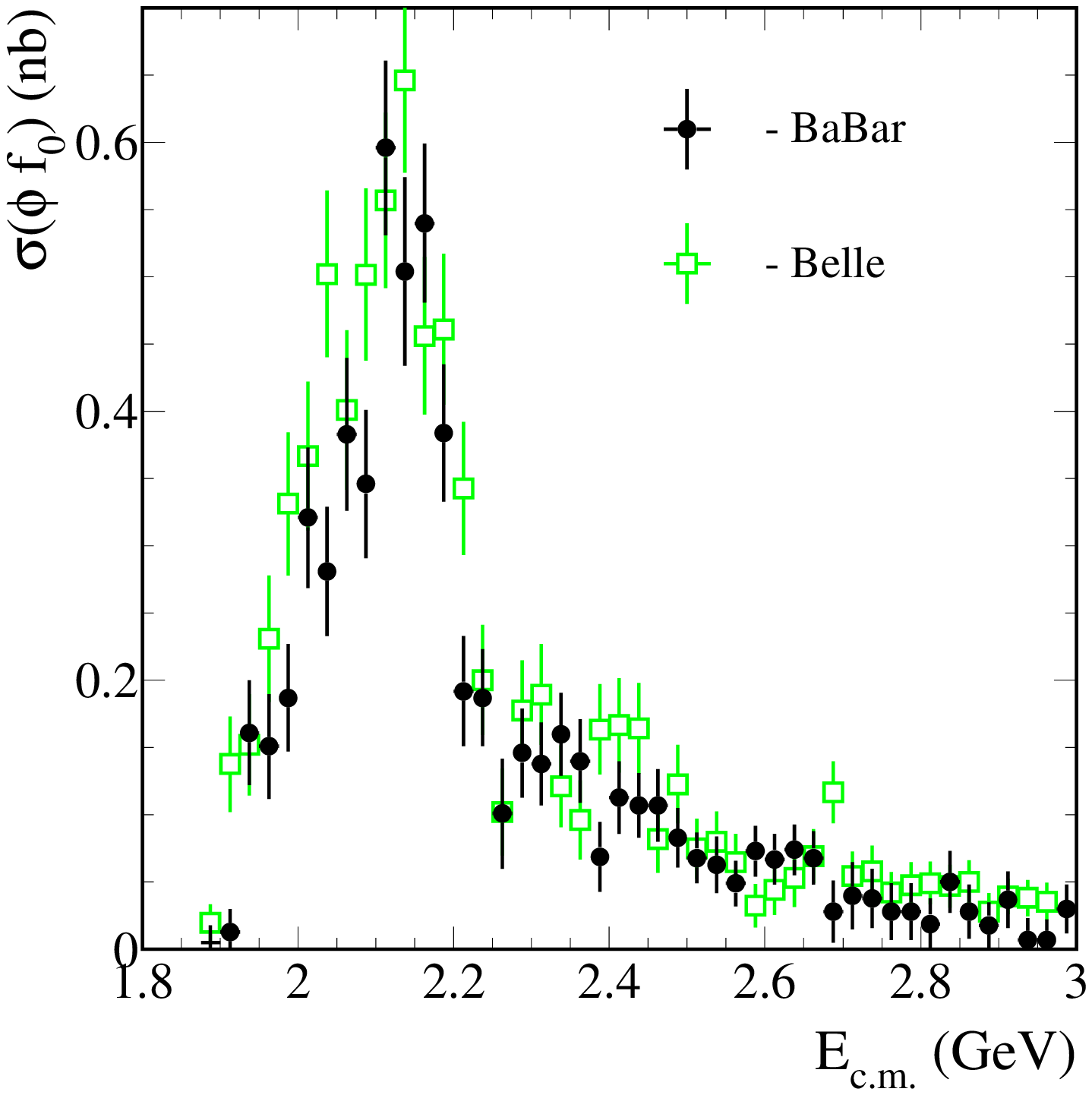}
\vspace{-0.3cm}
\caption{
The $\epem \!\!\to\! \phi(1020) f_{0}(980)$ cross section 
measurements from the \KKppch final state from ~\babar~(dots) and 
Belle~\cite{belle_phif0}(squares).
 }
\label{phif0xsbelle}
\end{figure}
\begin{figure*}[tbh]
\includegraphics[width=0.32\linewidth]{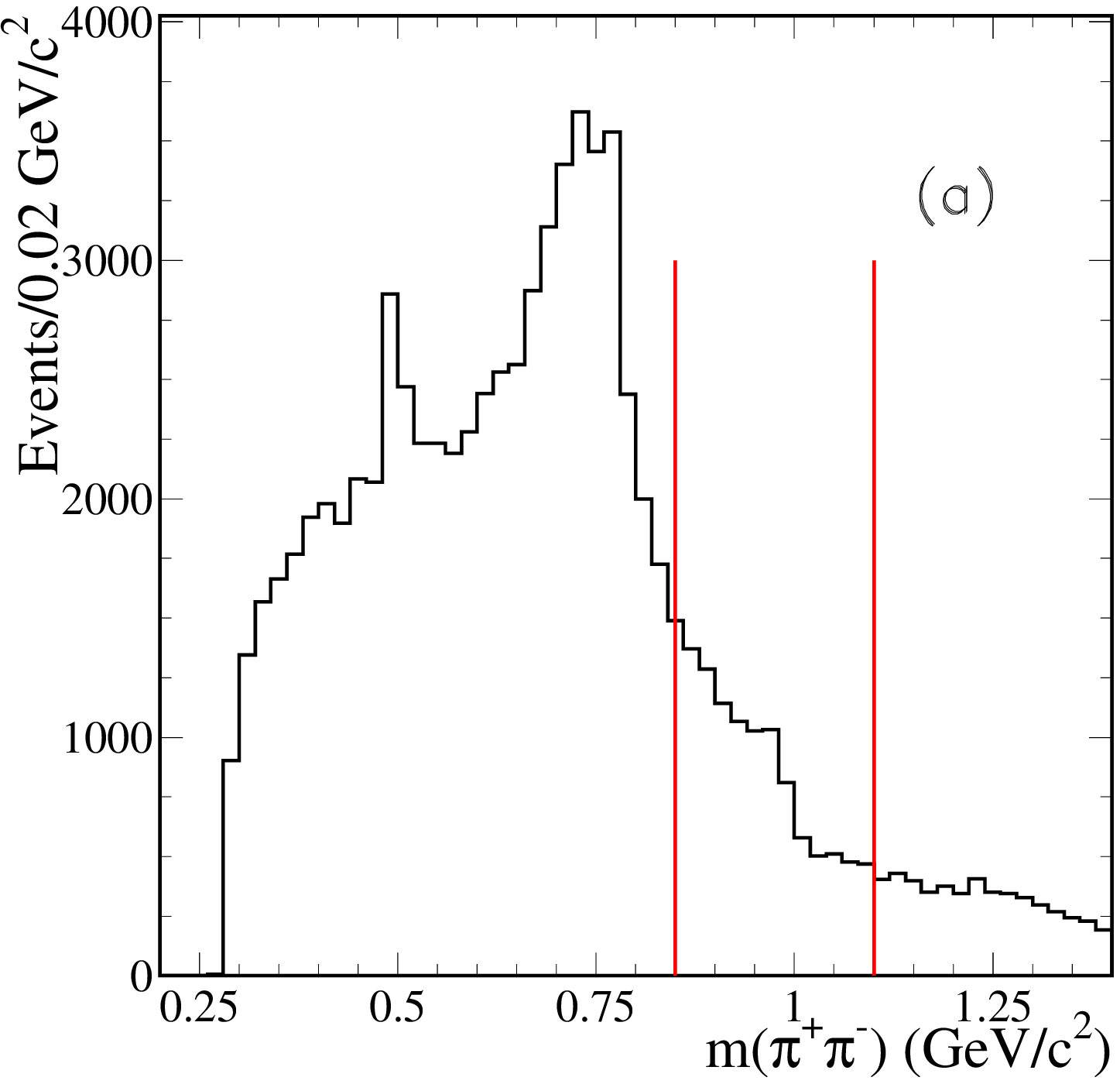}
\includegraphics[width=0.32\linewidth]{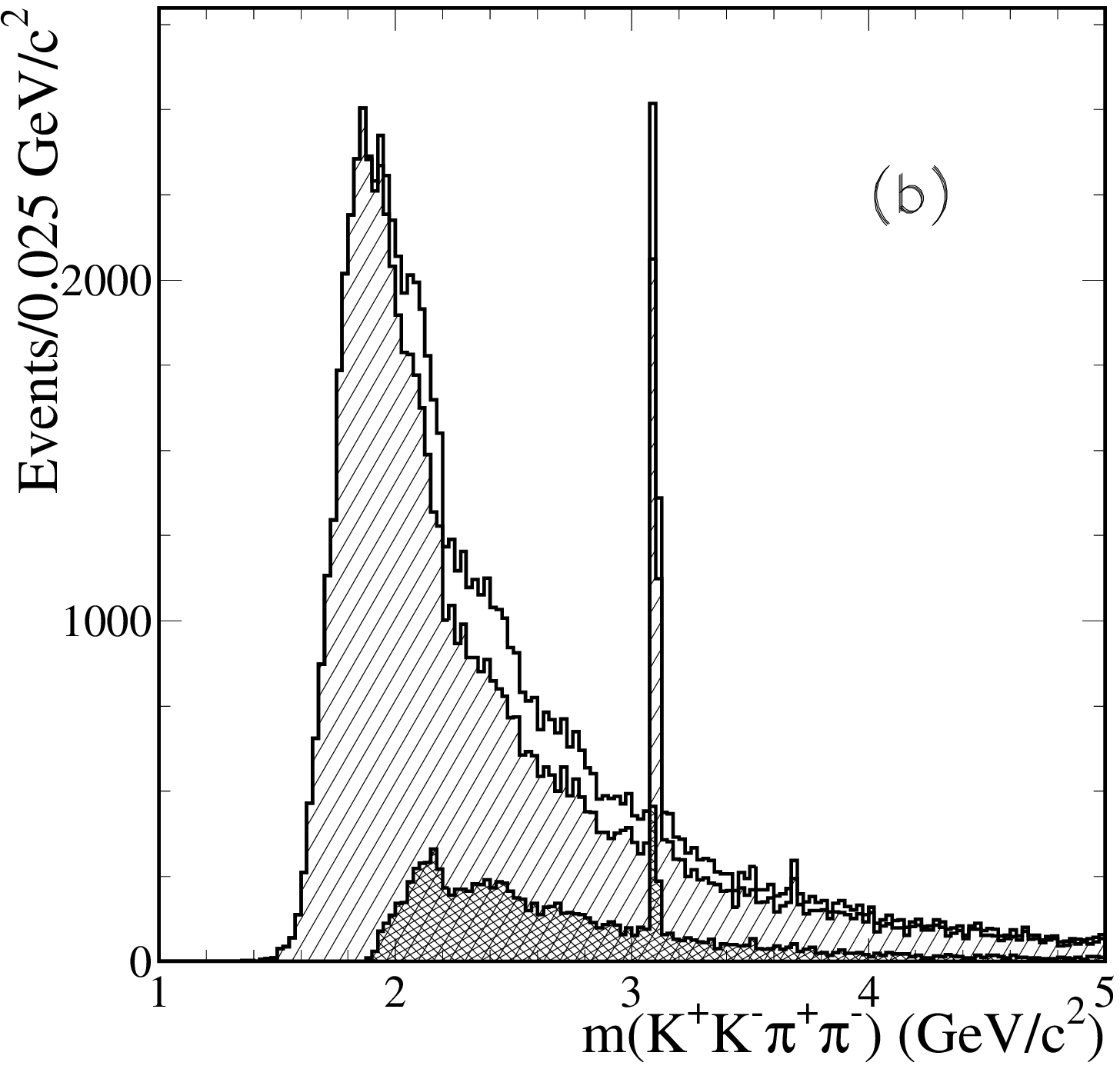}
\includegraphics[width=0.32\linewidth]{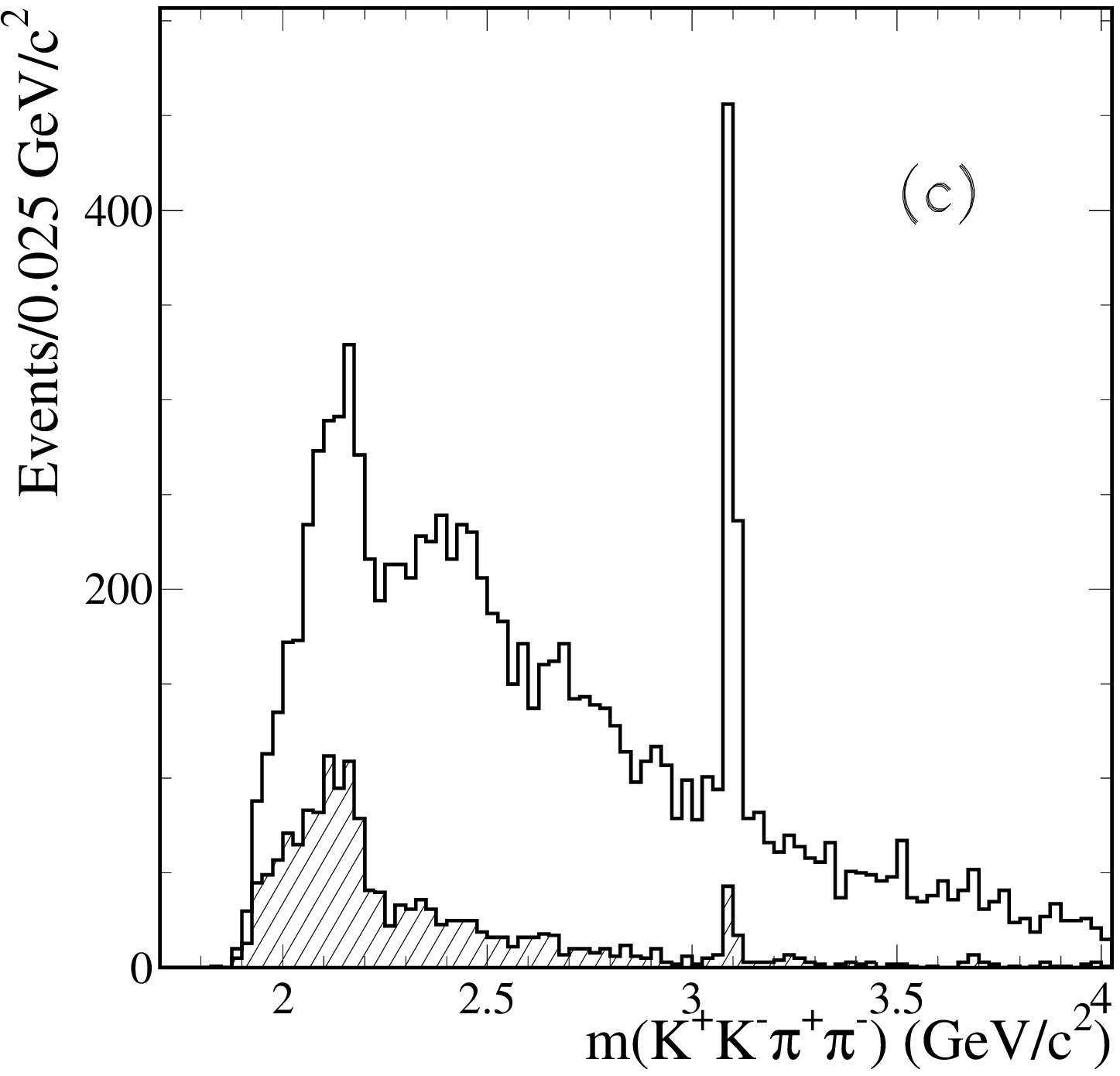}
\vspace{-0.3cm}
\caption{
(a) The $m(\pipi)$ distribution without background subtraction for
$K^+ K^-\pipi$ events. The vertical lines indicate the $f_0(980)$ region.
(b) All selected $K^+ K^-\pipi$ events (open histogram), selected
$K^+ K^- f_0(980)$ events (cross-hatched histogram), and all the rest
  (hatched histogram).
(c) The $K^+ K^- f_0(980)$ events (open histogram) in comparison with  the
$\phi(1020) f_0(980)$ sample (hatched histogram).  
 }
\label{select_2kf0ch}
\end{figure*}
\begin{figure*}[tbh]
\includegraphics[width=0.32\linewidth]{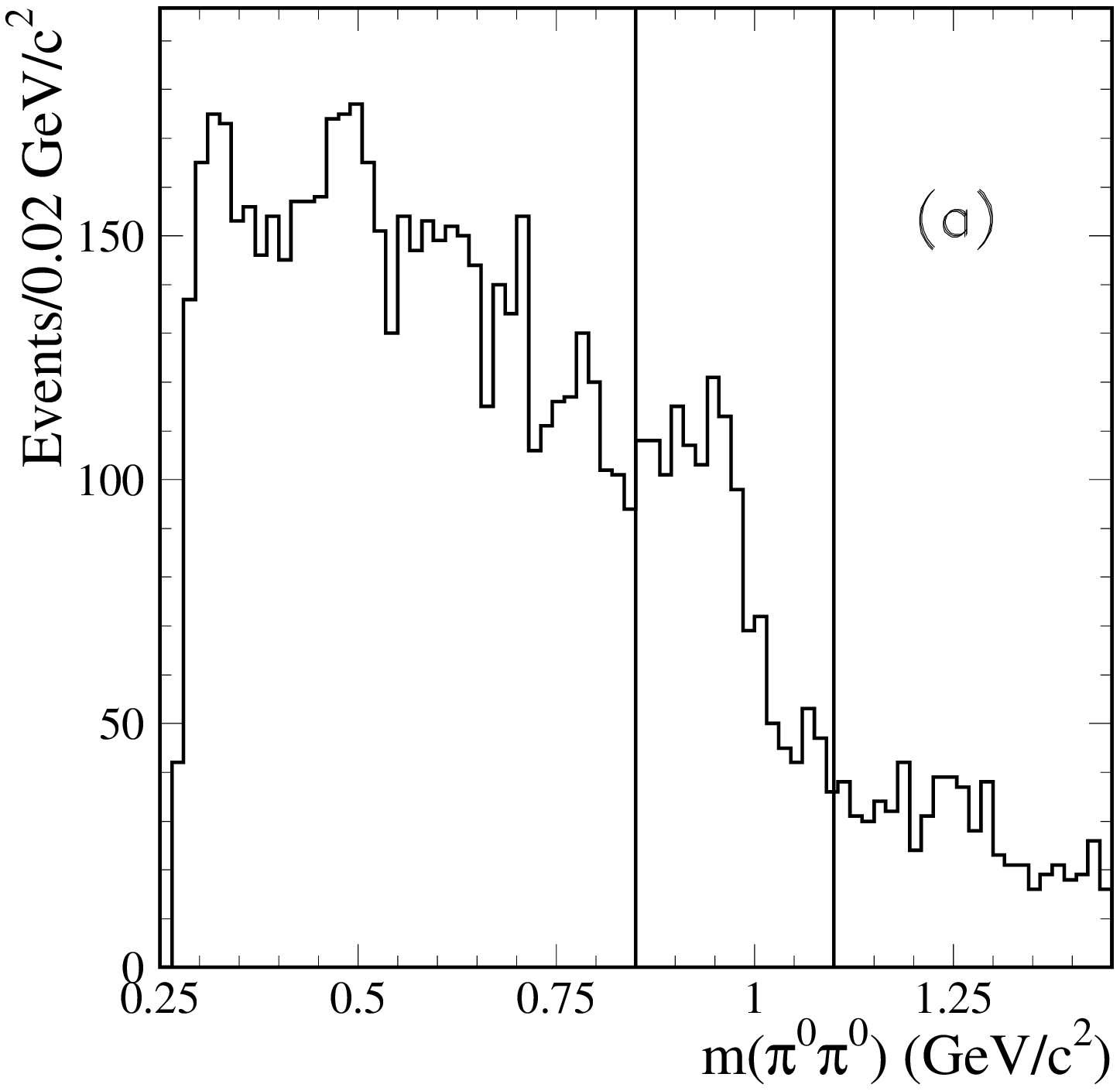}
\includegraphics[width=0.32\linewidth]{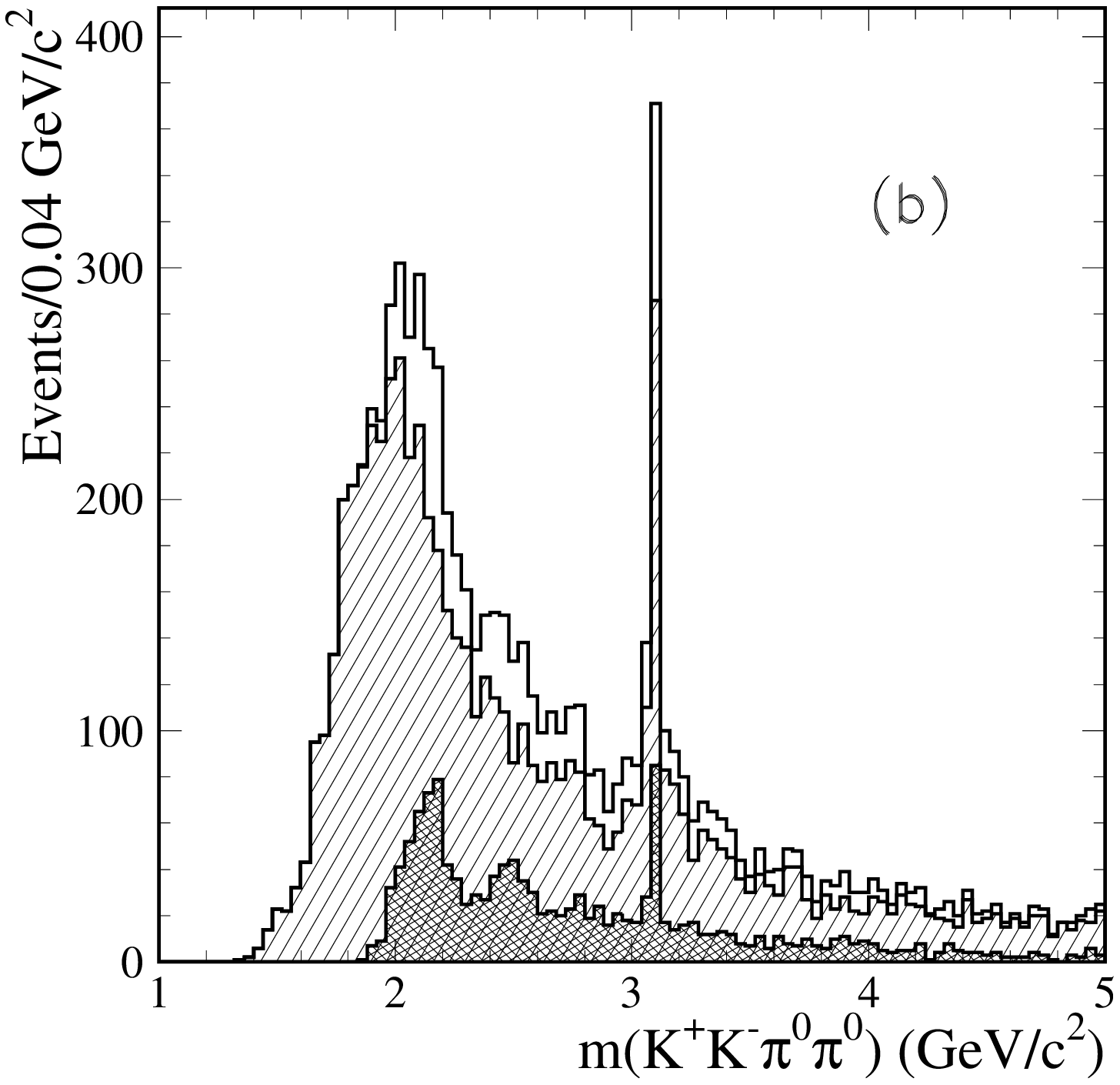}
\includegraphics[width=0.32\linewidth]{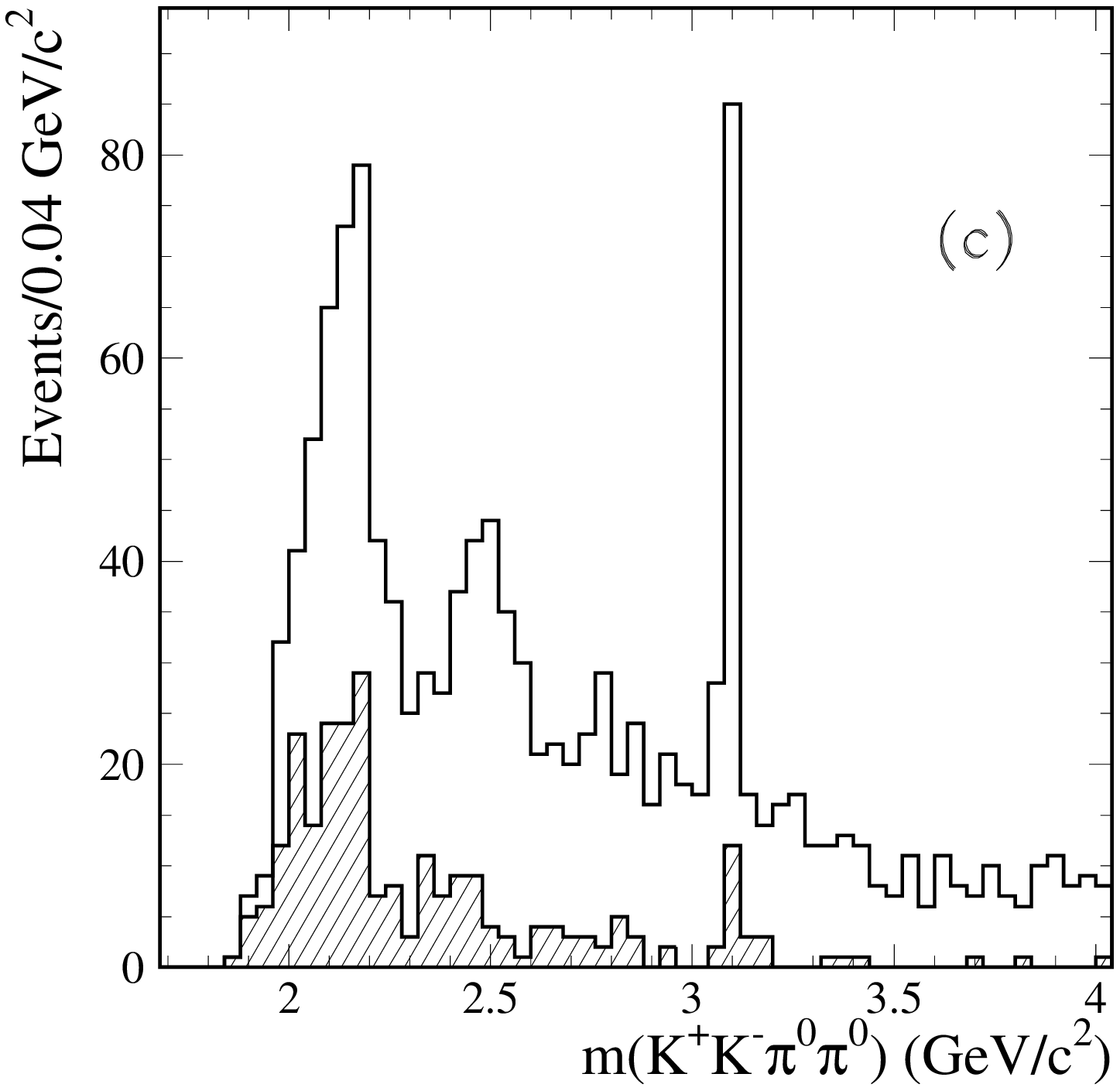}
\vspace{-0.3cm}
\caption{
(a) The $m(\ppz)$ distribution without background subtraction for
$K^+ K^-\ppz$ events. The vertical lines indicate the $f_0(980)$ region.
(b) All selected $K^+ K^-\ppz$ events (open histogram), selected
$K^+ K^- f_0(980)$ events (cross-hatched histogram), and all the rest
  (hatched histogram).
(c) The $K^+ K^- f_0(980)$ events (open histogram) in comparison with  the
$\phi(1020) f_0(980)$ sample (hatched histogram). 
 }
\label{select_2kf0nu}
\end{figure*}
\begin{figure*}[tbh]
\begin{center}
\includegraphics[width=0.32\linewidth]{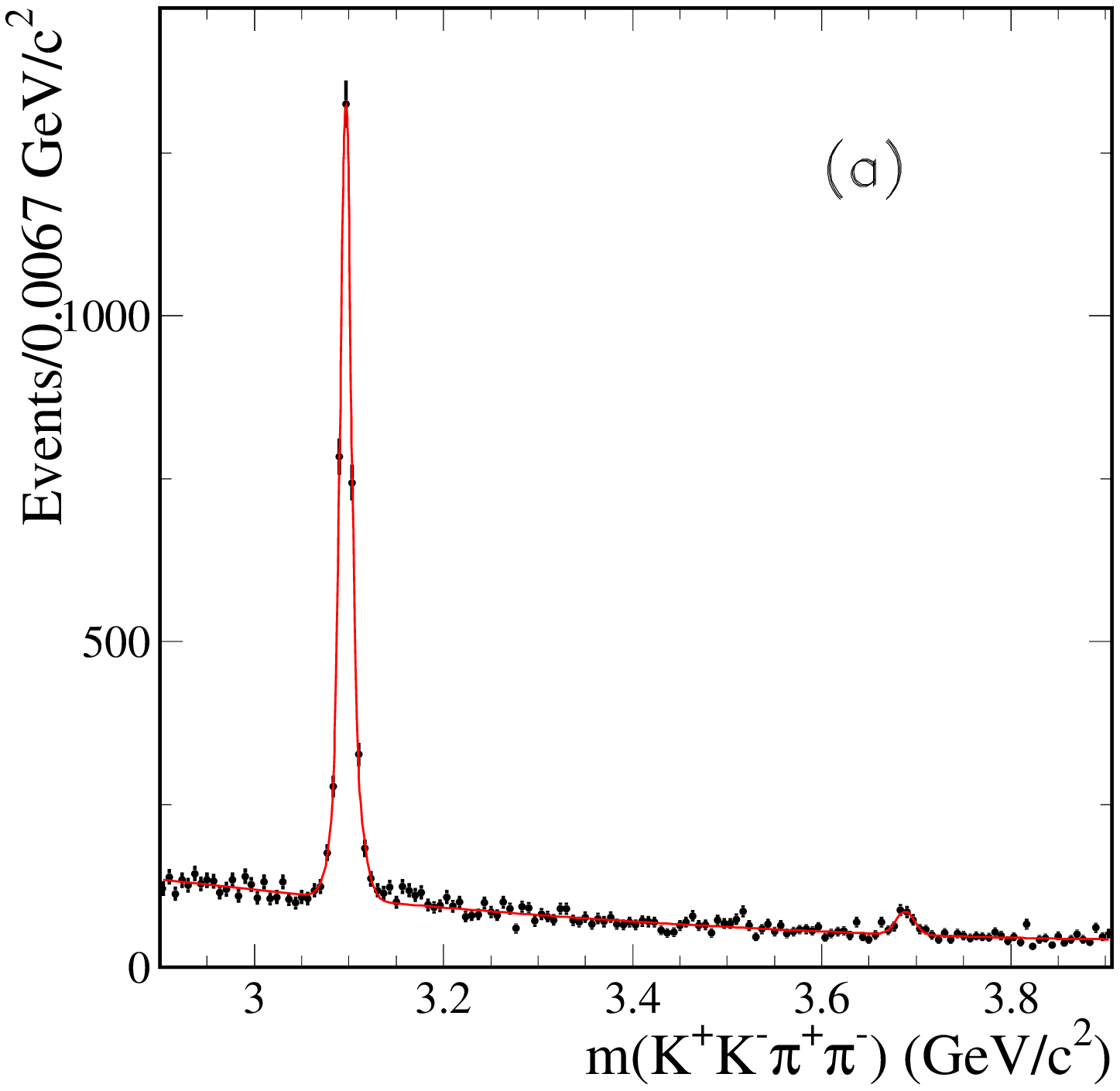}
\includegraphics[width=0.32\linewidth]{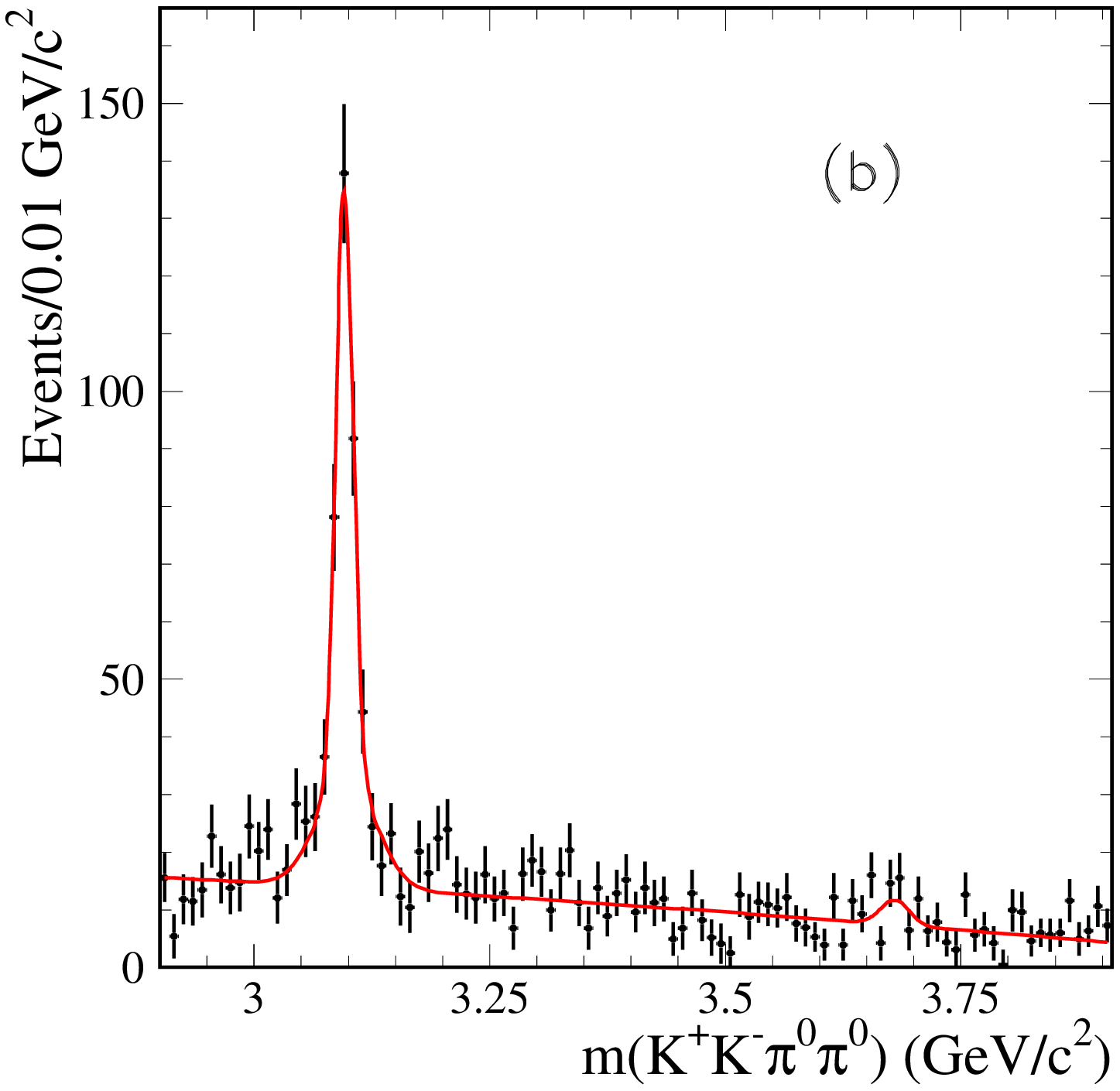}
\includegraphics[width=0.32\linewidth]{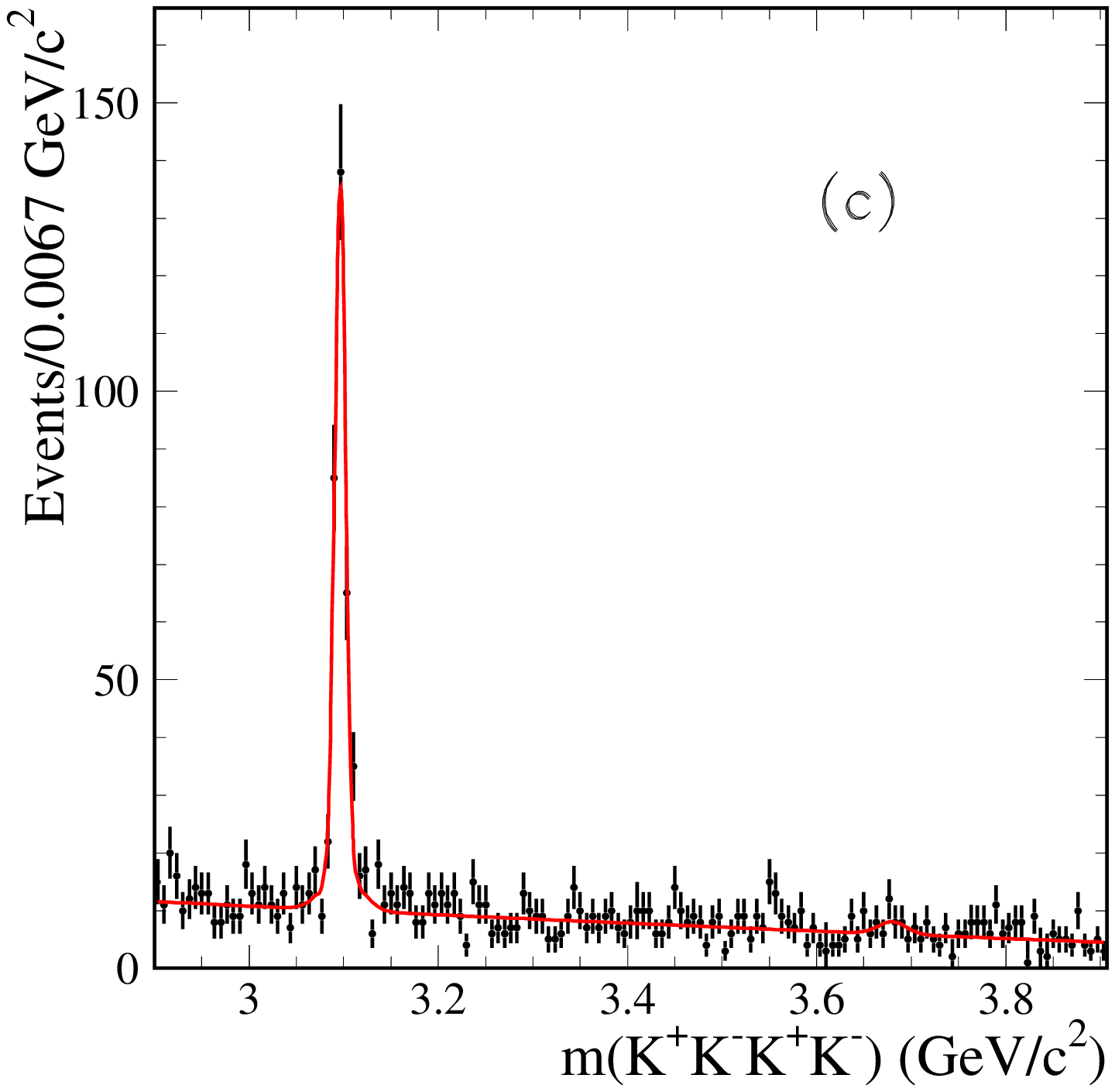}
\vspace{-0.4cm}
\caption{
  Raw invariant mass distribution for all selected events in the
  charmonium region for (a)
  $\epem\to\KKppch$, (b) $\epem\to K^+ K^- \ppz$, and
(c)  $\epem\to  K^+ K^- K^+ K^- $; in each figure the curve
represents the result of the fit described in the text.
}
\label{jpsi}
\end{center}
\end{figure*}
\section{\boldmath $\epem \!\to \phi f_0$ Near Threshold}
\label{phif0bump}

The behavior of the $\epem \!\!\to\! \phi f_0$ cross section near 
threshold shows a structure near 2.175~\gev,
and we have published this result in Ref.~\cite{isr2k2pi}.
Here we provide a more detailed study of the cross section for this channel in the
1.8--3~\gev region with the full \babar~ dataset.
In Fig.~\ref{phif0xsall} we superimpose the cross 
sections measured in the \KKppch 
and \KKppnt final states (shown in Figs.~\ref{phif0xs} and~\ref{phif0xs2});
they are consistent with each other.

We perform a combined fit to these cross section data using Eq.(\ref{2bwysig})
with the two-pion mass restricted to the region 0.85-1.1~\gevcc.
We fix the $\phi(1680)$ parameters for the $\phi(1020) f_0(600)$ decay mode
(which gives a small contribution in this mass range) and allow all
other parameters to float. 
The result of the fit is shown as the solid curve in 
Fig.~\ref{phif0xsall}.
As  demonstrated in Ref.~\cite{belle_phif0}, the observed
pattern can be a result of a constructive or destructive interference of
the narrow structure at 2.175~\gev with the coherent background.

The fit with constructive interference gives the resonance parameter values
\begin{eqnarray*}
 \sigma_{22} & = &  (0.093  \pm 0.021 \pm 0.010)~{\rm nb} ,   \\
      m_2    & = &  (2.180 \pm 0.008 \pm 0.008)~\gevcc ,    \\
 \Gamma_2    & = &  (0.077 \pm 0.015 \pm 0.010)~\gev ,    \\
   \psi_2    & = &  (-2.11  \pm 0.24 \pm 0.12)~{\rm rad} ,\\ 
 \sigma_{12} & = &  (0.140  \pm 0.009 \pm 0.010)~{\rm nb,} 
\end{eqnarray*}
and \chisq/n.d.f.$=\! 57/(61-6)$ (P(\chisq) =  0.33).
The statistical precision is improved compared to that of
Ref.~\cite{isr2k2pi}, for which the analysis was
based on half as much data.
For this state we estimate the product of electronic width and branching 
fraction to $\phi f_0$ as
$$
  \BR_{\phi f_0} \cdot \Gamma_{ee} = 
  \frac{\Gamma_2 \sigma_{22} m_2^2}{12\pi C } = (2.3\pm 0.3\pm 0.3)~\ev\ ,
$$
where we fit the product $\Gamma_2 \sigma_{22}$ to reduce correlations.
The second error is systematic, and corresponds to the normalization
uncertainty on the cross section.

The destructive interference yields exactly the same
overall curve with the same
parameters for the mass and width of the narrow state, but
significantly larger peak 
cross section with opposite sign of the mixing  angle:
$\sigma_{22} = (1.13  \pm 0.15 \pm 0.12)~{\rm nb}$, $ \psi_2 =(2.47
\pm 0.17 \pm 0.13)~{\rm rad} $.  To select between  two
solutions, we need more information 
on the decay rates to another modes, which are not available now.

If we assume no resonance structure other than the tail from 
$\phi(1680)\to\phi(1020) f_0(980)$, the fit yields 
\chisq/n.d.f.$= 150/(61-2)$ with P(\chisq)=$8\cdot 10^{-9}$.
The result of this fit
is shown as the dashed curve in Fig.~\ref{phif0xsall}.
It is a poor fit to the region below 2.3~\gev, but
gives a good description of the cross section behavior at higher
values of \Ecm. 
The fit, with or without the resonance at 2.18~\gevcc, gives a
maximum  value of the $\phi(1680)\to\phi f_0$
cross section of 0.3 nb  at $\Ecm\approx 2.1\gev$.
This is of independent theoretical interest, because it can be related
to  the $\phi\to f_0(980)\gamma$ decay studied at the
$\phi$-factory~\cite{phif0theory,phif0dispersion}. 

The significance of the structure calculated from the change in \chisq
between the fits with and without the resonance at 2.18~\gev
is $\sqrt{150 - 61} = 9.4$ standard deviations; the \chisq value, 61
for 61-2 n.d.f.,  yields the same probability as the \chisq value 57
for 61-6 n.d.f.. 

The cross section measurements from the \KKppch final state
     shown in Fig.~\ref{phif0xsall} are compared to those from
     Belle~\cite{belle_phif0} in Fig.~\ref{phif0xsbelle}. 
     There is good overall agreement between the results from the two
     experiments.
Overall agreement between the results of the fits to
the ~\babar~ and Belle data is also good.
\subsection{Structures in the $K^+ K^- f_0(980)$ final state}

We next search for other decay modes of the $Y(2175)$ state.
Figure~\ref{select_2kf0ch}(a) shows the ``raw''
(no background subtraction) two-pion mass distribution for all selected
$K^+ K^-\pipi$ events, and Fig.~\ref{select_2kf0nu}(a) shows the
same distribution for the $K^+ K^-\ppz$ sample. The $f_0(980)$ contribution is
relatively small for the charged-pion mode, and larger for the
neutral-pion mode.
If we select the region $0.85< m(\pi\pi)<1.1$~\gevcc, and plot the  $K^+
K^-\pi\pi$ mass distribution, the bump at 2.175~\gevcc is seen
much more clearly in spite of larger background
(Figs.~\ref{select_2kf0ch}(b) and ~\ref{select_2kf0nu}(b)), and a bump at
2.5~\gevcc is also seen; the rest
of events have no structures at 2.175~\gevcc or 2.5~\gevcc
(Figs.~\ref{select_2kf0ch}(b) and ~\ref{select_2kf0nu}(b) hatched histograms). 
The bumps are
seen only in  the $K^+ K^- f_0(980)$ sample (Figs.~\ref{select_2kf0ch}(c)
and ~\ref{select_2kf0nu}(c)), but if we select the
$\phi(1020)$ region, no bumps are seen at  2.5~\gevcc, as shown by the
hatched histograms in
Figs.~\ref{select_2kf0ch}(c) and ~\ref{select_2kf0nu}(c).

From these histograms we can conclude that the $Y(2175)$ resonance
has a $K^+ K^- f_0(980)$ decay mode when the $\KpKm$ system is not from $\phi$,
and that the decay rate is comparable to that for $\phi f_0(980)$. Also another
state at 2.5~\gev seems to exist; this decays to  $K^+ K^- f_0(980)$
(but seems not to couple
to  $\phi f_0(980)$) with width $\approx$0.06-0.08~\gev (see 
Ref.~\cite{isr2k2pi}). The large background does not
allow us to clearly separate this state.

%

\section{\boldmath The Charmonium Region}
\label{sec:charmonium}

For the \Ecm region above 3~\gev, our data  can be used to measure,
 or set limits on,  the decay branching fractions for the 
$J/\psi$ and $\psi(2S)$ (See Figs.~\ref{2k2pi_ee_babar},
 ~\ref{2k2pi0_ee_babar}, and ~\ref{4k_ee_babar}).
In addition, these signals allow checks of our mass
scale, and of our measurements of mass resolution.
Figure~\ref{jpsi}  shows the invariant
mass distributions for the selected \KKppch, \KKppnt, and \KKKK events,
respectively, in this region, using smaller mass intervals
than in the corresponding Figs.~\ref{2k2pi_babar}, 
~\ref{2k2pi0_babar}, and~\ref{4k_babar}.
We do not subtract any background from the \KKppch and \KKKK distributions,
since it is small and nearly uniformly distributed, but
we use the \chiKKppnt control region to subtract part of the ISR 
background from the \KKppnt distribution.
Production of the $J/\psi$ is apparent in all three distributions,
and a small, but clear, $\psi(2S)$ signal is visible in the \KKppch mode.

We fit each of these distributions using a sum of two Gaussian functions 
to describe the $J/\psi$ signal and incorporate a similar
representation of a $\psi (2S)$ signal, although there is no clear
evidence of the latter in Figs.~\ref{jpsi}(b) and \ref{jpsi}(c).  
In each case, a second-order-polynomial function is used to 
describe the remainder of the distribution.
We take the signal function parameter values from simulation, but let
the overall mean and width values vary in the fits, together with 
the coefficients of the polynomial. For the  \KKppnt and \KKKK
modes we fix the $\psi (2S)$ mass position~\cite{PDG}, and take the 
width from MC simulation.
The fits are of good quality, and are shown by the curves in
Fig.~\ref{jpsi}.
In all cases, the fitted mean value is within 1~\mevcc of the nominal
$J/\psi$ or $\psi (2S)$ mass position~\cite{PDG}
and the width is within 10\% of the simulated resolution 
discussed in Secs.~\ref{sec:xs2k2pi}, \ref{sec:2k2pi0xs}, and~\ref{sec:4kxs}.
\begin{figure}[tbh]
\includegraphics[width=0.9\linewidth]{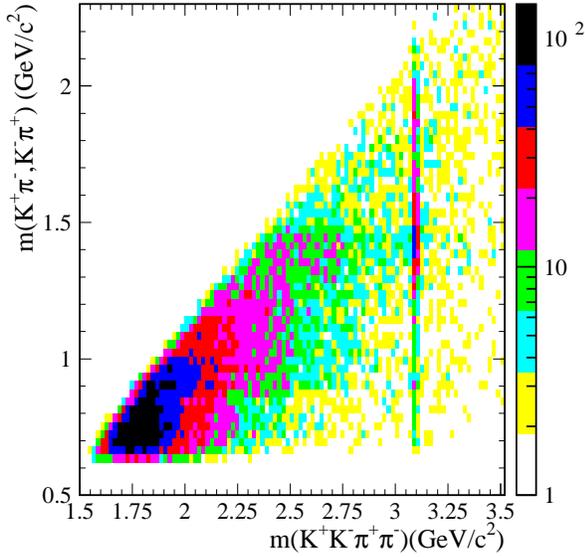}
\vspace{-0.6cm}
\caption{
  The $K^\pm\pi^\mp$ invariant mass versus \KKppch invariant mass for 
  events with the $K^\mp\pi^\pm$ combination in one of the
  $K^{*}(892)^{0}$ regions of Fig.~\ref{kkstar}(a);  for events in  overlap region,
  only one combination is chosen.
  }
\label{jpsi_kkstarvs2k2pi}
\end{figure}
\begin{figure}[tbh]
\includegraphics[width=0.9\linewidth]{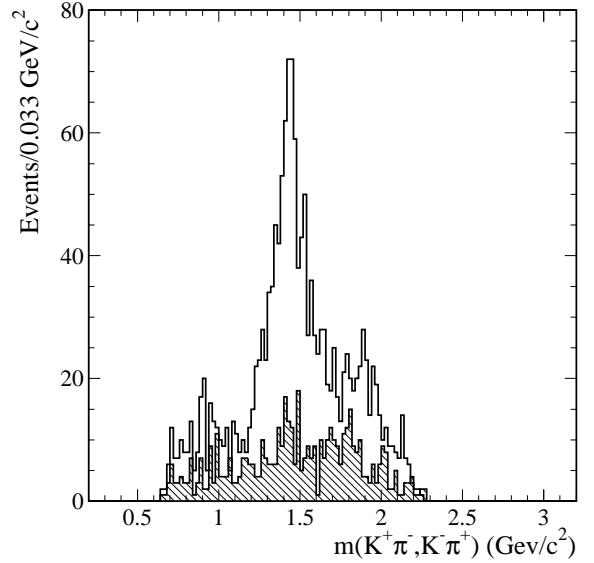}
\vspace{-0.6cm}
\caption{
  The $K^\pm\pi^\mp$ mass projection for events from Fig.~\ref{jpsi_kkstarvs2k2pi}
  with \KKppch invariant mass within 50~\mevcc of the nominal $J/\psi$ mass
  (open histogram),  and for events for which this mass value is 
50--100~\mevcc less than nominal (hatched).
  }
\label{jpsi_kk2}
\end{figure}
\begin{figure}[tbh]
\includegraphics[width=0.9\linewidth]{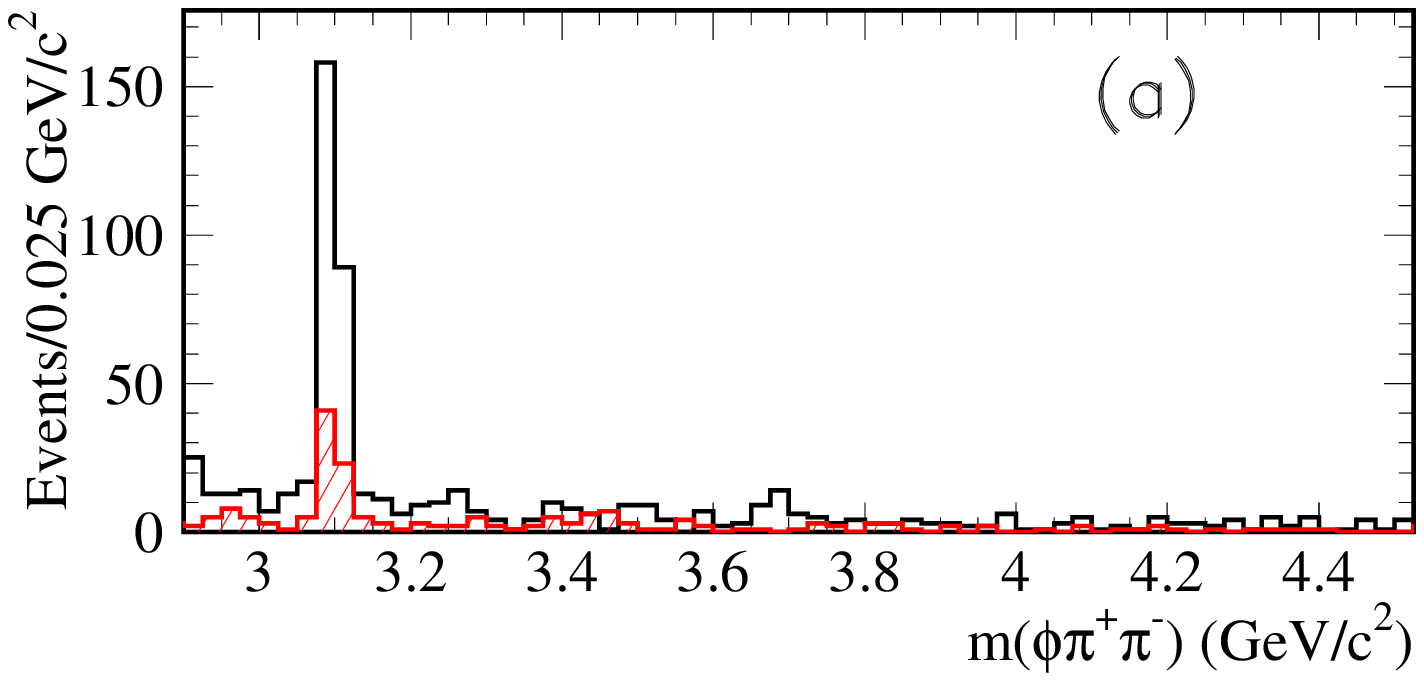}
\label{jjj}
\includegraphics[width=0.9\linewidth]{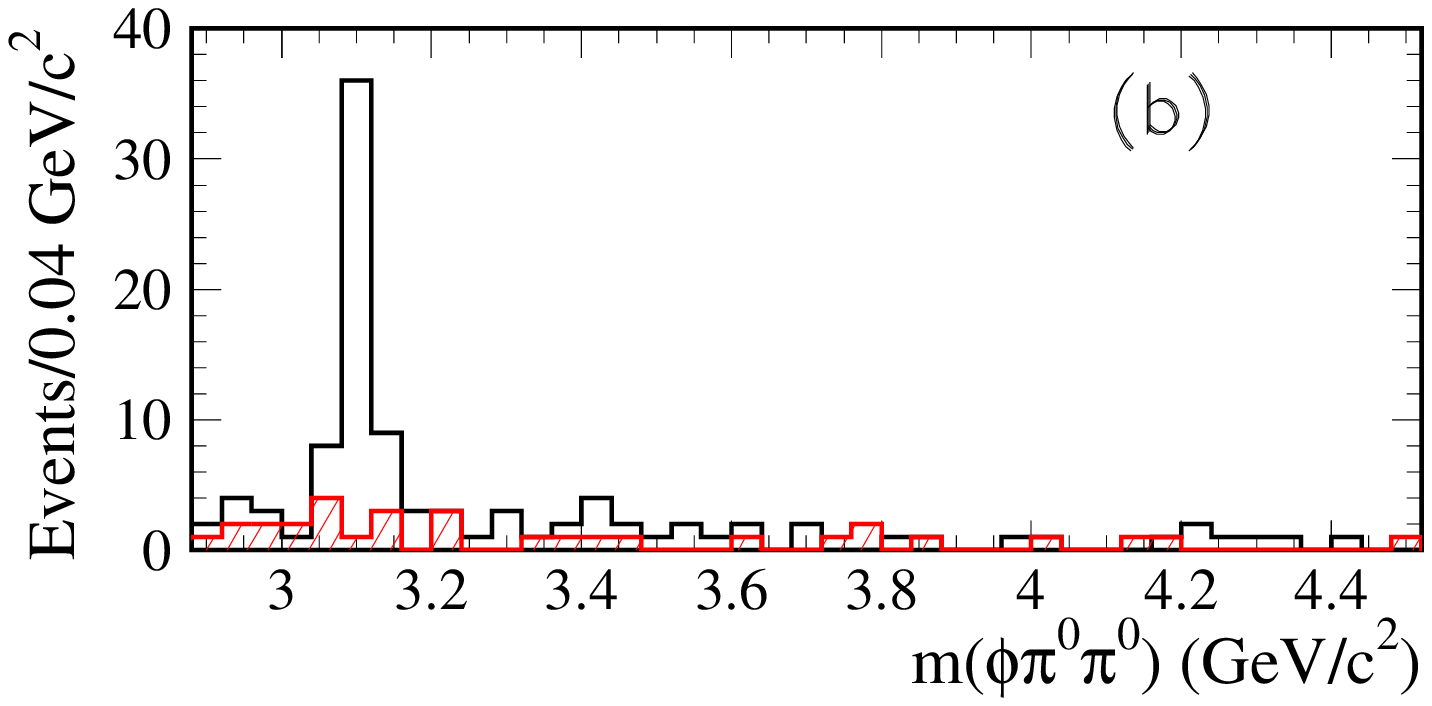}
\vspace{-0.3cm}
\caption{
Raw invariant mass distributions in the charmonium region for
 (a) candidate $\epem \!\!\to\! \phi\pipi$ events (open histogram), and for
  events in the $\phi$ sideband regions of Fig.~\ref{phif0_sel}(c)
  (hatched); (b) candidate $\epem \!\!\to\! \phi\ppz$ events (open histogram) and
  events in the \chiKKppnt control region (hatched).
 }
\label{jpsi_phi2pi}
\end{figure}

The fitted  $J/\psi$ signals for the  \KKppch, \KKppnt, and \KKKK final
states are found to contain
$3137\pm67$,
 $388\pm28$, and
 $287\pm24$ events, respectively.
From the number of events in each final state $f$,
$N_{J/\psi \!\to\! f}$, 
we calculate the product of the $J/\psi$ branching fraction to $f$ and
the $J/\psi$ electronic width using 
\begin{equation}
     \BR_{J/\psi \!\to\! f} \cdot \Gamma^{J/\psi}_{ee}  =
 \frac{N_{J/\psi \!\to\! f} \cdot m_{J/\psi}^2}
      {6\pi^2 \cdot d{\cal L}/dE \cdot \epsilon_f(m_{J/\psi}) \cdot C}
 ~~~, \\
\label{jpsicalc}
\end{equation}
where
$d{\cal L}/dE = 173.1\pm1.7~\invnb/\mev$, and $\epsilon_f(m_{J/\psi})$ 
are the ISR luminosity and corrected selection efficiency, respectively,
at the $J/\psi$ mass,
and $C$ is the conversion constant.  
We estimate $\epsilon_{\KKppch} \! =\! 0.198\pm0.006$, 
$\epsilon_{\KKppnt} \! =\! 0.079\pm0.004$, and $\epsilon_{\KKKK} \!
=\! 0.173\pm0.012$ using the corrections and errors discussed 
in Secs.~\ref{sec:xs2k2pi}, \ref{sec:2k2pi0xs}, and~\ref{sec:4kxs}.

We list the values of the product of the branching fraction(s) and
$\Gamma^{J/\psi}_{ee}$ in Table~\ref{jpsitab}, and using 
$\Gamma^{J/\psi}_{ee}\! =\! (5.55\pm0.14)~\kev$~\cite{PDG},
obtain the corresponding branching fraction values and list them
 together with their PDG values~\cite{PDG}.
The systematic uncertainties quoted include a 2.5\% uncertainty on $\Gamma^{J/\psi}_{ee}$.
Our measured branching fractions of \KKppch, \KKppnt, and \KKKK are
 more precise than the current PDG values,
which are dominated by our previous results  
((6.6$\pm$0.5)$\times$10$^{-3}$, (2.5$\pm$0.3)$\times$10$^{-3}$  and 
 (7.6$\pm$0.9)$\times$10$^{-4}$, respectively~\cite{isr2k2pi}).

These fits also yield 133$\pm$21 \KKppch events, 
17$\pm$9 \KKppnt events and 13$\pm$6 \KKKK events 
in the $\psi(2S)$ peak.
We expect 12 events from
$\psi(2S) \!\to\! J/\psi\pipi \!\!\to\! \KKppch$
from the relevant branching fractions~\cite{PDG},
which is less than the statistical error.
Subtracting this contribution and using the calculation analogous 
to Eq.(\ref{jpsicalc}), with $d{\cal L}/dE\! =\! 221.2\pm2.2~\invnb/\mev$,
we obtain the product of the branching fraction and electronic width for the decays
$\psi(2S) \!\!\to\! \KKppch$, $\psi(2S) \!\!\to\! \KKppnt$, and
$\psi(2S) \!\!\to\! \KKKK$.
Dividing by $\Gamma^{\psi(2S)}_{ee}=2.36\pm0.04$~\kev~\cite{PDG},
we obtain the branching fractions listed in Table~\ref{jpsitab}.
The \KKppch and \KKKK values are consistent with those in 
Ref.~\cite{PDG}. There is no entry in Ref.~\cite{PDG}
for the \KKppnt decay mode of the $\psi(2S)$.

As noted in Sec.~\ref{sec:kaons} and shown in Fig.~\ref{kkstar} and
Fig.~\ref{kstar_sel},  
the \KKppch final state is dominated by the $K^{*}(892)^{0} K^- \pi^+$ channels,
with a small contribution from the $K^{*}(892)^0 \Kbar_2^{*}(1430)^0$
channels.
Figure~\ref{jpsi_kkstarvs2k2pi} shows a  plot of the invariant
mass of a $K^\pm\pi^\mp$ pair versus that of the \KKppch system for events with
the mass of the $K^\mp\pi^\pm$ pair near the $K^{*}(892)^0$ mass,
i.e., within the bands in Fig.~\ref{kkstar}(a), but with only one
combination plotted in the overlap region.
There is a large concentration of entries in the $J/\psi$ band with 
$K^\pm\pi^\mp$ mass values near 1.43~\gevcc,
but a relatively small number of events in a horizontal band corresponding to
the $K_2^{*}(1430)^0$ production outside the $J/\psi$ region.  
We show the $K^\pm\pi^\mp$ mass projection for the subset of events
with \KKppch mass within 50~\mevcc of the nominal $J/\psi$ mass 
in Fig.~\ref{jpsi_kk2} as the open histogram.  
The hatched histogram is the projection for events with a \KKppch mass
between 50 and 100~\mevcc away from the nominal $J/\psi$ mass.

\begin{figure}[tbh]
\includegraphics[width=0.9\linewidth]{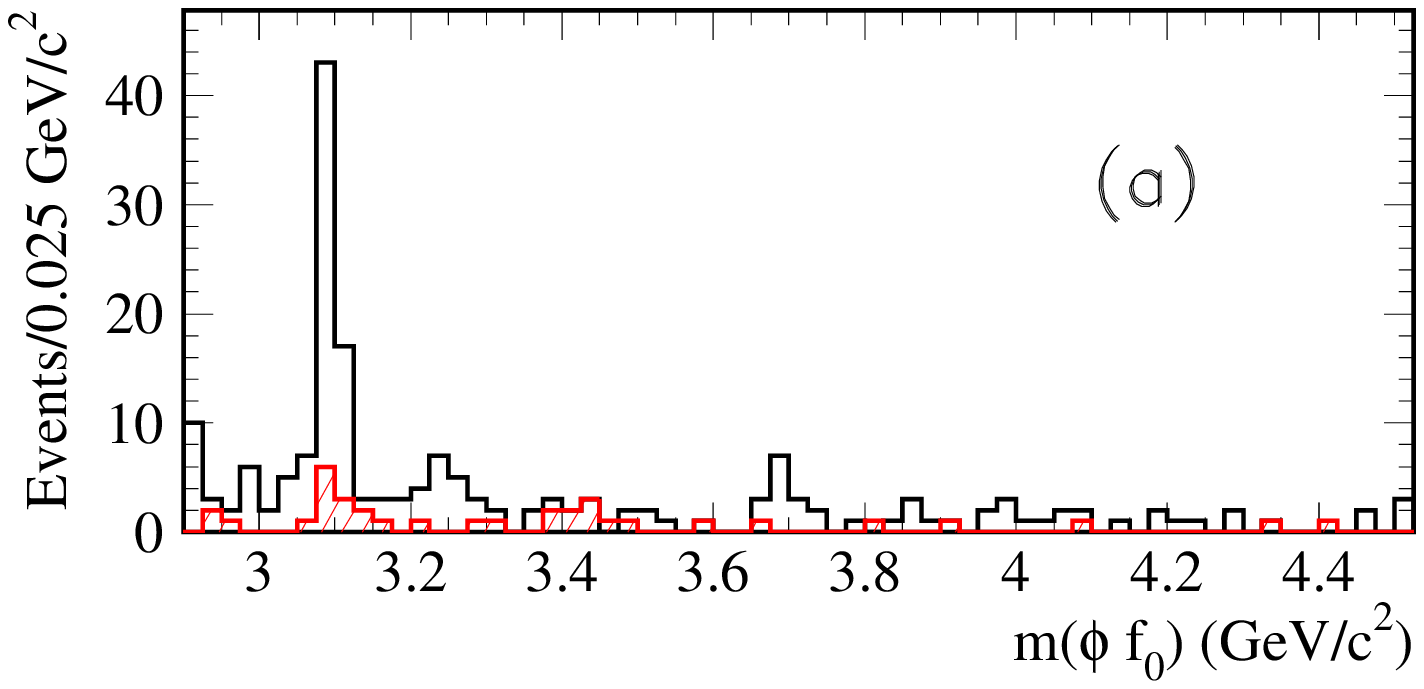}
\label{jjjj}
\includegraphics[width=0.9\linewidth]{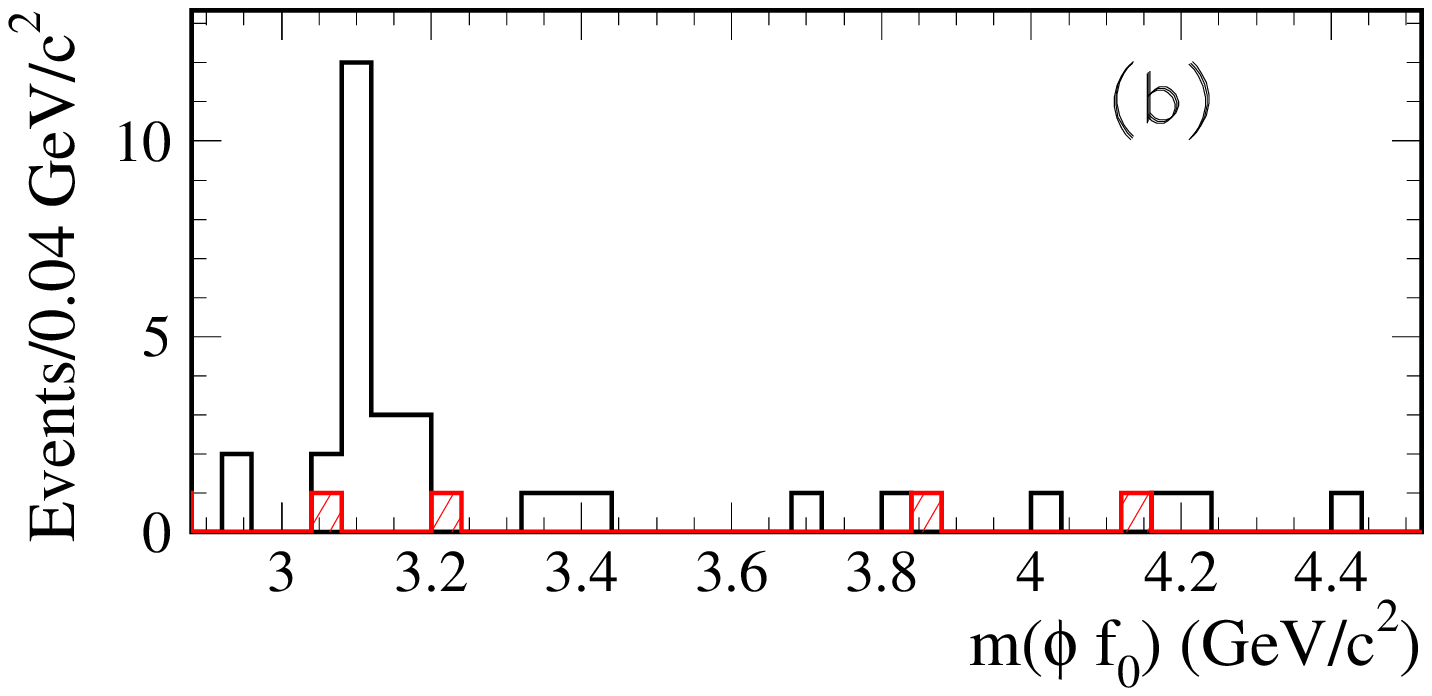}
\vspace{-0.3cm}
\caption{
  Raw invariant mass distribution  in the charmonium region (a)
for candidate $\phi f_0$,
  $f_0 \!\!\to\! \pipi$ events (open histogram), and for
  events in the $\phi$ sideband region (hatched), and
(b) for candidate $\phi f_0$,
  $f_0 \!\!\to\! \ppz$ events (open histogram) and for
  events in the \chiKKppnt control region (hatched).
  }
\label{jpsi_phif0}
\end{figure}
\begin{figure}[tbh]
\includegraphics[width=0.9\linewidth]{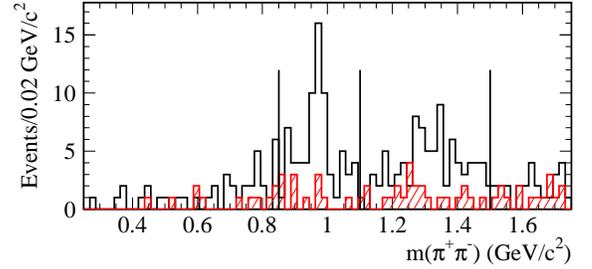}
\vspace{-0.3cm}
\caption{
  The $\pipi$ invariant mass distribution for $\phi \pipi$ events from
  the $J/\psi$ peak of Fig.~\ref{jpsi_phi2pi}(a) (open histogram), and
  for   events in the $\phi$ sideband region (hatched).
  }
\label{jpsi_phipipi}
\end{figure}
\begin{table*}[tbh]
\caption{
  Summary of the $J/\psi$ and $\psi(2S)$ parameters
  obtained in this analysis.
  }
\label{jpsitab}
\begin{ruledtabular}
\begin{tabular}{r@{$\cdot$}l  r@{.}l@{$\pm$}l@{$\pm$}l 
                              r@{.}l@{$\pm$}l@{$\pm$}l
                              r@{.}l@{$\pm$}l } 
\multicolumn{2}{c}{Measured} & \multicolumn{4}{c}{Measured}    &  
\multicolumn{7}{c}{$J/\psi$ or $\psi(2S)$ Branching Fraction (10$^{-3}$)}\\
\multicolumn{2}{c}{Quantity} & \multicolumn{4}{c}{Value (\ev)} &
\multicolumn{4}{c}{This work}    & 
\multicolumn{3}{c}{PDG2010} \\
\hline
$\Gamma^{J/\psi}_{ee}$  &  $\BR_{J/\psi  \to \KKppch}$  &
  37&94&0.81& 1.10  &   6&84 & 0.15 & 0.27  &    6&6 & 0.5  \\

$\Gamma^{J/\psi}_{ee}$  &  $\BR_{J/\psi  \to \KKppnt}$  &
  11&75&0.81& 0.90  &   2&12 & 0.15 & 0.18  &   2&45 & 0.31 \\

$\Gamma^{J/\psi}_{ee}$  &  $\BR_{J/\psi  \to \KKKK}  $  &
   4&00& 0.33& 0.29 &   0&72 & 0.06 & 0.05  &   0&76 & 0.09 \\[0.4cm]

$\Gamma^{J/\psi}_{ee}$  &  $\BR_{J/\psi\to K^{*0}\Kbar_{0,2}^{*0}}
                      \cdot \BR_{K^{*0}\to K^+\pi^-}
                      \cdot \BR_{\Kbar_{0,2}^{*0}\to K^-\pi^+}   $  & 
   8&59& 0.36 & 0.27  &   6&98  & 0.29  & 0.21   &   6&0  & 0.6 \\

$\Gamma^{J/\psi}_{ee} $ & $\BR_{J/\psi\to K^{*0}\Kbar^{*0}}
                        \cdot\BR_{K^{*0}\to K^+\pi^-} 
                         \cdot\BR_{\Kbar^{*0}\to K^-\pi^+}$ &
  0&57 & 0.15 & 0.03 &  0&23 & 0.06 & 0.01      &  0&23 & 0.07  \\

$\Gamma^{J/\psi}_{ee}$  &  $\BR_{J/\psi  \to \phi\pipi}
                      \cdot \BR_{\phi    \to \Kp \Km}$  &
   2&19& 0.23& 0.07 &   0&81 & 0.08 & 0.03  &   0&94 & 0.09   \\

$\Gamma^{J/\psi}_{ee}$  &  $\BR_{J/\psi  \to \phi\ppz} 
                      \cdot \BR_{\phi    \to \Kp \Km}$  &
   1&36& 0.27& 0.07 &   0&50 & 0.10 & 0.03  &   0&56 & 0.16   \\

$\Gamma^{J/\psi}_{ee}$  &  $\BR_{J/\psi  \to \phi\Kp\Km} 
                      \cdot \BR_{\phi    \to \Kp \Km}$  &
   2&26& 0.26& 0.16 &   1&66 & 0.19 & 0.12  &   1&83 & 0.24
\footnote{$\phi$ is selected as $|m_{\phi}-m(\Kp\Km)|<10$~\mev, $\BR_{J/\psi\to\phi\Kbar K}$ obtained as    $2\cdot\BR_{J/\psi\to\phi\Kp\Km}$.}\\

$\Gamma^{J/\psi}_{ee}$  &  $\BR_{J/\psi  \to \phi f_0}
                      \cdot \BR_{  \phi  \to \Kp\Km}
                      \cdot \BR_{  f_0   \to \pipi}  $  &
   0&69& 0.11& 0.05 &   0&25 & 0.04 & 0.02  &   0&18 & 0.04
\footnote{Not corrected for the $f_0\to\ppz$ mode. $f_0$ selected by $0.85<m(\ppz)<1.1$~\gevcc}  \\

$\Gamma^{J/\psi}_{ee}$  &  $\BR_{J/\psi  \to \phi f_0}
                      \cdot \BR_{  \phi  \to \Kp\Km}
                      \cdot \BR_{  f_0   \to \ppz}   $  &
   0&48& 0.12& 0.05 &   0&18 & 0.04&  0.02  &   0&17 & 0.07
\footnote{Not corrected for the $f_0\to\pipi$ mode. $f_0$ selected by $0.85<m(\pipi)<1.1$~\gevcc } \\

$\Gamma^{J/\psi}_{ee}$  &  $\BR_{J/\psi  \to \phi f_x}
                      \cdot \BR_{  \phi  \to K^+K^-}
                      \cdot \BR_{  f_x   \to \pipi}  $  &
   0&74& 0.12& 0.05 &   0&27 & 0.04 & 0.02  &   0&72 & 0.13
\footnote{We compare our $\phi f_x, f_x\to\pipi$ mode, selected by
  $1.1<m(\pipi)<1.5$ with  $\phi f_2(1270)$.}  \\[0.4cm]

$\Gamma^{\psi(2S)}_{ee}$  &  $\BR_{\psi(2S) \to \KKppch} $  &
   1&92& 0.30& 0.06 &   0&81  & 0.13  & 0.03   &   0&75 & 0.09  \\

$\Gamma^{\psi(2S)}_{ee}$  &  $\BR_{\psi(2S) \to \KKppnt} $  &
   0&60& 0.31& 0.03 &   0&25  & 0.13  & 0.02   &   \multicolumn{3}{c}{no entry}   \\

$\Gamma^{\psi(2S)}_{ee}$  &  $\BR_{\psi(2S) \to \KKKK} $  &
   0&22& 0.10& 0.02 &   0&09  & 0.04  & 0.01   &   0&060 & 0.014 \\ 

$\Gamma^{\psi(2S)}_{ee}$  &  $\BR_{\psi(2S) \to \phi\pipi}
                        \cdot \BR_{\phi     \to \Kp \Km} $  &
   0&27& 0.09& 0.02 &   0&23 & 0.08 & 0.01  &   0&117& 0.029   \\ 

$\Gamma^{\psi(2S)}_{ee}$  &  $\BR_{\psi(2S) \to \phi f_0}
                        \cdot \BR_{\phi     \to K^+K^-}
                        \cdot \BR_{ f_0     \to \pipi}   $  &
  \hspace*{0.5cm}    0&17  & 0.06  & 0.02 \hspace{0.6cm} &
  \hspace*{0.5cm}    0&15  & 0.05  & 0.01 \hspace{0.6cm} &
  \hspace*{0.5cm}    0&068 & 0.024
\footnote{ $\BR_{\psi(2S) \to \phi f_0}, f_0\to \pipi$ }         \hspace{0.6cm} \\

\end{tabular}
\end{ruledtabular}
\end{table*}
The $K \pi$ distribution from the $J/\psi$ is dominated by the
$K_2^{*}(1430)^0$ and $K_0^{*}(1430)^0$
 signals~\cite{PDG,beskstar}.
A small signal at the $K^{*}(892)^0$ 
indicates the presence of $K^{*}(892)^0\bar K^{*}(892)^0$ 
decay of the $J/\psi$; this is also seen
as an enhancement in the  $J/\psi$ band in Fig.~\ref{jpsi_kkstarvs2k2pi}. 
The enhancement at 1.9~\gevcc of Fig.~\ref{jpsi_kk2} 
may be due to the $^3 F_2$ ground state, or to the first 
radial excitation of the $K^*_2(1430)$, neither of 
which has been reported previously.
Subtracting the number of sideband events from the number in the
$J/\psi$ mass window,  
we obtain 710$\pm$30 events with $K^\pm\pi^\mp$ mass in the range 
1.2--1.7~\gevcc,
which we take as a measure of $J/\psi$ decay into
$K^{*}(892)^0 \Kbar_{0,2}^{*}(1430)^0$.
According to Ref.~\cite{beskstar},  there
  is an equal contribution from $K_0^{*}(1430)^0$ and $K_2^{*}(1430)^0$, which we cannot
  separate with our selection.
We obtain $47 \pm 12$ events in the
0.8--1.0~\gevcc window for 
$K^{*}(892)^0\bar K^{*}(892)^0$ decay, and
$185 \pm 21$ events for decay to $K^{*}(892)^0 K^- \pi^+$  
with $m(K\pi)$ in the 1.7--2.0~\gevcc region.
We convert these to branching fractions using Eq.(\ref{jpsicalc}), and 
divide by the known branching fractions of the $K^*$ states~\cite{PDG}.
The results are listed in Table~\ref{jpsitab}, which
are more precise than those in Ref.~\cite{PDG}.
For the 1.7--2.0~\gevcc mass region we obtain 
$\Gamma^{J/\psi}_{ee} B_{J/\psi\to K^{*}(892)^0 K^- \pi^+} = 
(2.24 \pm 0.25 \pm 0.15)\ev$.

We study decays into $\phi\pipi$ and  $\phi\ppz$ using the
mass distributions shown in Figs.~\ref{jpsi_phi2pi}(a),(b).
The open histograms are for events with $\Kp\Km$ mass within the
$\phi$ bands of Figs.~\ref{phif0_sel}(c) and~\ref{phif0_sel2}(c).
The hatched histogram in Fig.~\ref{jpsi_phi2pi}(a) is from the
$\phi$ sidebands of Fig.~\ref{phif0_sel}(c), and represents the
dominant background in the $\phi\pipi$ mode.
The hatched histogram in Fig.~\ref{jpsi_phi2pi}(b) is from the
\chiKKppnt control region, and represents the dominant background in
the $\phi\ppz$ mode.
Subtracting these backgrounds, and subtracting a small remaining
background using $J/\psi$ or $\psi(2S)$ sideband events,
we find 181$\pm$19 $J/\psi \!\to\! \phi\pipi$ events,
45$\pm$9 $J/\psi \!\to\! \phi\ppz$ events,
and 19$\pm$6 $\psi(2S) \!\to\! \phi\pipi$ events.
We convert these to branching fractions and,
after correcting for the modes other than $\phi\to\Kp\Km$,
 list them in Table~\ref{jpsitab}.
All are consistent with current PDG values,
of which the first two are dominated 
by our previous measurement.

We do not observe any evidence for  $Y(4260)$ decays to these modes, nor
do we see a $Y(4260)$ signal in any other mode studied here.

Figures ~\ref{jpsi_phif0}(a)(b)  show the
corresponding mass distributions for $\phi f_0(980)$ events,
i.e., the subsets of the events in Figs.~\ref{jpsi_phi2pi}(a) 
and~\ref{jpsi_phi2pi}(b) with a di-pion mass in the range 0.85--1.10~\gevcc.
Signals at the $J/\psi$ mass are visible in both cases.
From Fig.~\ref{jpsi_phif0}(b) we estimate 
$16 \pm 4$ $\phi f_0$ events in the $\ppz$ mode. 
 However, $\phi f_0(980)$ is not the dominant mode 
contributing to  $J/\psi \!\to\! \phi\pipi$ decay.
The open histogram of
Fig.~\ref{jpsi_phipipi} shows the $\pipi$ invariant mass
distribution for events in the $J/\psi$ peak of 
Fig.~\ref{jpsi_phi2pi}(a)
($|m(\KKppch)-m(J/\psi)| <0.05$~\gevcc); events in the
$J/\psi$ sidebands ($0.05<|m(\KKppch)-m(J/\psi)| <0.1$~\gevcc) 
are shown by the hatched histogram.
A two-peak structure is visible that is very similar to that 
studied by the BES Collaboration~\cite{bes4k} and
observed in $D^+_s\to\pipi\pi^+$ decay~\cite{D+s}. In both cases
the $\pipi$ system is believed to couple to an $s\bar s$ system;
both $\pipi$ distributions exhibit a clear $f_0(980)$ peak and a broad
bump in the 1.3-1.5~\gevcc region. The analysis of 
Refs.~\cite{bes4k, D+s}
shows that this bump is made up of $f_2(1270)$ and $f_0(1370)$
contributions; we denote this region by $f_x$.
By selecting  $f_0(980)$ in the  0.85--1.10~\gevcc range and
$f_x$ in the 1.1--1.5~\gevcc range, shown by vertical lines in
Fig.~\ref{jpsi_phipipi},  and subtracting $J/\psi$ sideband background  
we find $57\pm 9$ $J/\psi \!\to\! \phi f_0(980)$ events and
$61\pm 10$ $J/\psi \!\to\! \phi f_x$ events.

Using Eq.(\ref{jpsicalc}) and dividing by the appropriate branching
fractions, 
we obtain the $J/\psi$ branching fractions listed in Table~\ref{jpsitab}.
The measurements of $\BR_{J/\psi\to\phi f_0}$ in the \pipi and \ppz 
decay modes of the $f_0$ are consistent with each other and with the
PDG value, and combined they have roughly the same precision as given
in Ref.\cite{PDG}.

Note that, in contrast to $\phi(1680)\to\phi\pi\pi$ decay, there is
no indication of a $J/\psi\to\phi f_0(600)$ decay mode. Only
$J/\psi\to\phi f_0(980)$ is observed, as is true for the $Y(2175)$ state. 

We also observe $12\pm 4$ $\psi(2S) \!\to\! \phi f_0$, $f_0 \!\!\to\! \pipi$ 
events, which we convert to the branching fraction listed in 
Table~\ref{jpsitab}; 
it is consistent with the value in Ref.~\cite{PDG}, assuming
$\BR_{f_0\to\pipi} = 2/3$.  

The hatched histogram in Fig.~\ref{mkk_notphi}(a)  
shows the $K^+ K^-$ invariant mass distribution,
when the other kaon pair is in the $\phi$ region, 
for the \KKKK events in the $J/\psi$ peak, selected by requiring
$|m(\KKKK)-m(J/\psi)| < 0.05$~\gevcc. Subtracting  sideband events we
find 163 $\pm$ 19 events corresponding to  
$J/\psi\to\phi K^+ K^-$ decay.
Using our normalization we
obtain the branching fraction listed in Table~\ref{jpsitab}, 
which  agrees with that in Ref.~\cite{PDG} but
has better precision.
In obtaining these values, we 
have used
$B(\phi\to K^+ K^-)=0.489$~\cite{PDG}, and assume equal 
rates for $J/\psi\to\phi K^+ K^-$ and $J/\psi\to\phi K^0 \bar K^0$.


\section{Summary}
\label{sec:Summary}
\noindent

We use the excellent charged-particle tracking, track identification, and
photon detection of the \babar\ detector to fully reconstruct events
of the type 
$\epem \!\!\to\! \gamma\epem \!\!\to\! \gamma\KKppch$,  
$\gamma\KKppnt$, and  
$\gamma\KKKK$,  
where the $\gamma$ is radiated from the initial state $e^+$ or $e^-$.
Such events are equivalent to direct \epem annihilation at a
c.m.\@ energy corresponding to the mass of the hadronic system.
Consequently, we are able to use the full ~\babar~ dataset to
study annihilation into these three final states 
from their respective production thresholds up to 5~\gev c.m.~energy.
The \KKppch, \KKppnt and \KKKK measurements are consistent with, and supersede,
our previous results~\cite{isr2k2pi}.

The systematic uncertainties on the 
$\epem \!\!\to\! \KKppch$, \KKppnt and \KKKK cross section values are 
4\%, 7\% and 9\%, respectively, for $\Ecm \! <\! 3$~\gev, 
and increase, respectively, to 11\%, 16\% and 13\% in the 3--5~\gev range.
The values
obtained are considerably more precise than previous measurements, and 
cover this low-energy range completely. As such they
provide useful input to calculations of the hadronic corrections to
the anomalous magnetic moment of the muon, and of the fine structure
constant at the $Z^0$ mass.

These final states exhibit complex resonant substructures. 
For the \KKppch final state we measure the cross sections for
the specific channels
$\epem \!\!\to\! K^{*}(892)^0\Km\pip$, $\phi\pipi$, and $\phi f_0$,
and, for the first time, for the $\epem \!\!\to\! K_2^{*}(1430)^0\Km\pip$ and
$\epem \!\!\to\rho(770)^0\Kp\Km$ reactions.
We also observe signals for the $K_1(1270)$, $K_1(1400)$, and $f_2(1270)$ resonances.
It is difficult to disentangle these contributions to the final state,
and we make no attempt to do so in this paper. 
We note that 
the $\rho^0$ signal is consistent with being due entirely to $K_1$ decays,
and that while
the total cross section is dominated by the $K^{*}(892)^0\Km\pip$ channels,
only about 1\% of the events  correspond to the 
$\epem \!\!\to\! K^{*}(892)^0\Kbar^{*}(892)^0$ reaction.

For the \KKppnt final state we measure the cross section for 
$\epem \!\!\to\! \phi f_0$, 
and observe signals for the $K^{*}(892)^{\pm}$ and $K_2^{*}(1430)^{\pm}$ 
resonances.
Again, the total cross section is dominated by the
$K^{*}(892)^+ K^-\piz$ channel, but about 30\% of events 
are produced in the
$\epem \!\!\to\! K^{*}(892)^+ K^{*}(892)^-$ reaction.
For the \KKppnt final state 
we note that the cross section is roughly a factor of four smaller 
than that for \KKppch  over most of the \Ecm range,
consistent with  a factor of two isospin
suppression of the \ppz final state and another factor of two for the
relative branching fractions of the neutral and charged $K^*$ to
charged kaons.

With the larger data sample of the present analysis, 
we perform a more detailed study of 
the $\epem\to\phi(1020)\pi\pi$ reaction. The $\pipi$ and $\ppz$ 
invariant mass distributions
both show a clear $f_0(980)$ signal,
and a broad structure at lower mass interpreted as the
$f_0(600)$. We obtain parameter values for these resonances. 
The $\phi\pipi$ cross section measured in the \KKppch final state shows 
a structure around 1.7~\gev and some structures above 2.0~\gev.
The corresponding $\phi\ppz$ cross section in the \KKppnt 
final state shows similar behavior.
If the $f_0(980)$ is excluded from
the di-pion mass distribution, no structures above 2.0~\gev are seen. We fit
the observed cross section with the VMD model assuming $\phi(1680)\to\phi
f_0(600)$ and $\phi(1680)\to\phi f_0(980)$ decay; the latter appears
to be responsible
for the threshold increase of the cross section at 2.0~\gev.
Confirming our previous study~\cite{isr2k2pi}, our data require
an additional resonance at 2.175~\gev, which we call the $Y(2175)$, with decay to
$\phi f_0(980)$, but not to $\phi f_0(600)$.
Further investigation reveals consistent results for the \KKKK
final state,
and clear $Y(2175)$ signals in the $\Kp\Km f_0(980)$ channels, with 
$f_0(980) \!\!\to\! \pipi$ and \ppz.
This structure can be interpreted as a strange partner 
(with $c$-quarks replaced by $s$-quarks)
of the $Y(4260)$~\cite{y4260}, which has the analogous decay mode
$J/\psi\pipi$,
or perhaps as an $\ssbar\ssbar$ state that decays predominantly to $\phi f_0$.

In the \KKKK mode we find 
$\epem \!\!\to\! \phi\Kp\Km$ to be the dominant channel.
With the current data sample we can say little about other $\Kp\Km$
combinations. We observe an enhancement near threshold,
consistent with the $\phi f_0$ channel and if these events are
selected we have an indication of a $Y(2175)$ signal.  
Two other structures in the $\Kp\Km$ invariant mass spectrum are seen:
the smaller could be an indication of the $\phi f_0(1370)$ final
state, and the  larger
of the $\phi f_2'(1525)$ mode. If events corresponding to
the $\phi f_2'(1525)$ final state are selected, the \KKKK cross section
shows a resonance-like structure around 2.7~\gev, and a strong $J/\psi$ signal,
which has been studied in detail by the BES Collaboration~\cite{bes4k}.
In the \KKKK cross section we observe  a sharp peak at ~2.3~\gev, which
corresponds to the $\phi \Kp\Km$ channel with the
$\Kp\Km$ invariant mass in the 1.06--1.2~\gevcc
region.

We also investigate charmonium decays into the 
studied final states and through corresponding intermediate channels,
and measure the product of the electron 
width and the corresponding branching fraction. 
Some of the obtained $J/\psi$ branching fractions
listed in Table~\ref{jpsitab} are as
precise as, or more precise than, the current world averages, many of which
were obtained in our previous study~\cite{isr2k2pi}; the latter are superseded
by our new results.
We do not observe the $Y(4260)$ in any of the final states examined.
\section*{ACKNOWLEGMENTS}
\label{sec:Acknowledgments}
We are grateful for the 
extraordinary contributions of our \pep2\ colleagues in
achieving the excellent luminosity and machine conditions
that have made this work possible.
The success of this project also relies critically on the 
expertise and dedication of the computing organizations that 
support \babar.
The collaborating institutions wish to thank 
SLAC for its support and the kind hospitality extended to them. 
This work is supported by the
US Department of Energy
and National Science Foundation, the
Natural Sciences and Engineering Research Council (Canada),
Institute of High Energy Physics (China), the
Commissariat \`a l'Energie Atomique and
Institut National de Physique Nucl\'eaire et de Physique des Particules
(France), the
Bundesministerium f\"ur Bildung und Forschung and
Deutsche Forschungsgemeinschaft
(Germany), the
Istituto Nazionale di Fisica Nucleare (Italy),
the Foundation for Fundamental Research on Matter (The Netherlands),
the Research Council of Norway, the
Ministry of Science and Technology of the Russian Federation, and the
Particle Physics and Astronomy Research Council (United Kingdom). 
Individuals have received support from 
<<<<<<< acknowledgements.tex
CONACyT (Mexico), the Marie-Curie Intra European Fellowship program (European Union),
the A. P. Sloan Foundation, 
the Research Corporation,
and the Alexander von Humboldt Foundation.
=======
the Marie-Curie IEF program (European Union) and the A. P. Sloan Foundation (USA).

>>>>>>> 1.24

%

\end{document}